\documentclass[twocolumn]{aastex61}
\usepackage{amssymb}
\usepackage{amsmath}
\usepackage{color}
\usepackage{textcomp}
\usepackage{wasysym}

\submitjournal{ApJ}

\shorttitle{[oiii] emitters at z\ $\sim$\ 0.5}
\shortauthors{Li et al.}

\begin{document}

\title{Extreme [OIII] Emitters at z\ $\sim$\ 0.5}

\author{Zhihui Li}
\affiliation{Kavli Institute for Astronomy and Astrophysics, Peking University, Beijing 100871, China}
\affiliation{Department of Astronomy, School of Physics, Peking University, Beijing 100871, China}
\email{$^{1, 2}$zhihuili.astro@gmail.com}

\author{Matthew A. Malkan}
\affiliation{Physics and Astronomy Department, University of California, Los Angeles, CA 90095-1547}
\email{$^3$malkan@astro.ucla.edu}



\begin{abstract}

We have found a sample of extreme emission-line galaxies (EELGs) with strong [OIII]$\lambda$5007 emission at z $\sim$ 0.5. Using broadband photometric selection and requiring small uncertainties in photometry, we searched the 14th Data Release (DR14) of the Sloan Digital Sky Survey (SDSS) and found 2658 candidates with strong \emph{i}-band excess (\emph{i} - \emph{z} $\leqslant$ \emph{r} - \emph{i} - 0.7). We further obtained 649 SDSS spectra of these objects, and visually identified 22 [OIII] emitters lying at 0.40 $<$ z $<$ 0.63. Having constructed their UV-IR spectral energy distributions (SED) we found that they have fairly blue \emph{r}-W2 and red W1-W4 colors, indicative of strong warm dust emission. Their rest-frame [OIII]$\lambda$5007 equivalent widths are mostly 200-600 \AA, and their high [OIII]$\lambda$5007/H$\beta$ ratios put them at the boundary of star-forming galaxies and AGNs on line ratio classification diagrams. The typical \emph{E(B--V)} and electron temperature of [OIII] emitters are $\sim$ 0.1-0.3 mag and $\sim$ 10$^4$ K, respectively. The lowest metallicity of our [OIII] emitters with S/N$_{\rm [OIII]\lambda4363}$ $>$ 3 is 12 + log(O/H) = 7.98$^{+0.12}_{-0.02}$, with a median value of 8.24$^{+0.05}_{-0.04}$. Our [OIII] emitters exhibit remarkably high line luminosity -- 18/22 have L$_{\rm [OIII]\lambda5007}$ $>$ 5$\times$10$^{42}$ erg $\rm s^{-1}$ and 5/22 have L$_{\rm [OIII]\lambda5007}$ $>$ 10$^{43}$ erg $\rm s^{-1}$. Their estimated volume number density at z $\sim$ 0.5 is $\sim$ 2$\times$10$^{-8}$ Mpc$^{-3}$, with L$_{\rm [OIII]\lambda5007}$ down to $\sim$ 3$\times$10$^{42}$\ erg $\rm s^{-1}$. The cumulative number fraction distribution of EELGs across different redshifts is indicative of a strong redshift evolution at the bright end of the [OIII] luminosity function.

\end{abstract}




\section{Introduction} \label{sec:intro}

Extreme emission-line galaxies (EELGs) are dwarf galaxies undergoing vigorous star formation characterized by very strong emission lines such as H$\alpha$ and [OIII]. Lying above the typical `galaxy main sequence' (SFR-\emph M$_{\star}$ relation, \citealt{2007ApJ...660L..43N}), these relatively `primitive' objects provide complementary opportunities for studying both the early epochs of star formation, and chemical evolution across cosmic history. Although EELGs are rare in the local universe \citep{2009MNRAS.399.1191C}, 
they could have been far more important, or even dominant, in the younger universe. To illustrate, recent slitless spectroscopy surveys (e.g. the WISP survey, \citealt{2010ApJ...723..104A}) in the near-infrared have shown that the number density of EELGs with high rest-frame EW $>$ 200 \AA\ could increase by a factor of ten or more at z $\sim$ 1.5 (\citealt{2011ApJ...743..121A}, \citeyear{2014ApJ...789...96A}); \citet{2017ApJ...850....5M} concluded that even typical star-forming galaxies at z $\sim$ 3 show extremely strong [OIII] emission lines with rest-frame equivalent widths (EW) of 500-900 \AA. Moreover, the characteristic luminosity of EELGs (the `knee' in their luminosity function, LF) also shifts to higher values at higher redshifts \citep{2007ApJ...657..738L}; the bright end of their LF is yet to be determined accurately. 

A comprehensive survey of EELGs, which requires searching over a very wide sky area with well-defined criteria, is observationally challenging. These objects are fairly compact and faint in all broad bands (such as the infrared (IR) or the X-rays). Their stellar populations are very young and blue and they do not exhibit obviously different optical colors from quasars. The largest population of EELGs discovered so far is by deep narrow-band filter imaging \citep{2007ApJ...657..738L} but very large areas of sky have not yet been accessible to large-format cameras with such narrow-band imaging filters. This leaves us with the only practical method available \textemdash\ the combination of multiple broadband optical filter photometry.

In this work, we explore the properties of EELGs at z $\sim$ 0.5, which are basically the counterparts of `Green Peas' \citep{2009MNRAS.399.1191C} at a higher redshift. It is worth noting the many additional advantages of increasing the sample of bright confirmed EELGs at z $\sim$ 0.5. For example, our objects are particularly valuable for various follow-up observations, in that the brightest EELGs at this redshift can be spectroscopically observed in the rest-frame far-ultraviolet (UV), free from the absorption of intergalactic HI clouds. This may shed light on the Lyman leakage process, since we now have strong indications that the EELGs are the principal sources of ionizing photons \citep{2017ApJ...844..171Y} that escaped into the intergalactic medium and re-ionized it at z $\sim$ 6 or higher \citep{2013ApJ...766...91J}.

This paper is organized as follows. In Section \ref{sec:sample}, we describe our photometric sample selection criteria quantitatively and present our results. We also discuss the contamination by other objects, such as H$\alpha$ emitters. In Section \ref{sec:data}, we present spectra and SEDs of our [OIII] sample. We also derive an average spectrum for our [OIII]$\lambda$5007 emitters. In Section \ref{sec:results} we explore both photometric and spectroscopic properties of our [OIII]$\lambda$5007 emitters. We calculate the expected volume number density of [OIII] emitters at z $\sim$\ 0.5 in Section \ref{sec:LF}. In Section \ref{sec:sum} we summarize our results. Throughout this paper we adopt a flat $\Lambda$CDM cosmology with $\Omega_{m}$ = 0.308, $\Omega_{\Lambda}$ = 0.692 and $H_{0}$ = 67.8 km s$^{-1}$ Mpc$^{-1}$ \citep[]{2016A&A...594A..13P}.

\section{Sample} \label{sec:sample}
\subsection{Photometric Selection Criteria} \label{sec:select}
Our parent sample is constructed from the \emph{ugriz} photometry in \texttt{PhotoPrimary} catalog of the SDSS 14th Data Release (DR14, \citealt{2017arXiv170709322A}) using the CasJobs search tool\footnote{https://skyserver.sdss.org/CasJobs}. As the [OIII]$\lambda$5007 line is shifted to \emph{i}-band at z $\sim$ 0.5, we expect our target EELGs to have strong \emph{i}-band excess. The key selection criterion is:
\begin{equation}\label{eq1}
\    \ \ \ i-z \leqslant r-i-0.7     \     \ \ \ \ \ \ \ \ \ \ \ 
\end{equation}
which is essentially requiring an \emph{i}-band excess compared to the \emph{r} and \emph{z} bands. We further require fairly blue colors for our objects, since starbursts have young stellar populations:
\begin{equation}\label{eq2}
\    \ \ \    u-g \leqslant 0.3      \     \ \ \ \ \ \ \ \ \ \ \ 
\end{equation}
\begin{equation}\label{eq3}
\    \ \ \    g-r \leqslant 0.45      \     \ \ \ \ \ \ \ \ \ \ \ 
\end{equation}
\begin{equation}\label{eq4}
\    \ \ \    r-z \leqslant 0.8      \     \ \ \ \ \ \ \ \ \ \ \
\end{equation}
We also expect the \emph{r} and \emph{z} magnitudes of these compact and distant objects to be faint:
\begin{equation}\label{eq5}
\    \ \ \    r, z \geqslant 18.5      \     \ \ \ \ \ \ \ \ \ \ \ 
\end{equation}
To ensure the high precision of our data, we also require a small uncertainty of our \emph{u}-band, \texttt{err\_u} $<$ 0.25 (as \emph{u}-band is usually the faintest among the five bands and carries relatively larger error bars) and \texttt{err\_g}, \texttt{err\_r}, \texttt{err\_i}, \texttt{err\_z} $<$ 0.15. We use model magnitudes for all five bands. For more details, we refer readers to Appendix \ref{sec:criterion}, where the full SQL queries are presented.

We present our detailed selection procedure in Table \ref{table:select}. As a result, we obtained 968 objects in total as our parent sample of [OIII] candidates with \emph{i}-band excess. Next we searched for SDSS spectra of these objects in DR14. We found 200 observations, in which 75 objects are classified as `galaxy' (based on image analysis) instead of `star' or `QSO' by the SDSS pipeline. With visual inspection of their spectra we identified 17 [OIII]$\lambda$5007 emitters, whose redshifts lie in the range 0.40 $<$ z $<$ 0.63. We have also tested searching in the \texttt{PhotoObjAll} catalog (where multiple photometry of a certain object could exist) and it yielded a list of 2658 unique objects, of which 649 have observed SDSS spectra. In this sample we identified five more [OIII] emitters. However, this sample is less `clean', in the sense that it contains more unreliable photometry (\texttt{clean} = 0) and is more contaminated by local stars and QSOs. We refer to these two samples as the `primary sample' and the `total sample' respectively hereafter, and the `primary sample' is a subset of the `total sample'. For the sake of comprehensiveness, in the following analysis we study the 22 [OIII] emitters altogether.

\begin{figure}[htbp]
\includegraphics[width=\linewidth, clip]{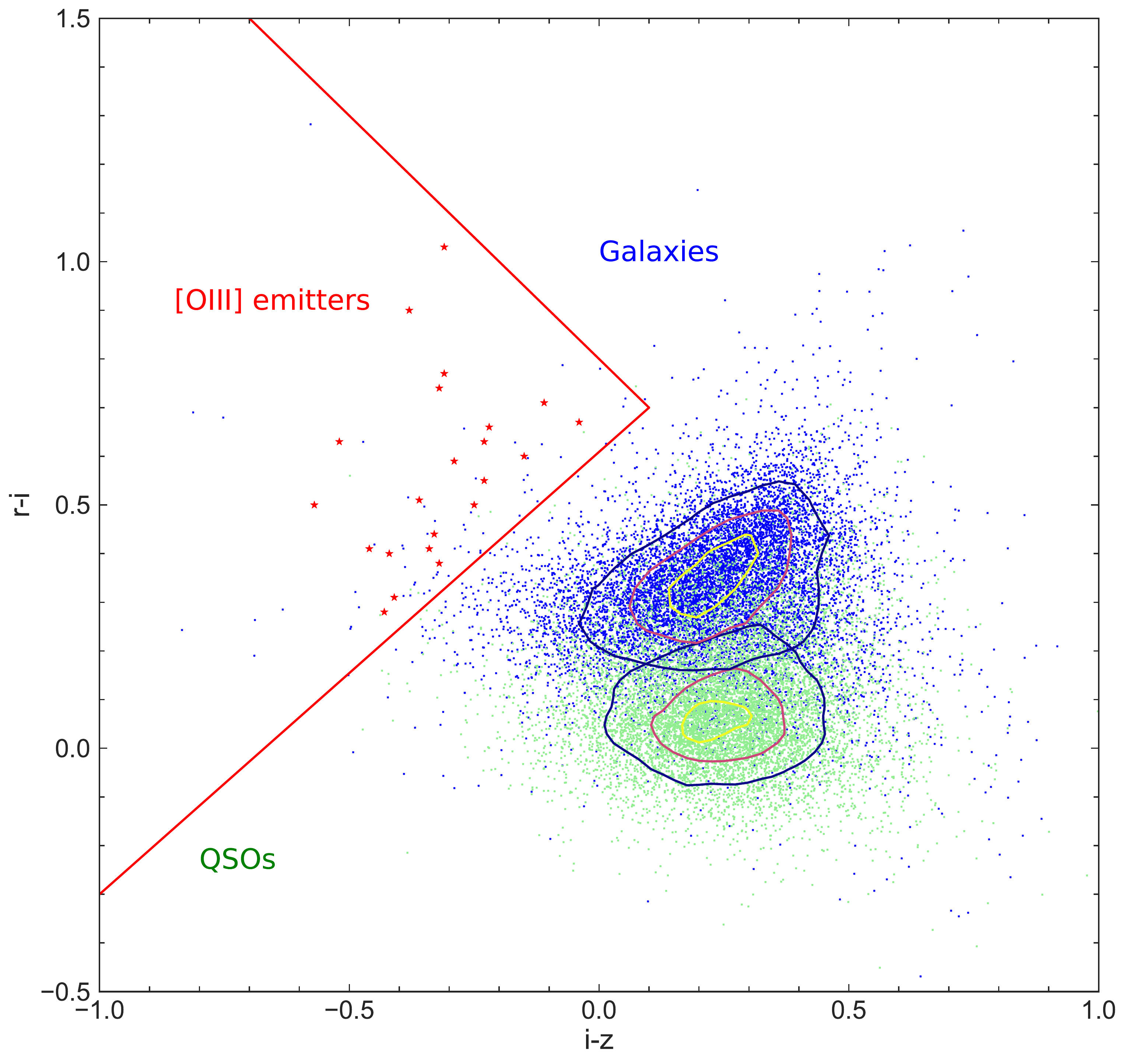}
\includegraphics[width=\linewidth, clip]{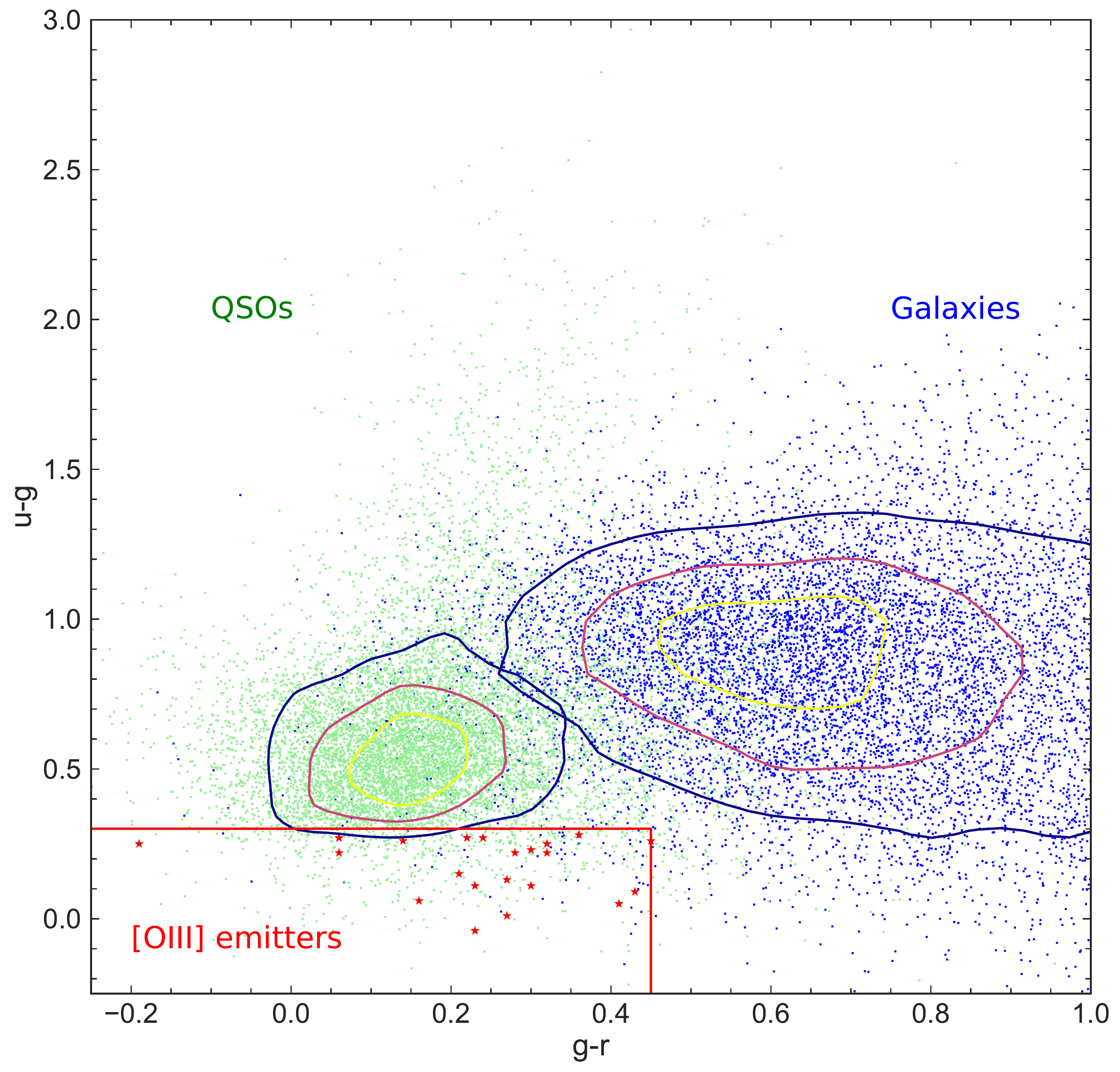}
\caption{Color-color diagrams for 22 [OIII] emitters (red stars), a sample of galaxies (blue points) and a sample of QSOs (green points) with similar \emph{r}-band magnitudes and photometry errors. The series of contours (from yellow to dark blue) overplotted onto each comparison sample delineate the regions where the point densities are the highest.}
\label{fig:selection}
\end{figure}

The effectiveness of our selection criteria is illustrated in Figure \ref{fig:selection}. We plotted colors of our 22 [OIII] emitters together with 10000 galaxies and 10000 QSOs. The galaxies and the QSOs are selected from DR14 and the SDSS 12th Data Release Quasar Catalog (DR12Q, \citealt{2017A&A...597A..79P}) respectively by requiring that they have similar \emph{r}-band magnitudes (19 $<$ \emph{r} $<$ 21) and photometry errors ($<$ 0.25 mag for all five SDSS bands) to [OIII] emitters. We use model magnitudes for galaxies and [OIII] emitters, and PSF magnitudes for QSOs. As can be seen in Figure \ref{fig:selection}, our candidates have successfully, if not perfectly, avoided the large populations of galaxies and QSOs.

To demonstrate that the \emph{i}-band excess of the 22 emitters is mainly caused by [OIII] emission, we examine their observed [OIII] and H$\beta$ EWs. Note that for each emission line in the \emph{i}-band we have
\begin{equation}\label{ew1}
{\rm EW_{obs}} =\ {\rm W_{eff}(10^{-0.4\Delta \emph{i}}-1)}
\end{equation}
where ${\rm EW_{obs}}$ is the observed EW, ${\rm W_{eff}}$ = 1044.55 \AA\ is the equivalent bandwidth of the \emph{i}-band, and $\Delta \emph{i}$ is the $\emph{i}$-band excess compared to the $\emph{i}$-band continuum, which roughly equals $\emph{i}-0.5(\emph{r}+\emph{z})$. By definition ${\rm EW_{obs}}$ = f$_{\rm line}$/f$_{\rm \lambda, \emph{i}}$ (f$_{\rm \lambda, \emph{i}}$ is the \emph{i}-band continuum flux), and considering that three dominant emission lines are in the \emph{i}-band filter (H$\beta$, [OIII]$\lambda\lambda$4959, 5007), we have
\begin{equation}\label{ew2}
{\rm \frac{\Sigma f_{\rm line}}{f_{\rm \lambda, \emph{i}}}} =\ {\rm W_{eff}(10^{-0.4\Delta \emph{i}}-1)}
\end{equation}
where $\Sigma$ represents the flux summation of H$\beta$ and [OIII] doublets.

Considering our selection criterion (\ref{eq1}), the right side of Equation (\ref{ew2}) should be $\geqslant$ 397 \AA. All 22 [OIII] emitters satisfy this condition except OIII-20, possibly due to its noisy continuum and large uncertainty in $\emph{z}$-band magnitude (0.13 mag). The maximum flux ratio of H$\beta$ to [OIII]$\lambda\lambda$4959, 5007 is 19\% except for OIII-20 (33\%). Thus, we conclude that the [OIII] emission is, indeed, the main reason for $\emph{i}$-band excess.

\begin{figure}
\includegraphics[width=\linewidth, clip]{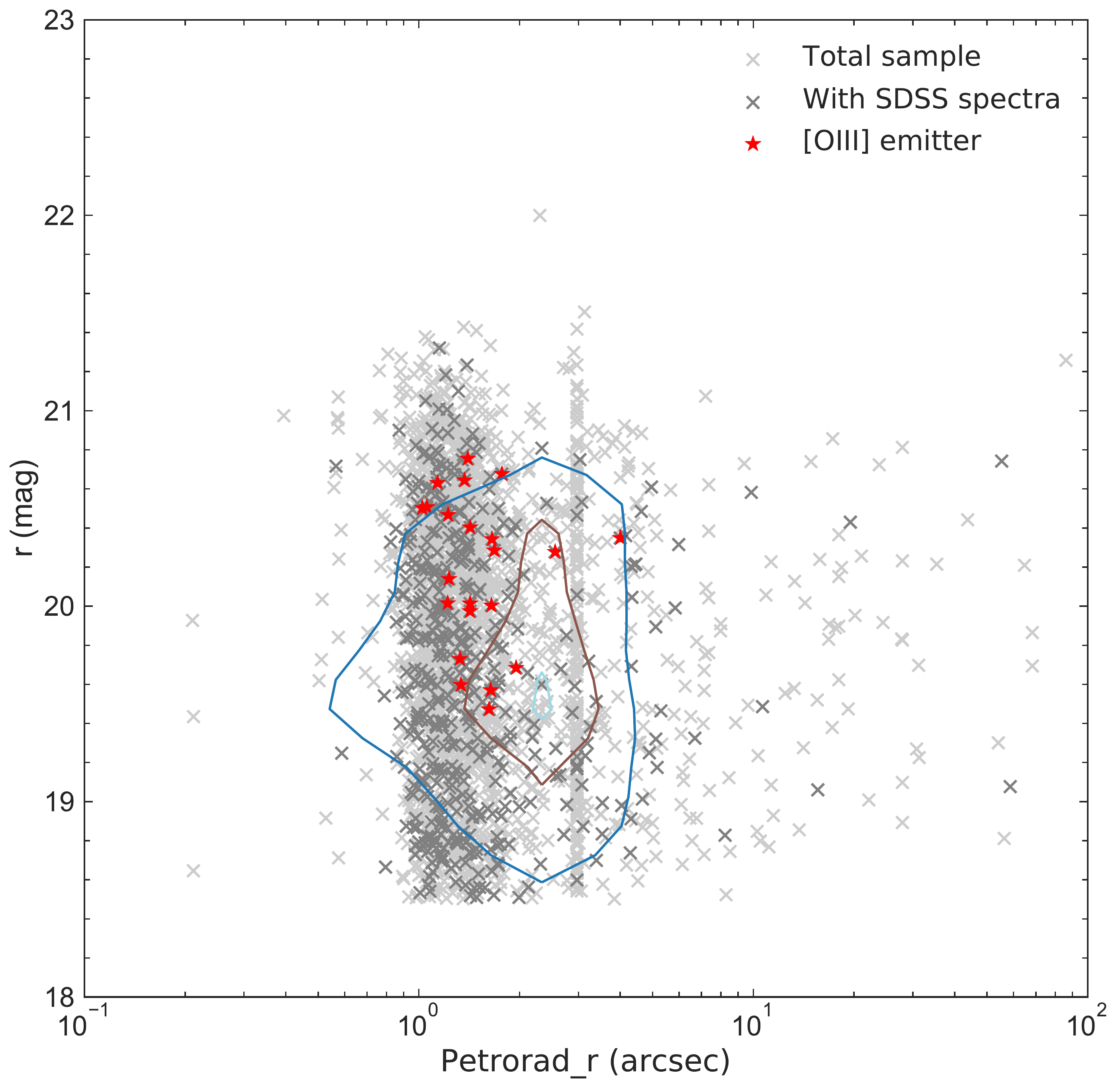}
\caption{Distribution of the sizes (\emph{r}-band Petrosian radius) v.s. \emph{r}-band magnitude of 2658 [OIII] candidates, 649 objects with SDSS spectra, and 22 [OIII] emitters. The light gray crosses represent our total sample and the dark grey crosses represent the objects which have SDSS spectra. The red stars represent the 22 [OIII] emitters in our total sample. The series of contours (from light blue to blue) overplotted delineate the regions where the point densities of our total sample are the highest.}
\label{fig:size}
\end{figure}

The positions, \emph{r}-band magnitudes, Petrosian radii \texttt{petrorad\_r}, and clean photometry flags \texttt{clean} of our `total sample' are tabulated in Table \ref{table:parent}. The distribution of the Petrosian radii versus \emph{r}-band magnitudes of 2658 [OIII] candidates, 649 objects with SDSS spectra, and 22 [OIII] emitters is illustrated in Figure \ref{fig:size}. The mean and median \texttt{petrorad\_r} of our `total sample' are 2.5 and 1.4 arcsec, respectively, and 95\% of these [OIII] candidates have \texttt{petrorad\_r} $<$ 5 arcsec.

\begin{deluxetable*}{ccccc}
\tabletypesize{\footnotesize}
\tablewidth{0.1\textwidth}
\tablecaption{{\centering}Selection procedure used to construct our samples\label{table:select}}

\tablehead{\colhead{Step}                          &        
           \colhead{Selection Criteria}              &
           \multicolumn{2}{c}{Number of Objects}  &\\
           \colhead{} &
           \colhead{} &
           \colhead{Primary Sample}  &
           \colhead{Total Sample}       &\\
           \colhead{(1)}                          &
           \colhead{(2)}              &
           \colhead{(3)}              &
           \colhead{(4)}              &    
           }
\startdata 
1&	Eq. (1)-(5)&968	&2658 \\
2&	Requiring SDSS spectra & 200&649\\
3&	Removing spectroscopically identified QSOs and stars &75&140\\
4&	Picking [OIII] emitters by visual check on spectra &17&22\\
\enddata



\end{deluxetable*}

\begin{deluxetable}{cccccccc}
\tabletypesize{\scriptsize}
\tablewidth{0.5\textwidth}
\tablewidth{\textwidth}
\tablecaption{{\centering}Positions, \emph{r}-band magnitudes, Petrosian radii, and clean photometry flags of our total sample of [OIII] candidates\label{table:parent}}

\tablehead{\colhead{RA}                          &        
           \colhead{Dec}              &
           \colhead{\emph{r}}       &
           \colhead{\emph{petrorad\_r}}       &
           \colhead{\texttt{clean}}       &
           \colhead{Primary flag}       &\\
            \colhead{(J2000 deg)}                              &
           \colhead{(J2000 deg)}              &
           \colhead{(mag)}                              &
           \colhead{(arcsec)}                              &
           \colhead{}                              &\\
           \colhead{(1)}                          &
           \colhead{(2)}              &
           \colhead{(3)}              &
           \colhead{(4)}              &
           \colhead{(5)}              &
           \colhead{(6)}              &   
           }
\startdata 
0.06925 & 14.73523 & 19.70&2.97&0&0\\
0.11109 & 6.20795 & 19.60&2.29&0&0\\
0.11965 & 0.86930 & 18.78&8.20&0&0\\
0.30502 & 31.65254 & 19.90&0.62&0&0\\
0.46296 & 33.16742 & 20.01&1.67&0&0\\
0.51115 & 7.99360 & 21.15&0.90&1&1\\
1.01737 & 14.77080 & 19.86 & 0.92&1&1\\
1.33597 & 36.29656 & 20.95&0.57&0&1\\
1.48225 & 18.14386 & 18.97&4.82&0&0\\
1.48347 & 4.99813 & 20.18&1.30&1&1\\
\enddata


\tablenotetext{}{\textbf{Notes.} (1)-(2) RA and Dec. (3) \emph{r}-band model magnitude. (4) \emph{r}-band Petrosian radius. (5) Clean photometry flag, where \texttt{clean} = 1 indicates reliable photometry and \texttt{clean} = 0 indicates unreliable photometry. (6) Flag indicating whether the object is in the `primary' sample (1 if yes and 0 if not). Only the first ten objects in this sample are shown. Full version of this table is available in a machine-readable format online.}
\end{deluxetable}

\subsection{Contamination by H$\alpha$ Emitters, QSOs and stars} \label{sec:contamination}
By examining the SDSS spectra of the objects in our photometric sample, we find the contamination comes from mainly three types of objects: (a) z $\gtrsim$ 1 QSOs; (b) z $\lesssim$ 0.2 H$\alpha$ emitters; (c) local stars. We first find that many of these objects have \emph{r} - \emph{z} $<$ -0.3, which is unphysical and indicative of unreliable \emph{z}-band photometry. Having excluded these objects, we find that the \emph{i}-band excess of the remaining z $\gtrsim$ 1 QSOs is caused by the Mg II emission line. For the local stars, we examined them in our `primary sample' (11 in total) and found all of them have unreliable photometry (\texttt{clean} = 0) and do not exhibit \emph{i}-band excess in their spectra. 

We also examined the composition of the 140 spectroscopically observed objects in our `total sample', all of which are classified as `galaxies' in SDSS DR14. When we require clean photometry (\texttt{clean} = 1), most objects are excluded and only 22 identified [OIII] emitters and 22 H$\alpha$ emitters remained. The redshifts of these H$\alpha$ emitters range from 0.07 to 0.14, with a median value of 0.095, which accounts for their \emph{i}-band excess. 
We noticed that apart from strong H$\alpha$ emission, some of the H$\alpha$ emitters also exhibit comparably strong [OIII]$\lambda$5007 emission to our [OIII] emitters. We will further compare the emission line properties of H$\alpha$ and [OIII] emitters in Section \ref{sec:results}.

\section{Multi-wavelength Observations of [OIII] emitters} \label{sec:data}
\subsection{Photometry}\label{sec:photometry}
We have constructed UV-IR spectral energy distributions (SED) for each of our [OIII] emitters by collecting \emph{GALEX}, SDSS, UKIDSS and WISE data from their corresponding online catalogs. For \emph{GALEX} data, we searched for NUV and FUV detections using the \emph{GALEX} \textbf{All-Sky Survey Source Catalog (GASC)} and \textbf{Medium Imaging Survey Catalog (GMSC)}\footnote{http://galex.stsci.edu/galexview/}. We reject data which have a Window or Dichroic flag. When multiple observations exist we select the deepest one, and when no detections are available we adopt the 3-$\sigma$ upper limit. For SDSS data, we adopt the model magnitudes from DR14. To ensure all the magnitudes are on the AB system, we added -0.04 to measured \emph{u}-band magnitudes and 0.02 to \emph{z}-band magnitudes\footnote{http://www.sdss.org/dr14/algorithms/fluxcal/}.

For near-infrared data, we searched in \textbf{UKIDSS Large Area Survey (LAS)} from \texttt{UKIDSSDR10PLUS} database\footnote{http://wsa.roe.ac.uk/dbaccess.html} and chose the \texttt{aperMag6} magnitude, which corresponds to a $\sim$5.6 arcsec diameter aperture. For WISE data, we collected \texttt{wxmpro} profile-fit magnitudes from the \textbf{ALLWISE Source Catalog} \citep{2013wise.rept....1C}, as our [OIII] emitters are all identified as point sources \texttt{(ext\_flg = 0)}. We further added small zero-point corrections of 0.03, 0.04, 0.03, -0.03 to W1-W4 magnitudes respectively following \citet{2012AJ....144...68J}. All SEDs for [OIII] emitters are presented in Appendix \ref{sec:SED}.

\subsection{Spectra}\label{sec:spectra}
\begin{figure}
\includegraphics[width=\linewidth, clip]{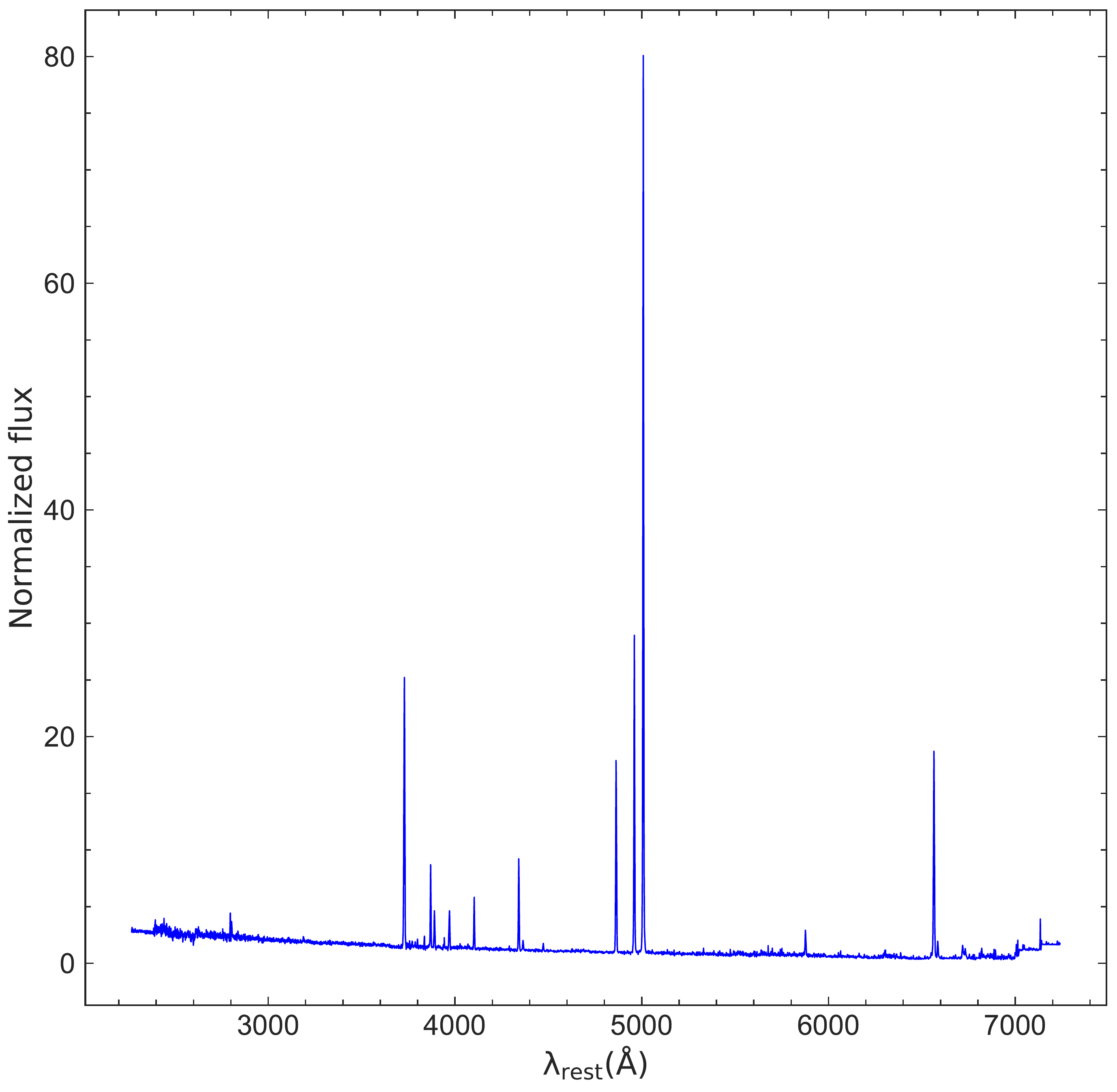}
\caption{Average rest-frame spectrum of 22 [OIII] emitters. This spectrum is available in a machine-readable format online.}
\label{fig:average}
\end{figure}

For our 22 [OIII] emitters, we have compiled their spectra and line measurements from SDSS. In Figure \ref{fig:average}, we present the rest-frame mean spectrum of our [OIII] emitters, which is the average of 22 normalized (using \emph{g}-band continuum, 4500 \AA) spectra. We also present the rest-frame spectrum of each [OIII] emitter in Appendix \ref{sec:SED}. Strong [OIII]$\lambda$5007 emission at z $\sim$ 0.5 can be seen in each spectrum.

\section{Results}\label{sec:results}

\subsection{WISE Color}\label{sec:color}
\begin{figure}
\includegraphics[width=\linewidth, clip]{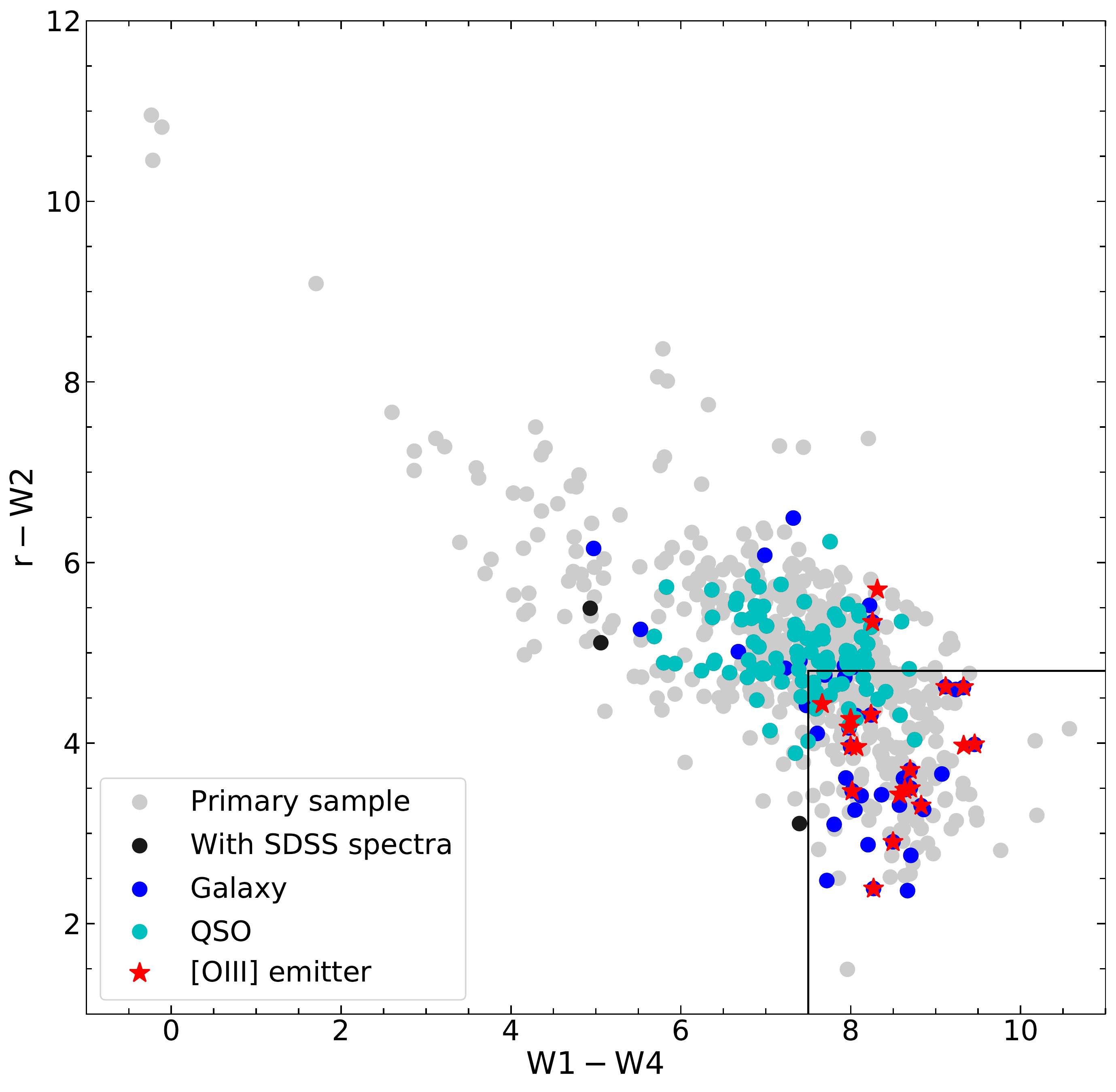}
\caption{Distribution of \emph{r}-W2 v.s. W1-W4 color for 22 [OIII] emitters, our primary sample and its different subsamples. All of the objects plotted have all four WISE bands detected. The grey points represent our primary sample and the black points represent the objects which have SDSS spectra. The blue and cyan points (overplotted onto the majority of the black points) represent `galaxies' and `QSOs' classified by SDSS pipeline respectively. The red stars represent the 22 [OIII] emitters in our total sample.}
\label{fig:wisecolor}
\end{figure}

In addition to examining the effectiveness of our selection criteria by plotting color-color diagram of SDSS colors, we also studied the WISE colors. As illustrated in Figure \ref{fig:wisecolor}, our [OIII] emitters have bluer \emph{r}-W2 and redder W1-W4 colors compared to most other objects, especially quasars. 

We further compare the WISE colors of our [OIII] emitters with 96 metal-deficient star-forming galaxies from \citet{2014A&A...561A..33I}. As the typical redshift of our [OIII] emitters is z $\sim$ 0.5, we use NUV, \emph{i}, W2, (W2+W3)/2, and W4 magnitudes that approximately cover the same rest-frame wavelengths as FUV, \emph{r}, W1, W2 and W3 magnitudes of star-forming galaxies (mainly z $\lesssim$ 0.2) in Izotov's sample. 

\begin{figure}
\raggedleft
\includegraphics[width=\linewidth, clip]{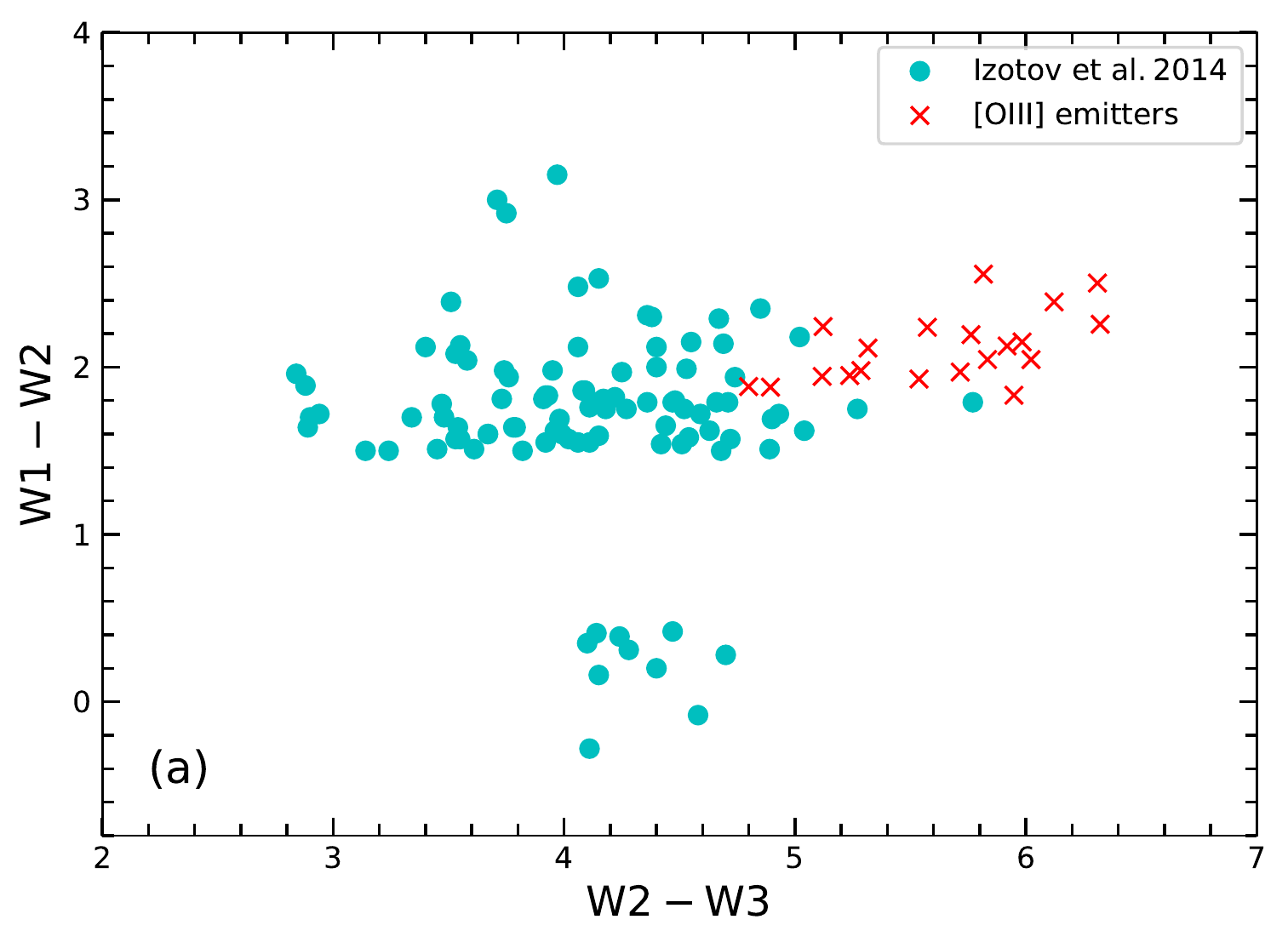}
\includegraphics[width=\linewidth, clip]{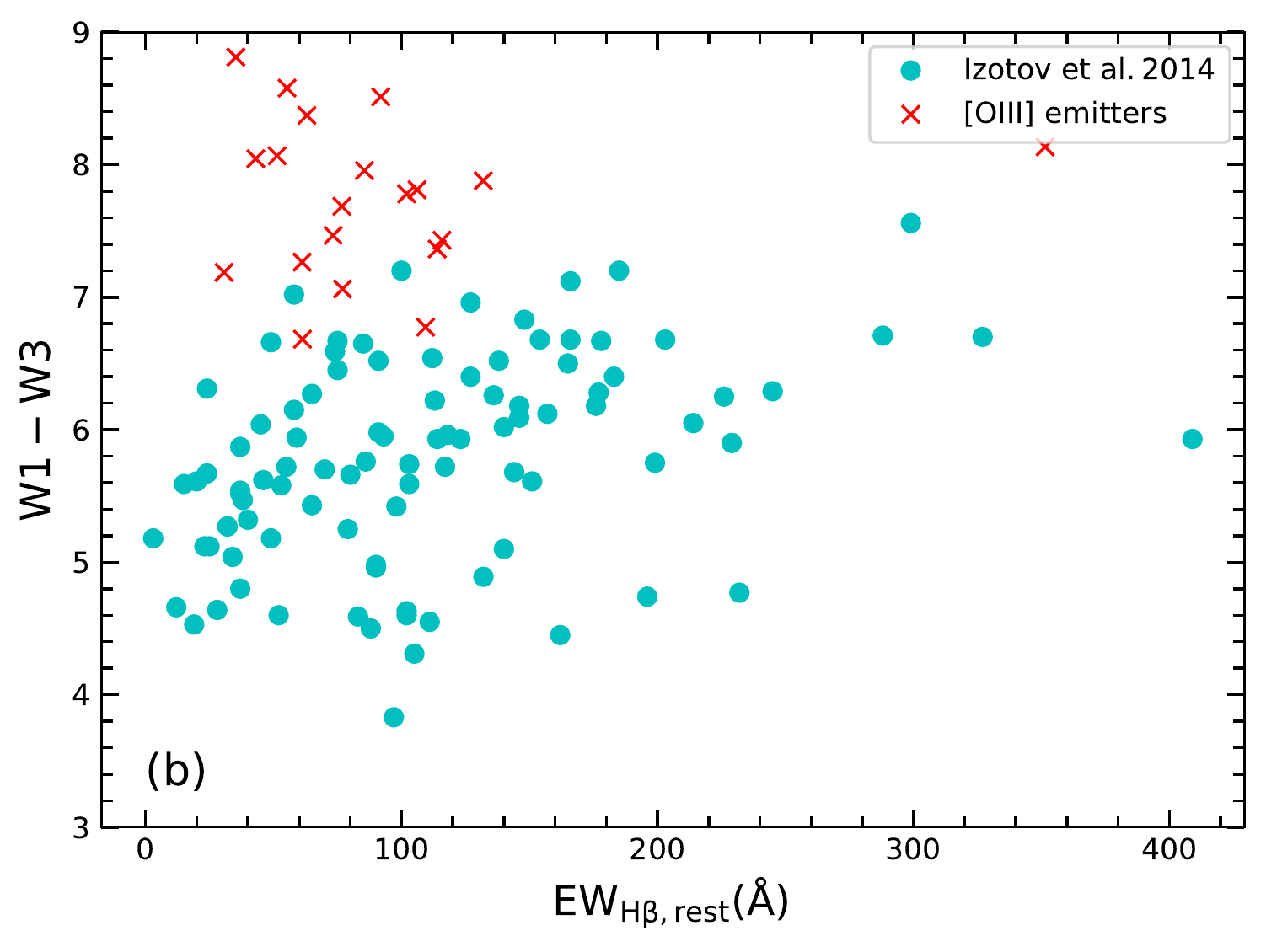}
\includegraphics[width=\linewidth, clip]{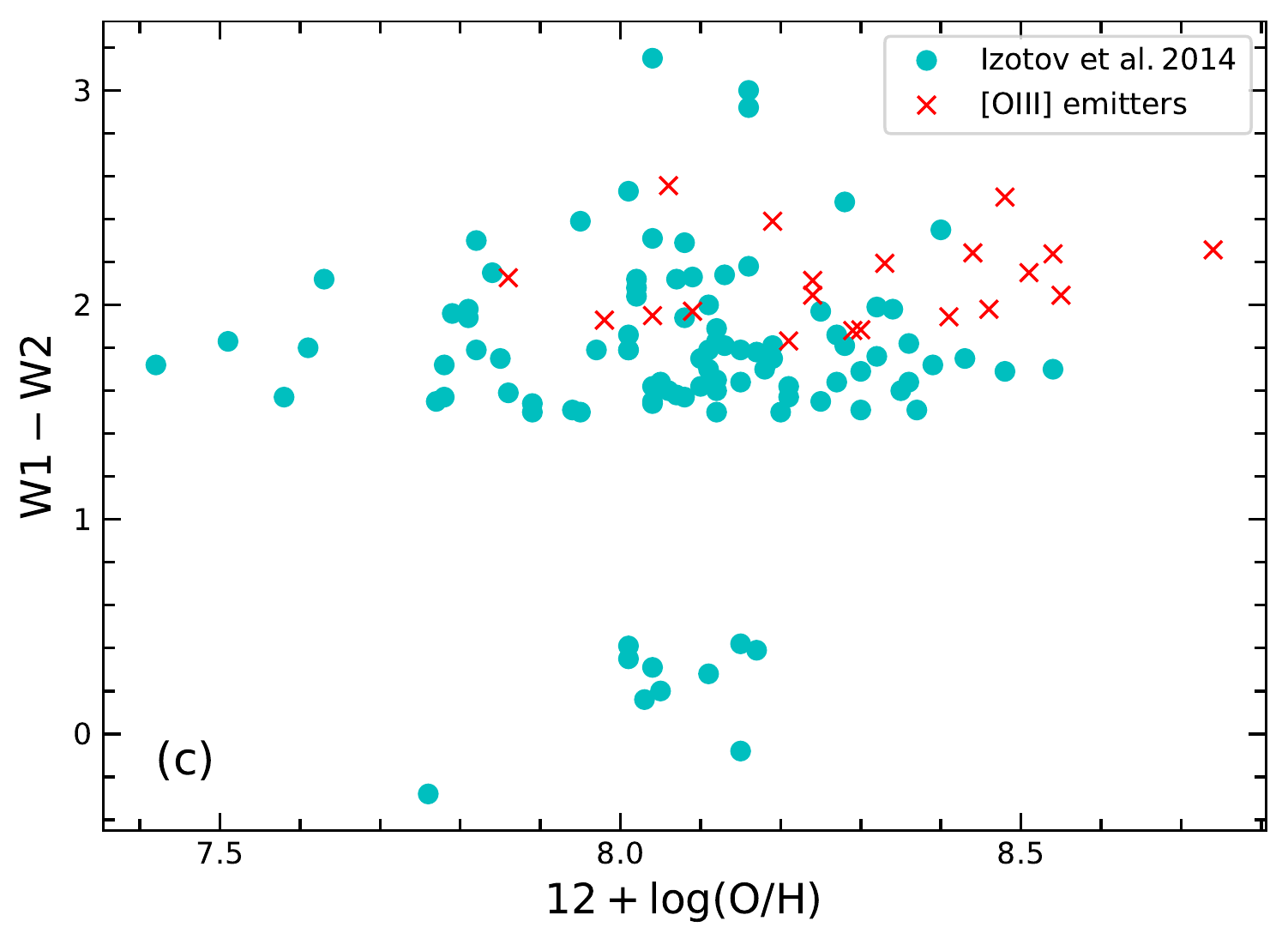}
\caption{Comparison plots of WISE colors for [OIII] emitters (red crosses) and star-forming galaxies (cyan solid circles) in \citet{2014A&A...561A..33I}. (a) W1-W2 v.s. W2-W3 color-color diagram. (b) W1-W3 v.s. EW$_{\rm H\beta,rest}$ diagram. (c) W1-W2 v.s. 12+log(O/H) diagram. The metallicities of 22 [OIII] emitters are calculated in Section \ref{sec:metal}.}
\label{fig:wise}
\end{figure}

From Figure \ref{fig:wise}a, we can see that all of our [OIII] emitters detected in WISE have W1-W2 color redder than 1.6 mag. 
Such red W1-W2 colors are strongly indicative of hot dust emission of 400-600 K grains \citep{2011A&A...536L...7I}. Moreover, our [OIII] emitters have redder W1-W3 and W2-W3 colors than Izotov's sources, indicating that their strong 12$\mu$m emission is at or beyond the extreme seen in Izotov's galaxies.  The likely explanation is the extreme star formation in our sources.

From Figure \ref{fig:wise}b we can see that our [OIII] emitters have redder W1-W3 colors than Izotov's sample, although the rest-frame EW(H$\beta$) of both samples are similar. The lowest values of rest-frame EW(H$\beta$) and H$\beta$ luminosity for our [OIII] emitters are 31 \AA\ and 6.9$\times$10$^{41}$ erg $\rm s^{-1}$, respectively. The very intense burst of young star formation is evidently also effective in heating dust.

Figure \ref{fig:wise}c shows that our [OIII] emitters with high H$\beta$ luminosity and red W1-W2 colors have relatively high metallicity (12 + log(O/H) $>$ 7.8, calculated in Section \ref{sec:metal}). This might be because higher metallicity corresponds to larger amounts of dust, which emit more mid-infrared emission, heated in young star-forming regions. 

\subsection{EW of [OIII]$\lambda$5007 }\label{sec:EW}
\begin{figure}
\includegraphics[width=\linewidth, clip]{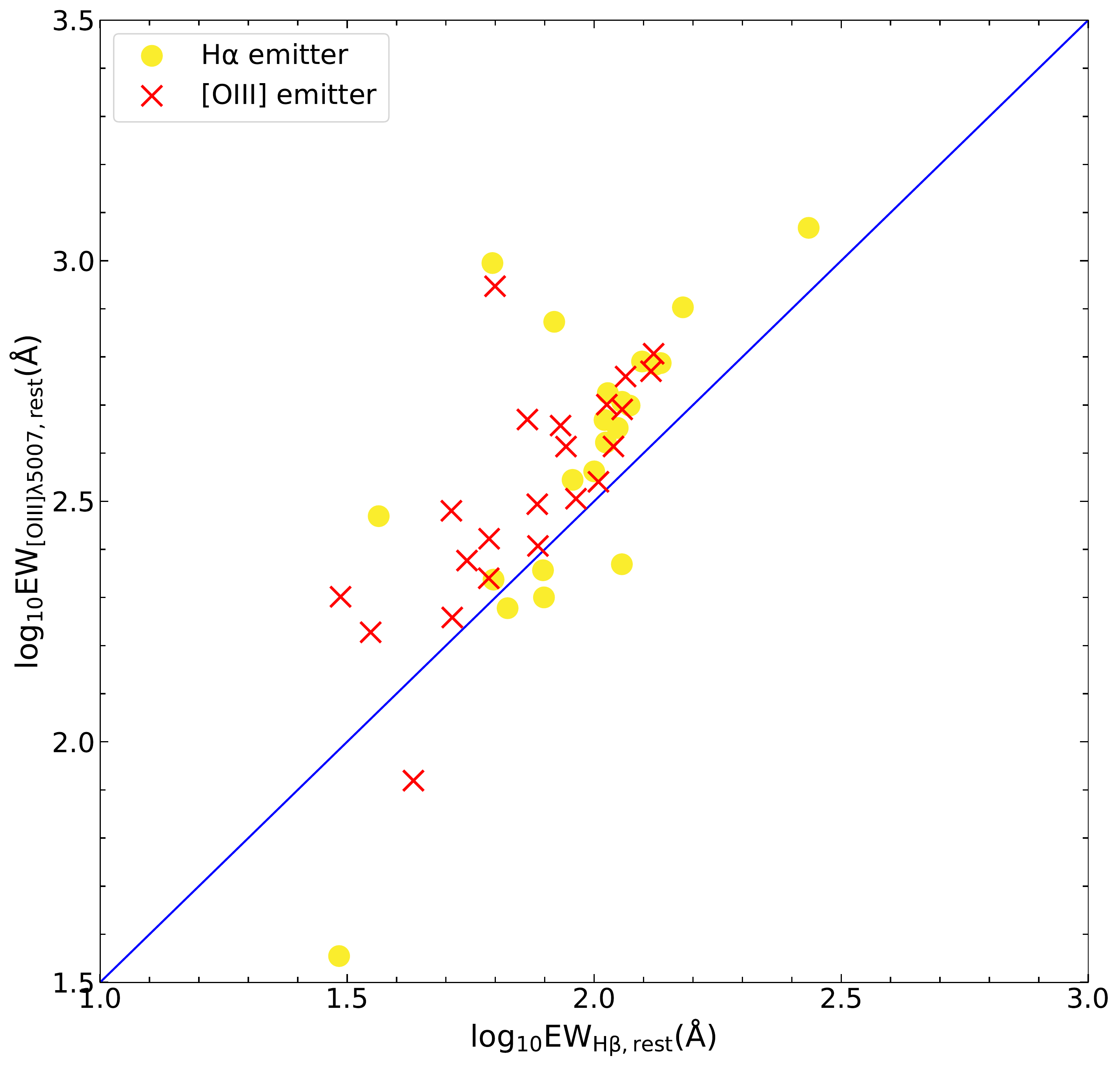}
\caption{Comparison plot of rest-frame EW$_{\rm [OIII]\lambda5007}$ v.s. EW$_{\rm H\beta}$ of [OIII] emitters (red crosses) and H$\alpha$ emitters (yellow solid points). The blue straight line is drawn from EW$_{\rm [OIII]\lambda5007,rest}$ = 10$^{0.5}$EW$_{\rm H\beta,rest}.$}
\label{fig:EW}
\end{figure}

We compiled the rest-frame EWs of [OIII]$\lambda$5007 for both our [OIII] emitters and H$\alpha$ emitters. From Figure \ref{fig:EW} we can see that the rest-frame EW of [OIII] emitters mainly range from 200 \AA\ to 600 \AA, which is also the case for H$\alpha$ emitters. It is noteworthy that several H$\alpha$ emitters and [OIII] emitters exhibit extremely high rest-frame EW of $\sim$ 1000 \AA. Almost all [OIII] emitters have high EW$_{\rm [OIII]\lambda5007,rest}$/EW$_{\rm H\beta,rest}$ ratios ($>$10$^{0.5}$), indicative of the high ionization levels of their interstellar gas.

\subsection{BPT Diagram}\label{sec:BPT}
\begin{figure}
\includegraphics[width=\linewidth, clip]{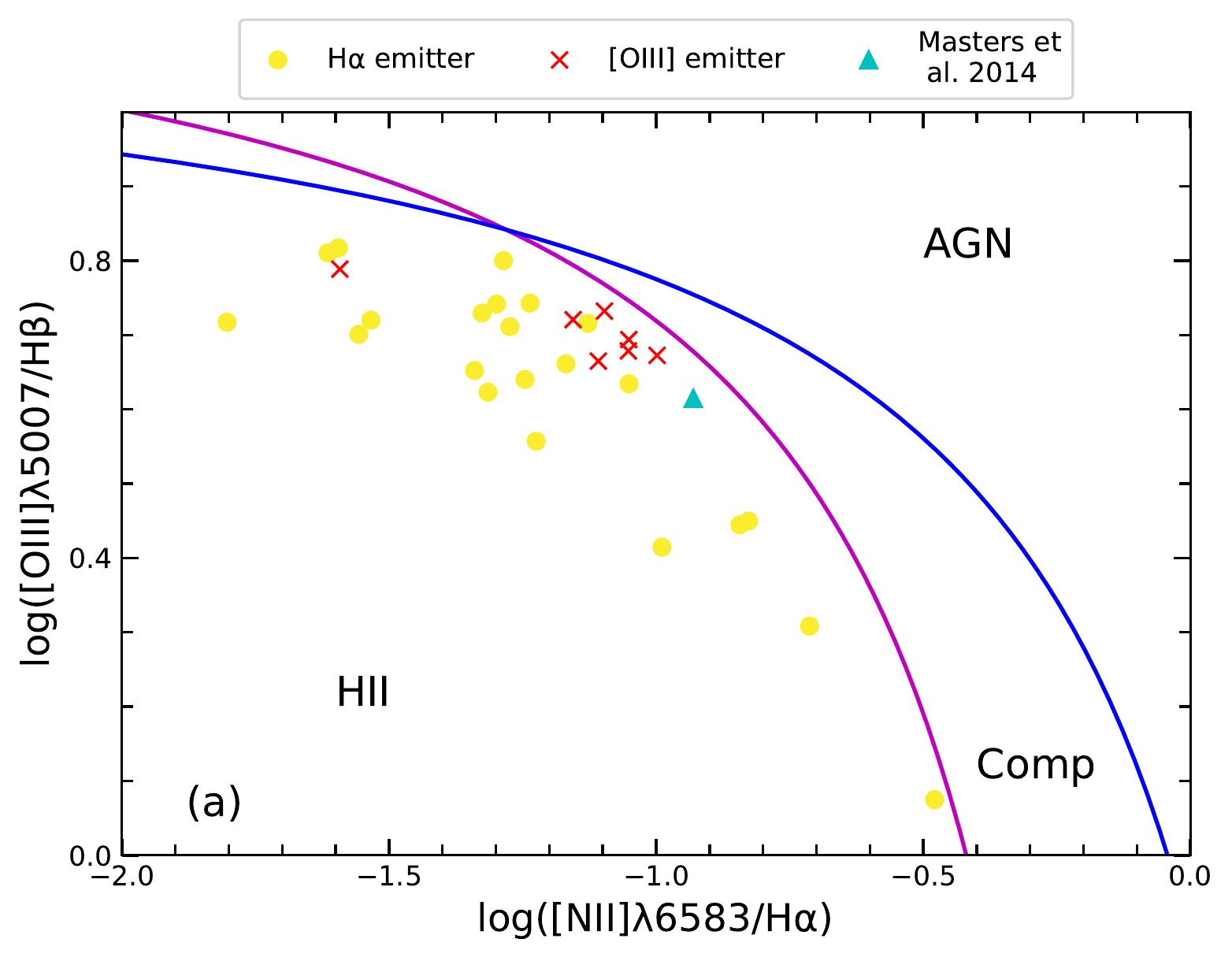}
\includegraphics[width=\linewidth, clip]{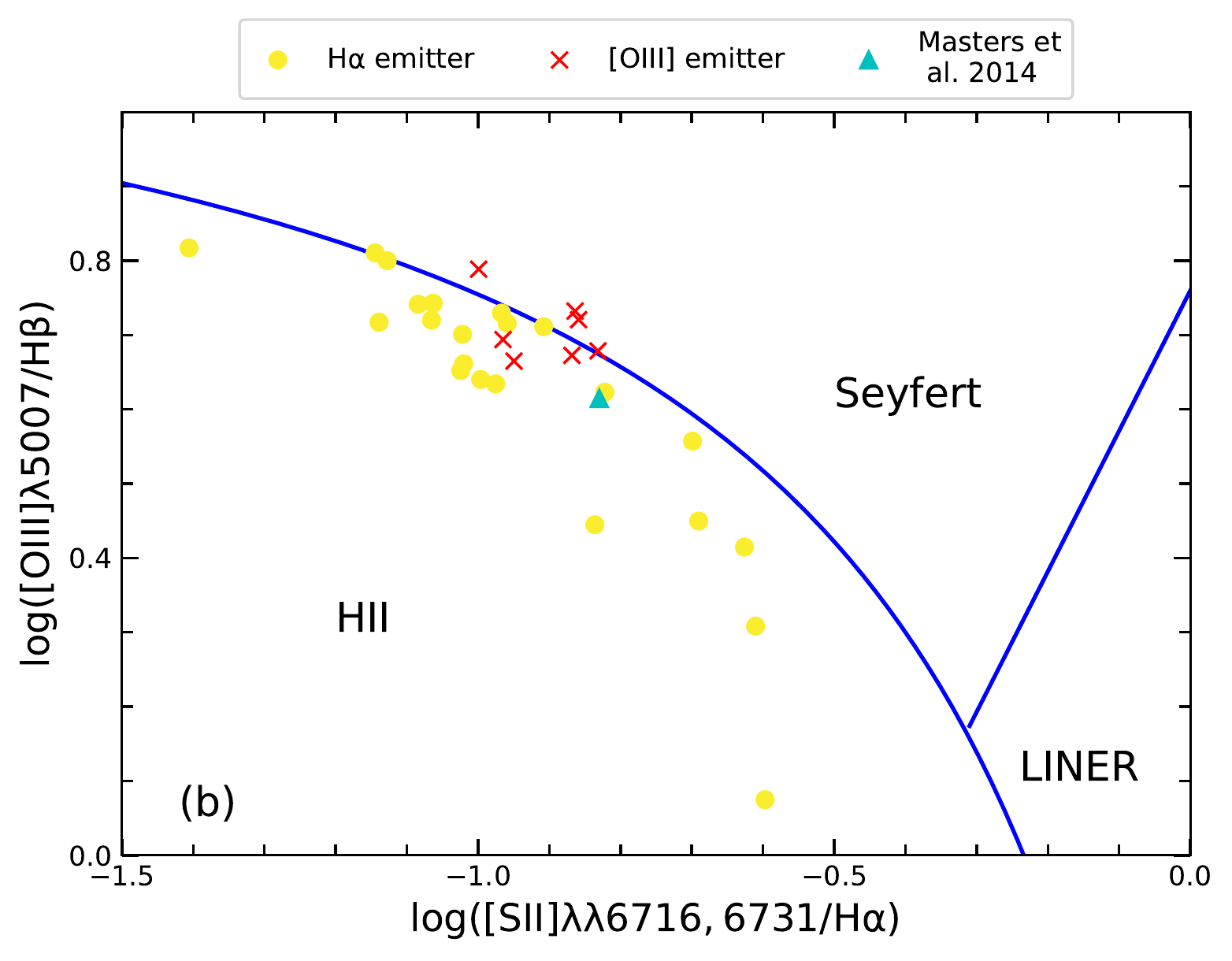}
\includegraphics[width=\linewidth, clip]{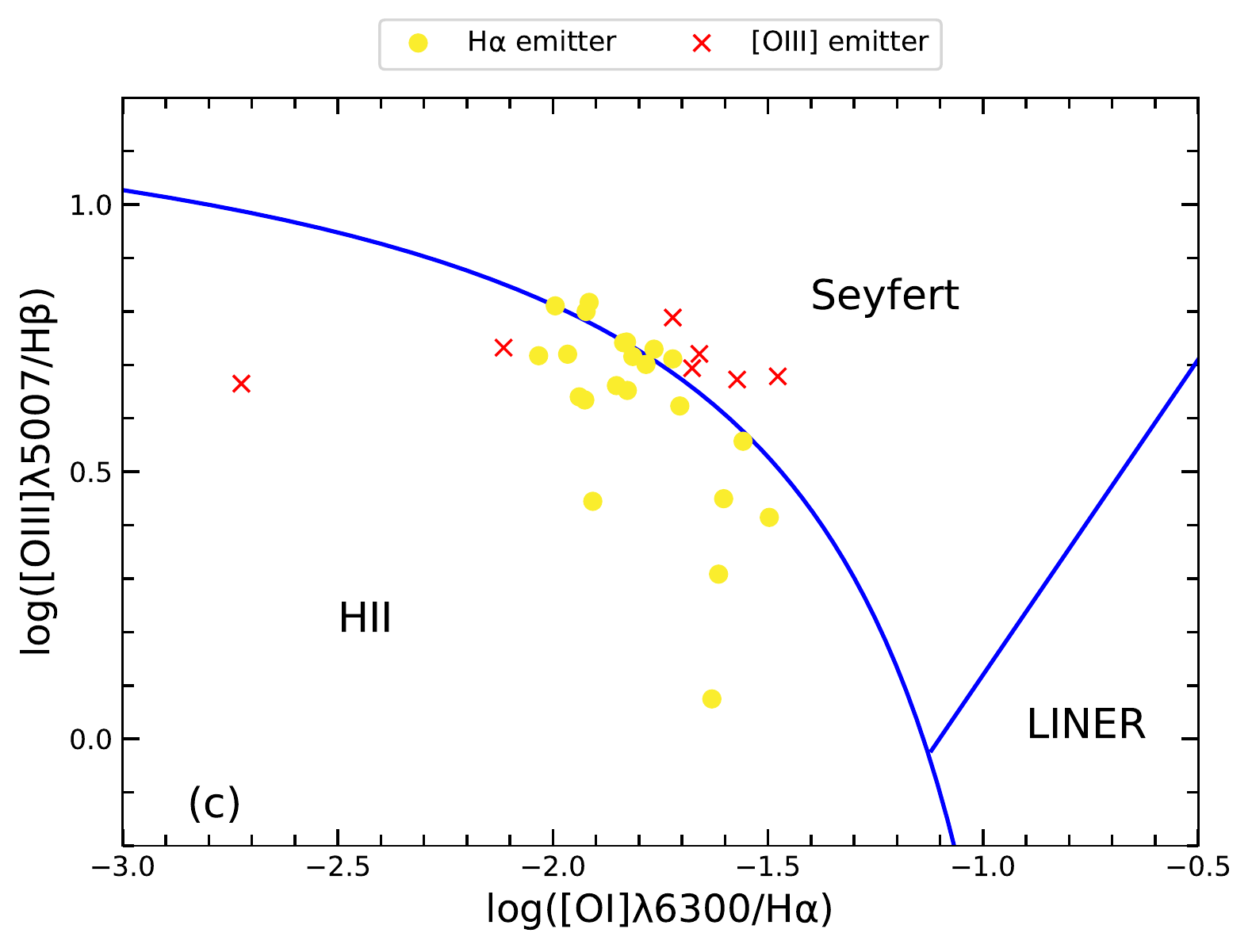}
\caption{The location of 22 [OIII] emitters (red crosses) and 22 H$\alpha$ emitters (yellow solid circles) on BPT diagrams. `HII' = normal star-forming galaxies. `Comp' = composite galaxies. (a) The [NII]-BPT diagram, in which the AGN/starburst boundary lines are from \citet{2003MNRAS.346.1055K} (shown in magenta) and \citet{2006MNRAS.372..961K} (shown in blue). (b)-(c) The [SII]-BPT and [OI]-BPT diagrams, in which the boundary lines are from \citet{2006MNRAS.372..961K}, which separate normal star-forming galaxies from Seyferts and LINERs. The cyan triangle plotted in (a) and (b) represents the line ratios of the composite spectrum of 26 emission-line galaxies at z $\sim$ 2 from \citet{2014ApJ...785..153M}.}
\label{fig:BPT}
\end{figure}

We used the Baldwin-Phillips-Terlevich (BPT hereafter, \citealt{1981PASP...93....5B}) diagnostic line-ratio diagrams to infer the photoionization mechanisms in our [OIII] emitters. The most widely-used three BPT diagrams are relations of [OIII]$\lambda$5007/H$\beta$ v.s. [NII]$\lambda$6583/H$\alpha$, [SII]$\lambda\lambda$6716, 6731/H$\alpha$ and [OI]$\lambda$6300/H$\alpha$. For the [NII]-BPT diagram (shown in Figure \ref{fig:BPT}a), we included AGN/starburst boundary lines from \citet{2003MNRAS.346.1055K} (shown in magenta) and \citet{2006MNRAS.372..961K} (shown in blue). For the [SII]-BPT and [OI]-BPT diagrams (shown in Figures \ref{fig:BPT}b and \ref{fig:BPT}c) we adopted the boundary lines from \citet{2006MNRAS.372..961K}. 

It can be seen from the BPT diagrams that our [OIII] emitters mainly lie just in the star-forming region, but near the AGN boundary. On the [NII]-BPT diagram, all [OIII] emitters with H$\alpha$ measurements lie in the HII region zone. On the [SII]-BPT and [OI]-BPT diagrams, we found three and five [OIII] emitters just across the boundary into Seyferts, respectively, while the rest lie with the HII regions. As a comparison, we also plotted the line ratios of the composite spectrum of 26 emission-line galaxies at z $\sim$ 2 from \citet{2014ApJ...785..153M}, which lie near our [OIII] emitters and the HII regions on the [NII]-BPT and [SII]-BPT diagrams.

\subsection{Dust Extinction}\label{sec:extinction}
We have estimated the amount of dust extinction from the observed Balmer decrements. Assuming the hydrogen nebular emission follows Case B recombination, the intrinsic Balmer flux ratios are (H$\alpha$/H$\beta$)$_{0}$ = 2.86, (H$\gamma$/H$\beta$)$_{0}$ = 0.468 and (H$\delta$/H$\beta$)$_{0}$ = 0.259 for $T_{e}$ = 10$^{4}$ K. 

We adopt the reddening curve of \citet{1989ApJ...345..245C}, which gives
\begin{equation}\label{eq6}
E (B -V) =\ 2.33{\rm log\left(\frac{H{\alpha}}{H{\beta}}\right)_{obs}}-1.06
\end{equation}
and
\begin{equation}\label{eq7}
E (B -V) = -4.46{\rm log\left(\frac{H{\gamma}}{H{\beta}}\right)_{obs}}-1.47
\end{equation}

Since only 7/22 [OIII] emitters have H$\alpha$ line flux measurements, while all 22 emitters have  H$\gamma$ line flux measurements, we calculated \emph{E(B--V)} in both ways using Equation (\ref{eq6}) and (\ref{eq7}) and present our results in Table \ref{table:e_bv}. To estimate the uncertainties, we generated realizations of the emission line fluxes based on their S/N ratios, and constructed probability distributions of the emission line fluxes. The error bars given in Table \ref{table:e_bv} correspond to 68\% confidence levels. In our calculation we have not corrected the higher Balmer emission lines for the underlying stellar absorption that might be present. This is because the rest-frame equivalent widths of H$\beta$, EW$_{\rm H\beta,rest}$, are so high (19/22 have EW$_{\rm H\beta,rest}$ $>$ 50 \AA) that the effects of underlying absorption from A stars would be small. It is possible, however, that some of our \emph{E(B--V)} have been slightly over-estimated.

The typical value of \emph{E(B--V)} for our [OIII] emitters is $\sim$0.1-0.3 mag. The median value of the extinction inferred from H$\gamma$/H$\beta$ (\emph{E(B--V)}$_{\rm H\gamma, H\beta}$) and that from H$\alpha$/H$\beta$ (\emph{E(B--V)}$_{\rm H\alpha, H\beta}$) are 0.23 mag and 0.21 mag, respectively. We find that although \emph{E(B--V)}$_{\rm H\gamma, H\beta}$ tends to be slightly higher than \emph{E(B--V)}$_{\rm H\alpha, H\beta}$, they are basically consistent considering their uncertainties. In the following calculations we adopted \emph{E(B--V)}$_{\rm H\alpha, H\beta}$ when they are available, and \emph{E(B--V)}$_{\rm H\gamma, H\beta}$ in other cases to correct for dust extinction (Section \ref{sec:metal}).

\subsection{O$_{32}$ and R$_{23}$}\label{sec:o32}
To better understand the ionization and excitation situation of our [OIII] emitters, we also calculated the O$_{32}$ and R$_{23}$ indices, defined as:

\begin{equation}\label{o32}
{\rm O_{32}} =\ {\rm log\frac{[OIII]\lambda\lambda4959, 5007}{[OII]\lambda\lambda3727, 3729}}
\end{equation}

\begin{equation}\label{r23}
{\rm R_{23}} =\ {\rm log\frac{[OIII]\lambda\lambda4959, 5007+[OII]\lambda\lambda3727, 3729}{H\beta}}
\end{equation}

\begin{figure}
\includegraphics[width=\linewidth, clip]{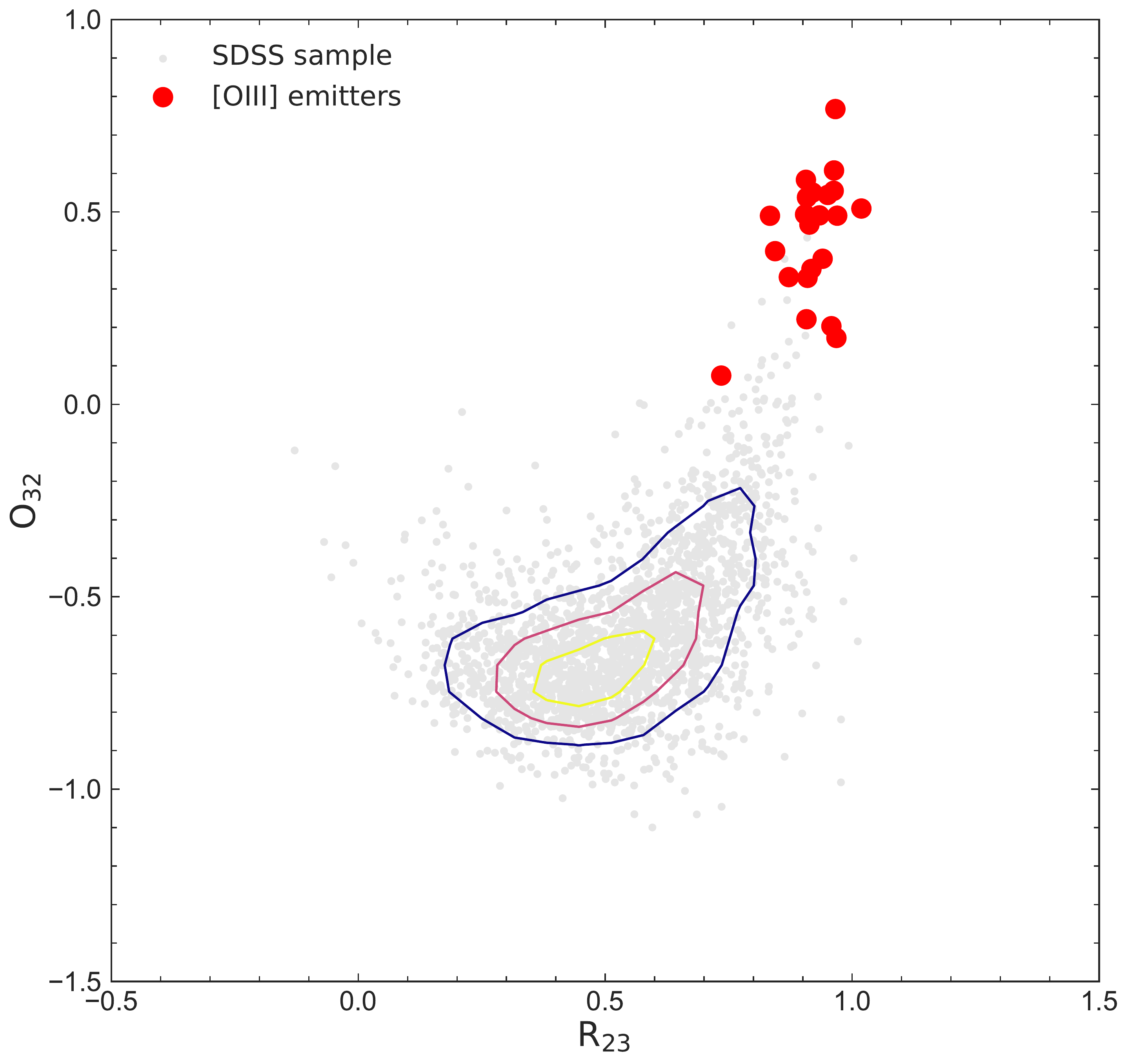}
\caption{The distribution of O$_{32}$ and R$_{23}$ ratios of [OIII] emitters (red solid circles) and the SDSS star-forming galaxies (light gray circles) at z $\sim$ 0. All ratios have been corrected for internal dust extinction. The series of contours (from yellow to dark blue) overplotted onto the SDSS comparison sample delineate the regions where the point densities are the highest.}
\label{fig:O32}
\end{figure}

O$_{32}$ and R$_{23}$ are often used to independently characterize the degree of ionization in HII regions \citep{2017ApJ...836..164S}. Figure \ref{fig:O32} shows the distribution of our [OIII] emitters on the O$_{32}$-$\rm R_{23}$ plane compared with z $\sim$ 0 normal star-forming galaxies selected from the MPA-JHU\footnote{http://www.sdss3.org/dr10/spectro/galaxy\_mpajhu.php} catalog. This SDSS comparison sample is composed of 2000 z $<$ 0.1 objects, classified as star-forming galaxies from BPT diagrams with a `reliable'
flag. All the O$_{32}$ and R$_{23}$ ratios have been corrected for internal dust extinction derived from Balmer decrements (see Equation (\ref{eq6}), (\ref{eq7})). 

As is shown in Figure \ref{fig:O32}, our [OIII] emitters mainly occupy the region with the highest O$_{32}$ and R$_{23}$ compared to the SDSS sample, indicative of the extreme high degrees of ionization of these objects. The maximum value of R$_{23}$  for both samples is $\sim$1, as expected for star-forming galaxies with a maximum hardness of ionizing spectrum. Hard ionizing radiation fields and sub-solar oxygen abundances are required to reach the upper limit of R$_{23}$ \citep{2002ApJS..142...35K, 2003ApJ...597..730L, 2017ApJ...836..164S}.

\subsection{Electron Temperature and Metallicity}\label{sec:metal}
Following \citet{2016ApJS..226....5L}, we calculated electron temperatures (\emph{T}$_e$) and gas-phase metallicities for our 22 [OIII] emitters. \emph{T}$_e$ can be estimated from auroral-to-nebular [OIII] flux ratio \emph{R}:

\begin{equation}
R\equiv \frac{F({\rm[OIII]\lambda4363)}}{F({\rm[OIII]\lambda4959)}+F({\rm[OIII]\lambda5007)}}
\end{equation}
and
\begin{equation}
T_e =a(-{\rm log}\textsl(R)-b)^{-c}
\end{equation}
where \emph{a} = 13205, \emph{b} = 0.92506, and \emph{c} = 0.98062 \citep{2014ApJ...790...75N}. The typical value of our derived \emph{T}$_e$ is $\sim$10$^4$ K. Correcting the nebular-to-auroral [OIII] flux ratio for internal dust extinction will slightly increase our estimates of \emph{T}$_e$ ($\sim$0.05 dex at most, see Table \ref{table:metal}).  

Next we determine the gas-phase ionic oxygen abundances using two emission-line flux ratios, [OII]$\lambda$3727/H$\beta$ and [OIII]$\lambda\lambda$4959,5007/H$\beta$ \citep{2006A&A...448..955I}:
\begin{equation}
\begin{split}
12+{\rm log\left(\frac{O^{+}}{H^{+}}\right)}= {\rm log\left(\frac{[OII]}{H\beta}\right)}+5.961+\frac{1.676}{t_2} \\
-0.4{\rm log}{t_2}-0.034{t_2}+{\rm log}(1+1.35x)
\end{split}
\end{equation}

\begin{equation}
\begin{split}
12+{\rm log\left(\frac{O^{++}}{H^{+}}\right)}= {\rm log\left(\frac{[OIII]}{H\beta}\right)}+6.200+\frac{1.251}{t_3} \\
-0.55{\rm log}{t_3}-0.014{t_3}
\end{split}
\end{equation}
where \emph{t}$_3$ = \emph{T}$_e$([OIII])/10$^4$ K, \emph{t}$_2$ = \emph{T}$_e$([OII])/10$^4$ K and \emph{x} = 10$^{-4}$\emph{n}$_e$\emph{t}$_3^{-0.5}$. We adopted the standard two-zone temperature model with \emph{t}$_2$ = 0.7\emph{t}$_3$ + 0.17 \citep{2013ApJ...765..140A}. We note that for very extreme star-forming regions, a slightly higher value of \emph{t}$_2$ is recommended \citep{1986MNRAS.223..811C}. However, as the majority of oxygen atoms are doubly ionized, such a higher \emph{t}$_2$ would only result in a difference of our
derived oxygen abundances of no more than 2\%, which is minor compared with the observational uncertainties.
For the electron density \emph{n}$_e$, we assume \emph{n}$_e$ = 100 cm$^{-3}$, but the results are almost independent of this assumption.
The total oxygen abundance is given by O/H = (O$^{+}$+O$^{++}$)/H$^{+}$ since the most abundant ions of oxygen in HII regions are O$^{+}$ and O$^{++}$. 

We also corrected ${\rm [OIII]\lambda\lambda4959, 5007}/{\rm [OIII]\lambda4363}$ and ${\rm {[OII]4363}/{H\beta}}$ for dust extinction, and derived corresponding electron temperatures and metallicities. Both the dust-corrected and the dust-uncorrected results are shown in Table \ref{table:metal}. For 15/22 [OIII] emitters with S/N $>$ 3 detections of ${\rm [OIII]\lambda4363}$ (referred as `[OIII]$\lambda$4363-detected') we generated realizations of the emission line fluxes and constructed probability distributions. Their propagated error bars given in Table \ref{table:metal} correspond to 68\% confidence levels. For the other 7/22 [OIII] emitters with S/N $<$ 3 detections of ${\rm [OIII]\lambda4363}$ (referred as `[OIII]$\lambda$4363-non-detected') we calculated the upper limits of \emph{T}$_e$ and lower limits of metallicity.

As a result, for our [OIII]$\lambda$4363-detected sources, the lowest and median metallicities are 12 + log(O/H) = 7.98$^{+0.12}_{-0.02}$ and 8.24$^{+0.05}_{-0.04}$. For our [OIII]$\lambda$4363-non-detected sources, the lowest and median metallicities are 12 + log(O/H) = 7.86 and 8.48. Our results are comparable to the values from \citet{2016ApJ...828...67L} (median 12 + log(O/H) = 8.19$^{+0.16}_{-0.19}$ for 0.3 \texttt{$<$} z $\leqslant$ 0.5, log \emph M$_\star$ = 9.25 $\pm$\ 0.25 and 12 + log(O/H) = 8.49$^{+0.16}_{-0.00}$ for 0.5 \texttt{$<$} z $\leqslant$ 1.0, log \emph M$_\star$ = 10.00 $\pm$\ 0.25).
\subsection{Stellar Masses and Star Formation Rates}\label{sec:SFR}
We estimate the stellar masses (\emph M$_{\star}$) of our [OIII] emitters by performing UV-MIR SED fitting using CIGALE (Code Investigating GALaxy Emission, \citealt{2009A&A...507.1793N}). The \emph{i}-band photometry has been excluded from SED fitting as it contains strong [OIII]$\lambda$5007 emission. We adopt a double-exponential star formation history model, composed of a main stellar population and a recent burst. The priors of the mass fraction and the age of the recent burst are set to be 0.01-0.99 and 10-100 Myr, respectively. We further incorporate BC03 \citep{2003MNRAS.344.1000B} and DL07 \citep{2007ApJ...657..810D} models to fit the stellar and dust emission. A Chabrier initial mass function (IMF) \citep{2003PASP..115..763C} is assumed here, and the metallicities are fixed, using Eq. (18)-(20)  from \citet{2016ApJS..226....5L}. For internal dust extinction, we adopt a \citet{2000ApJ...533..682C} law with a free \emph{E(B--V)} varying from 0 to 2 mag. We also assume both the young and the old stellar population suffer from the same amount of extinction.

Our SED fitting results are summarized in Table \ref{table:SFR}. The derived stellar masses of our [OIII] emitters range from 10$^{9.2}$ \emph{M$_{\odot}$} to 10$^{10.15}$ \emph{M$_{\odot}$}, with an average value of 10$^{9.71}$ \emph{M$_{\odot}$} and a median value of 10$^{9.78}$ \emph{M$_{\odot}$}.

Next we derive the star formation rates (SFR) of our [OIII] emitters. We convert H$\alpha$ luminosity to SFR using the metallicity-dependent relation in \citet{2016ApJS..226....5L}. When H$\alpha$ is not measured we use H$\beta$ luminosity instead, assuming (H$\alpha$/H$\beta$)$_{0}$ = 2.86. The internal dust extinction has been corrected for all objects. The results are presented in Table \ref{table:SFR}. The SFR of our [OIII] emitters range from 9 \emph{M$_{\odot}$} yr$^{-1}$ to 129 \emph{M$_{\odot}$} yr$^{-1}$, with an average value of 32.4 \emph{M$_{\odot}$} yr$^{-1}$ and a median value of 30.2 \emph{M$_{\odot}$} yr$^{-1}$.

\begin{figure}
\includegraphics[width=\linewidth, clip]{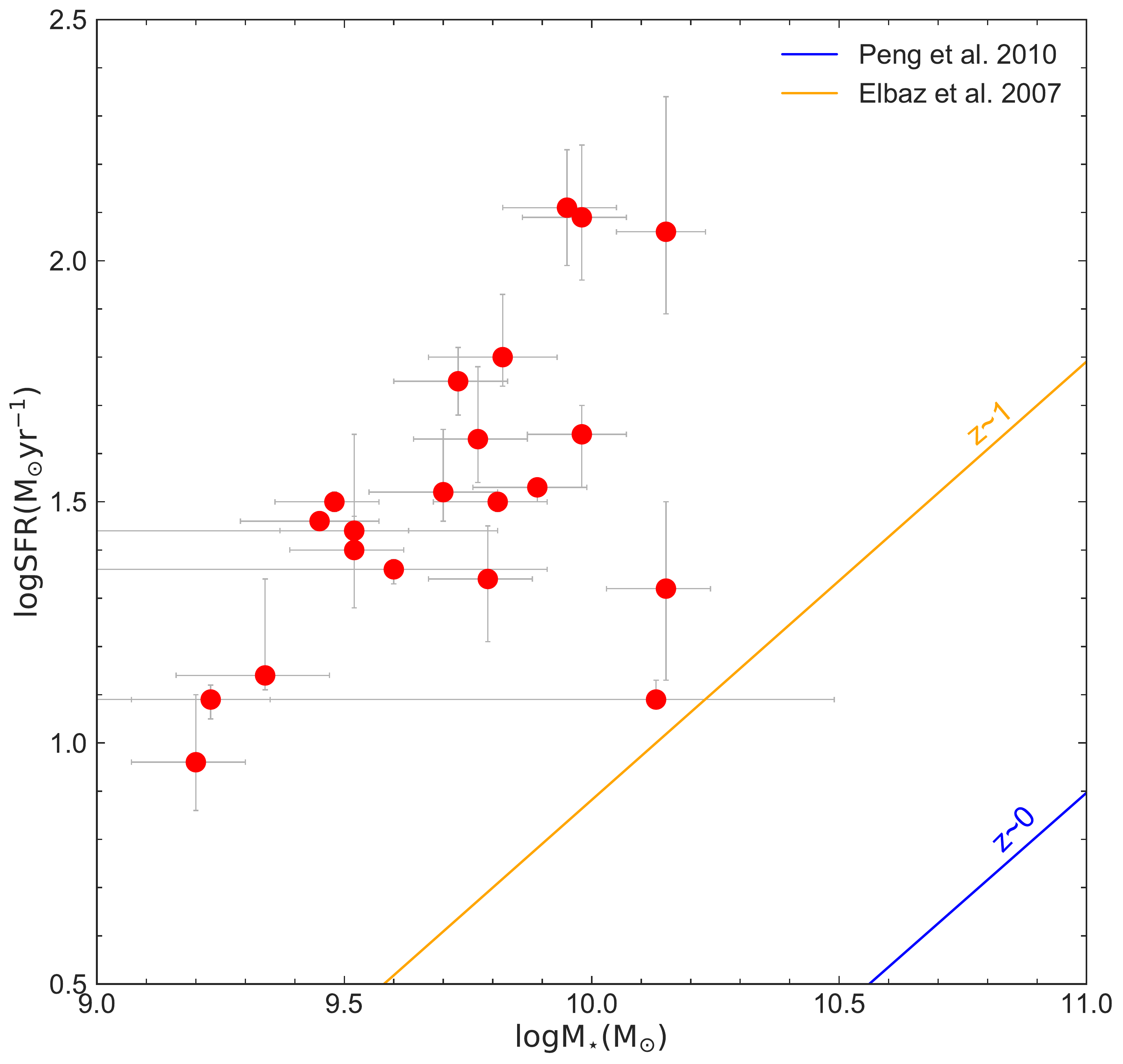}
\caption{SFR v.s. stellar mass of 22 [OIII] emitters (red solid circles). To compare our results with previous works, the sequences observed in \citet{2010ApJ...721..193P} at z $\sim$ 0 and \citet{2007A&A...468...33E} at z $\sim$ 1 are shown in blue and cyan colors, respectively.}
\label{fig:main}
\end{figure}

\begin{figure}
\includegraphics[width=\linewidth, clip]{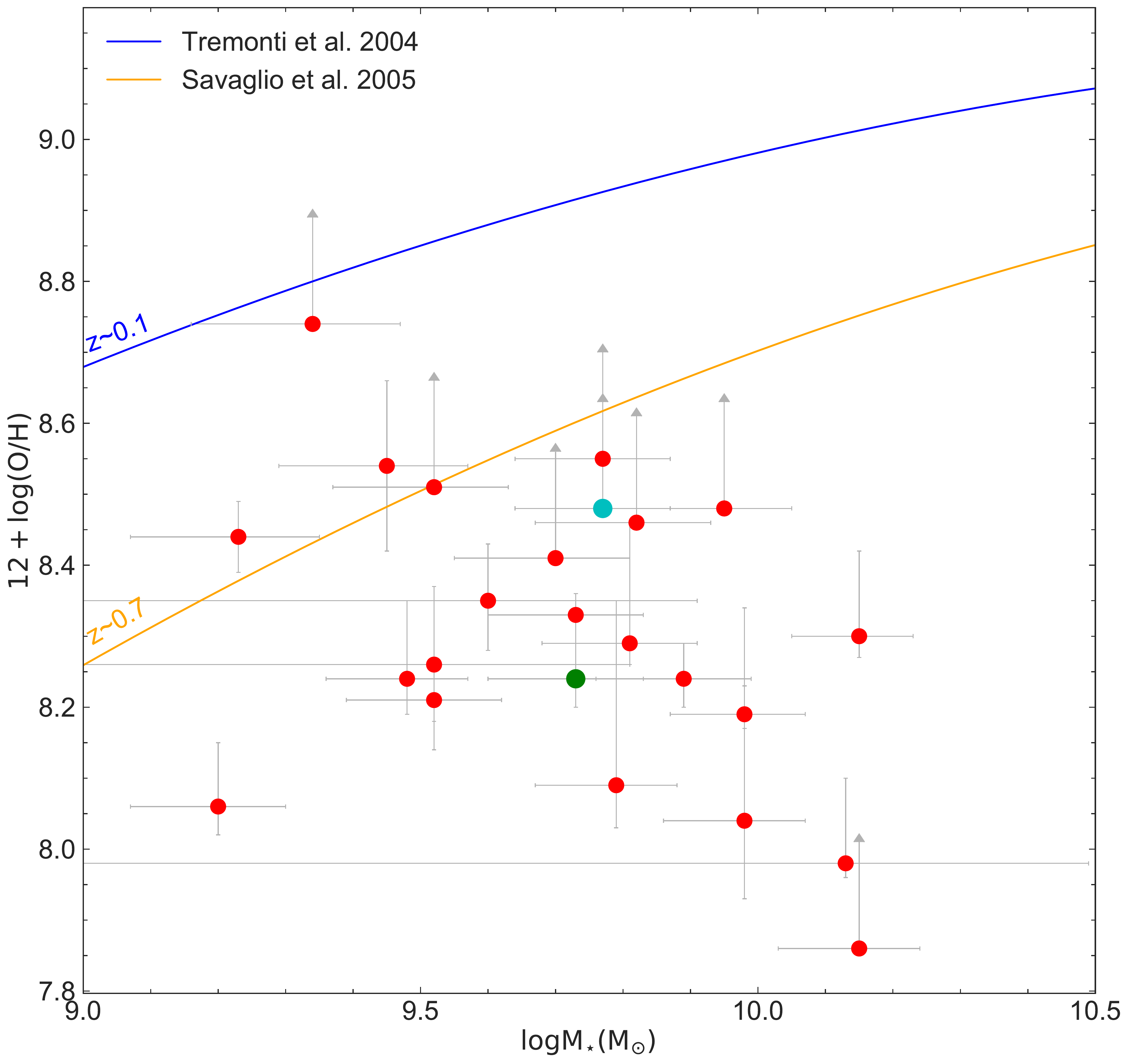}
\caption{Metallicity v.s. stellar mass of 22 [OIII] emitters (red solid circles). To compare our results with previous works, the fitting curves derived  in \citet{2004ApJ...613..898T} at z $\sim$ 0.1 and \citet{2005ApJ...635..260S} at z $\sim$ 0.7 are shown in blue and orange colors, respectively. The green point is an artificial point with median metallicity and stellar mass of our [OIII]$\lambda$4363-detected sample. Similarly, the cyan point represents the median metallicity and stellar mass of our [OIII]$\lambda$4363-non-detected sample.}
\label{fig:metal-mass}
\end{figure}

We then plot the SFR-\emph M$_{\star}$ relation in Figure \ref{fig:main} and the mass-metallicity relation in Figure \ref{fig:metal-mass} for the [OIII] emitters. Compared with SFR-\emph M$_{\star}$ sequences in \citet{2010ApJ...721..193P} at z $\sim$ 0 and \citet{2007A&A...468...33E}  at z $\sim$ 1, our [OIII] emitters have much higher SFR, indicative of more recent starbursts of $\sim$1 order of magnitude. As for the mass-metallicity relation, after shifting the relations in \citet{2004ApJ...613..898T} at z $\sim$ 0.1 and \citet{2005ApJ...635..260S} at z $\sim$ 0.7 to Chabrier IMF, we find that our galaxies have fairly low metallicity compared to other star-forming galaxies. Previous works (e.g. \citealt{2008ApJ...672L.107E, 2010MNRAS.408.2115M} and the \emph M$_{\star}$-metallicity-SFR fundamental plane in \citealt{2010A&A...521L..53L}) have revealed that galaxies with high SFR and low stellar masses (i.e., high specific SFR) have systematically low gas-phase metallicities, which is consistent with our finding here.
\newcommand{\msun}{$M_{\odot}$}
\begin{deluxetable}{ccccccc}
\tabletypesize{\footnotesize}
\tablewidth{\textwidth}
\tablecaption{{\centering}Stellar masses, burst ages and color excesses derived from SED fitting and H$\alpha$-derived SFRs of 22 [OIII] emitters\label{table:SFR}}

\tablehead{\colhead{ID}                          &        
           \colhead{log \emph M$_{\star}$}              &
           \colhead{burst age}              &
           \colhead{\emph{E(B -- V)}}              &
           \colhead{SFR}       &\\
            \colhead{}                              &
           \colhead{(\msun)}              &
           \colhead{(Myr)}                              &
           \colhead{(mag)}                              &
           \colhead{(\msun\ yr$^{-1}$)}&\\
           \colhead{(1)}                          &
           \colhead{(2)}              &
           \colhead{(3)}              &
           \colhead{(4)}              &
           \colhead{(5)}              &  
           }
\startdata 
 OIII-1&9.23$^{+0.12}_{-0.16}$ & 51$\pm${9}& 0.20$\pm${0.01}& 12.3$^{+0.9}_{-1.0}$\\
 OIII-2&9.20$^{+0.10}_{-0.13}$ & 57$\pm${17}& 0.10$\pm${0.01}& 9$^{+4}_{-2}$ \\
 OIII-3&9.52$^{+0.10}_{-0.13}$ & 80$\pm${38}& 0.15$\pm${0.05}& 25$^{+4}_{-6}$\\
 OIII-4&9.34$^{+0.13}_{-0.18}$ & 70$\pm${38}& 0.16$\pm${0.05}& 14$^{+8}_{-1}$\\
 OIII-5&10.13$^{+0.36}_{-\inf}$ & 54$\pm${16}& 0.22$\pm${0.04}& 12.3$^{+1.2}_{-0.3}$\\
 OIII-6&10.15$^{+0.08}_{-0.10}$ & 50$\pm${5}& 0.30$\pm${0.02}& 115$^{+104}_{-37}$\\
 OIII-7&9.52$^{+0.29}_{-1.32}$ & 97$\pm${51}& 0.17$\pm${0.06}& 28$^{+1}_{-2}$\\
 OIII-8&9.48$^{+0.09}_{-0.12}$ & 53$\pm${11}& 0.20$\pm${0.01}& 32$^{+2}_{-1}$\\
 OIII-9&9.81$^{+0.10}_{-0.13}$ & 63$\pm${30}& 0.29$\pm${0.02}& 31.6$^{+0.7}_{-1.4}$\\
 OIII-10&9.89$^{+0.10}_{-0.13}$ & 56$\pm${18}& 0.20$\pm${0.01}& 33.9$^{+0.1}_{-2.3}$\\
 OIII-11&9.45$^{+0.12}_{-0.16}$ & 97$\pm${32}& 0.10$\pm${0.005}& 28.8$^{+0.7}_{-1.3}$\\
 OIII-12&9.52$^{+0.11}_{-0.15}$ & 148$\pm${52}& 0.10$\pm${0.02}& 28$^{+16}_{-2}$\\
 OIII-13&9.77$^{+0.10}_{-0.13}$ & 135$\pm${58}& 0.11$\pm${0.03}& 43$^{+18}_{-8}$\\
 OIII-14&9.70$^{+0.11}_{-0.15}$ & 51$\pm${6}& 0.29$\pm${0.03}& 33$^{+12}_{-4}$\\
 OIII-15&9.60$^{+0.31}_{-\inf}$ & 72$\pm${36}& 0.16$\pm${0.06}& 23$^{+1}_{-2}$\\
 OIII-16&9.82$^{+0.11}_{-0.15}$ & 75$\pm${27}& 0.10$\pm${0.005}& 63$^{+22}_{-8}$\\
 OIII-17&9.73$^{+0.10}_{-0.13}$ & 87$\pm${30}& 0.09$\pm${0.02}& 56$^{+10}_{-8}$\\
 OIII-18&9.79$^{+0.09}_{-0.12}$ & 55$\pm${15}& 0.20$\pm${0.01}& 22$^{+6}_{-6}$\\
 OIII-19&9.95$^{+0.10}_{-0.13}$ & 76$\pm${25}& 0.10$\pm${0.005}& 129$^{+41}_{-31}$\\
 OIII-20&10.15$^{+0.09}_{-0.12}$ & 134$\pm${48}& 0.17$\pm${0.05}& 21$^{+11}_{-7}$\\
 OIII-21&9.98$^{+0.09}_{-0.12}$ & 55$\pm${15}& 0.20$\pm${0.01}& 123$^{+51}_{-32}$\\
 OIII-22&9.98$^{+0.09}_{-0.11}$ &100$\pm${5}& 0.05$\pm${0.01}& 44$^{+6}_{-10}$\\
\enddata


\tablenotetext{}{\textbf{Notes.} Fitting results derived from CIGALE and H$\alpha$-derived SFRs. (1) Object ID. (2) Stellar mass. (3) Age of the recent burst in Myr. (4) Color excess. (5) H$\alpha$-derived SFR. Error bars correspond to 68\% confidence levels.}
\end{deluxetable}

\subsection{UV Photometry v.s. Redshift}\label{sec:UV}
\begin{figure}
\includegraphics[width=\linewidth, clip]{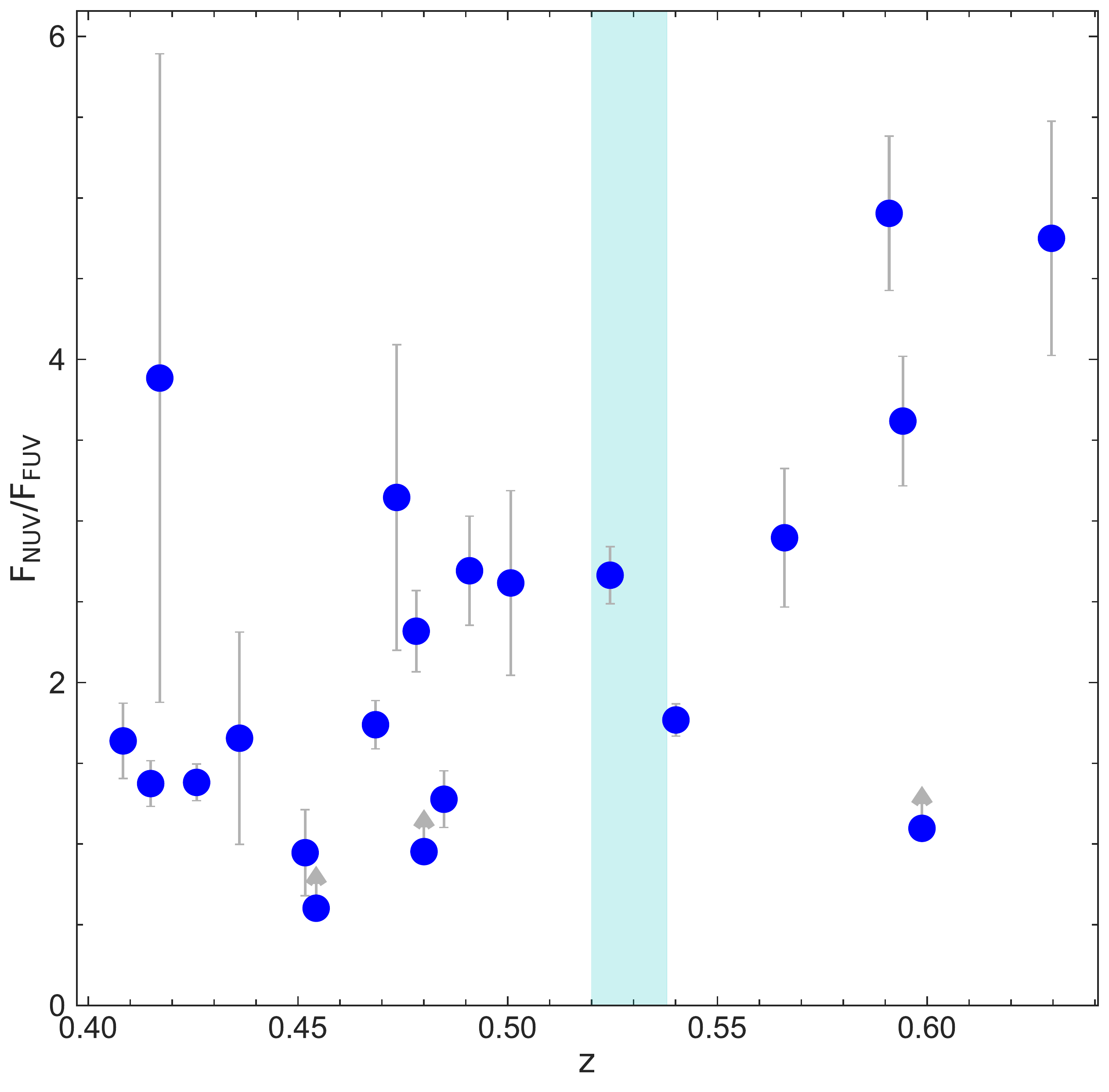}
\caption{NUV to FUV ratio of 22 [OIII] emitters (blue solid circles). The error bars (light gray lines) correspond to 68\% confidence levels. The objects which only have 3-$\sigma$ flux upper limits are marked with gray arrows. The cyan patch represents the O$_2$ A-band absorption by the earth's atmosphere at 7623-7699 \AA\ \citep{2001PASA...18..252W}, suggesting potential incompleteness of our sample.}
\label{fig:FUV/NUV}
\end{figure}

We calculated the ratios of NUV flux to FUV flux versus redshift (see Figure \ref{fig:FUV/NUV}). A positive correlation between the \emph{F}$_{\rm NUV}$/\emph{F}$_{\rm FUV}$ ratio and redshift is observed, which we believe is the result of Lyman break absorption. To quantify the statistical significance of this relation, we performed survival analysis\footnote{http://stsdas.stsci.edu/cgi-bin/gethelp.cgi?survival} using the \texttt{asurv} package \citep{1985ApJ...293..192F,1986ApJ...306..490I}. The derived probability that a correlation exists is 99.5\%, or 2.8$\sigma$ significance equivalently. All galaxies with z $>$ 0.55 have a drop in FUV flux of at least a factor of $\sim$3 compared to the NUV. This provides us with the possibility of predicting the redshift of [OIII] candidates, simply by calculating their \emph{F}$_{\rm NUV}$/\emph{F}$_{\rm FUV}$ ratio.

\subsection{Line Luminosity and Number Density}\label{sec:LF}
\begin{figure}
\includegraphics[width=\linewidth, clip]{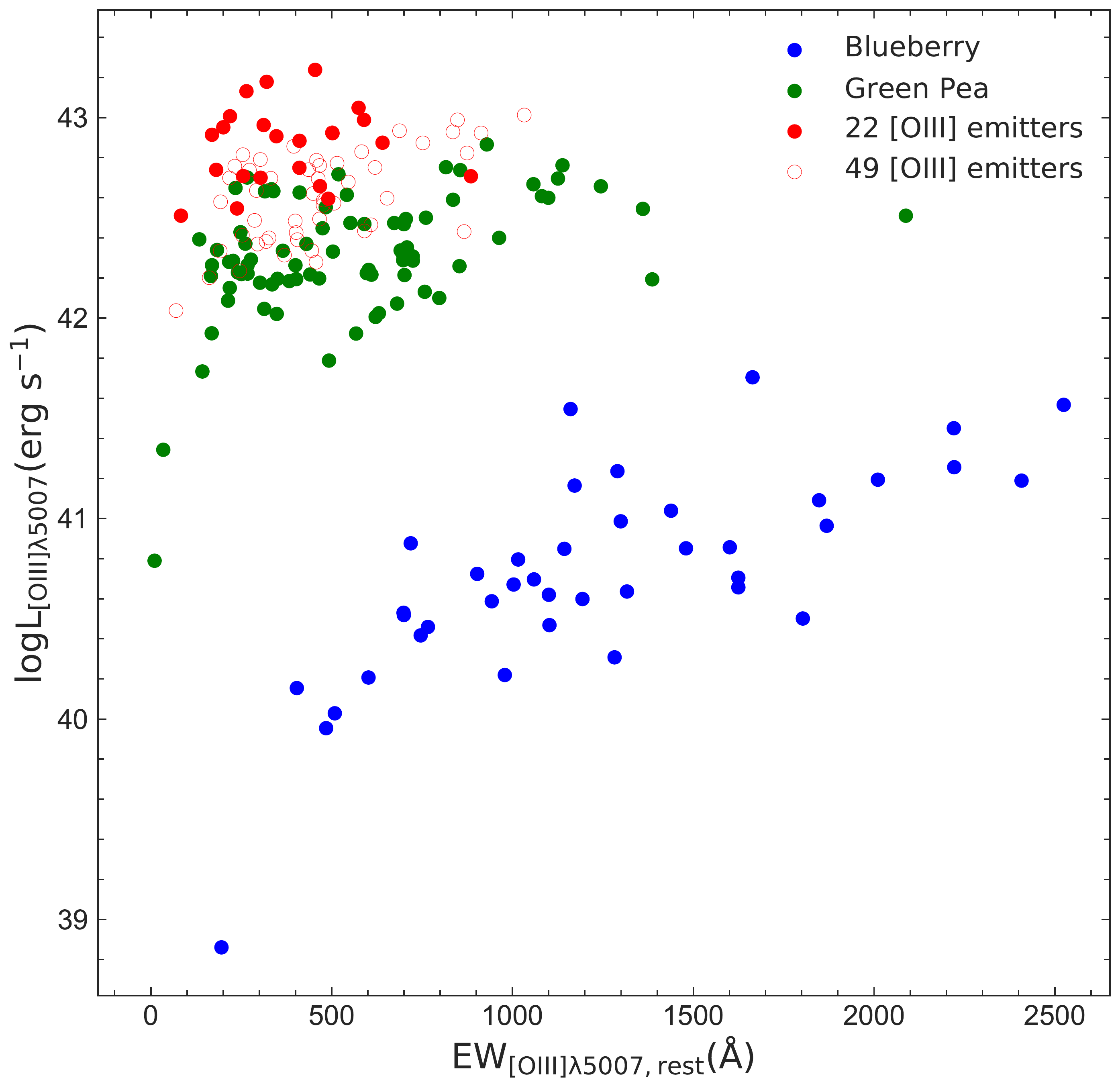}
\caption{The distribution of L$_{\rm [OIII]\lambda5007}$ and rest-frame EW$_{\rm [OIII]\lambda5007}$ for 22 [OIII] emitters in our total sample (shown in red solid circles) and 49 [OIII] emitters with 0.15 mag $<$ \texttt{err\_z} $<$ 0.25 mag (shown in red open circles) with EELGs at z $\lesssim$ 0.05 (`Blueberries', blue circles) and 0.112 $\lesssim$ z $\lesssim$ 0.360 (`Green Peas', green circles).}
\label{fig:distribution}
\end{figure}

We calculated the [OIII]$\lambda$5007 line luminosity for our [OIII] emitters (provided in Table \ref{table:OIII}) from L$_{\rm [OIII]\lambda5007}$ = 4$\pi$d$^2_L$F$_{\rm [OIII]\lambda5007}$, where F$_{\rm [OIII]\lambda5007}$ is the $\rm [OIII]\lambda5007$ line flux and d$_L$ is the luminosity distance. We found that our [OIII] emitters have remarkably high line luminosity: 18/22 have L$_{\rm [OIII]\lambda5007}$ $>$ 5$\times$10$^{42}$ erg $\rm s^{-1}$ and 5 of them have L$_{\rm [OIII]\lambda5007}$ $>$ 10$^{43}$ erg $\rm s^{-1}$. The mean and median L$_{\rm [OIII]\lambda5007}$ are 8.0$\times$10$^{42}$\ erg $\rm s^{-1}$ and 7.9$\times$10$^{42}$ erg $\rm s^{-1}$, respectively. We compare the distribution of L$_{\rm [OIII]\lambda5007}$ and rest-frame EW$_{\rm [OIII]\lambda5007}$ for our 22 [OIII] emitters with EELGs at z $\lesssim$ 0.05 (`Blueberries', \citealt{2017arXiv170602819Y}) and 0.112 $\lesssim$ z $\lesssim$ 0.360 (`Green Peas', \citealt{2009MNRAS.399.1191C}) in Figure \ref{fig:distribution}.

We estimated the volume number density of [OIII] emitters. First, we calculated the corresponding comoving volume searched by SDSS DR14 photometry: 
\begin{equation}
{\rm comoving\ volume}=\frac{\Omega}{4\pi} \frac{4\pi}{3} (R^{3}_{\rm z_{max}}-R^{3}_{\rm z_{min}})
\end{equation}
where $\Omega$ is the sky coverage solid angle of SDSS DR14,  $R_{\rm z_{max}}$ and $R_{\rm z_{min}}$ are the maximum and minimum comoving distance of our [OIII] emitters. With $\Omega$ = 14555 deg$^{2}$, $z_{\rm max}$ = 0.630 and $z_{\rm min}$ = 0.415 we derived a comoving volume = 1.288$\times$10$^{10}$ Mpc$^{3}$.

Based on our photometric selection criteria, our search in the SDSS DR14 \texttt{PhotoPrimary} catalog yielded 968 objects (our `primary sample'), of which 200 have observed SDSS spectra and 17 are [OIII] emitters at z $\sim$ 0.5. Our search in the SDSS DR14 \texttt{PhotoObjAll} catalog yielded 2658 unique objects (our `total sample'), in which 514 are spectroscopically observed (some of them have more than one spectrum) and 22 are [OIII] emitters at z $\sim$ 0.5. We assume the fraction of [OIII] emitters is the same for those objects which have not yet been spectroscopically observed. Then we estimate the total number of [OIII] emitters to be 82 or 114 (deduced from two samples respectively) in the corresponding comoving volume we searched photometrically. That means the number density of [OIII] emitters (with L$_{\rm [OIII]\lambda5007}$ down to $\sim$ 3$\times$10$^{42}$\ erg $\rm s^{-1}$) should be (6.0 $\sim$ 9.0)$\times$10$^{-9}$ Mpc$^{-3}$. 

To test the comprehensiveness of our selection criteria, we did a search in the SDSS DR14 \texttt{emissionLinesPort}\footnote{Emission line kinematics results for SDSS and BOSS galaxies. See: https://skyserver.sdss.org/dr10/en/help} spectroscopy catalog simply by requiring observed EW$_{\rm [OIII]\lambda5007}$ $>$ 300\AA\ and 0.4 $<$ z $<$ 0.6. This search yielded a sample of 564 objects with strong [OIII] emission. After checking all these spectra visually, we find most objects are actually QSOs or Seyfert 1.9 galaxies, because their spectra exhibit broad H$\alpha$ wings and [Ne V] lines, although some of those are misclassified as `starbursts' by SDSS. 
In total, we find 111 real [OIII] starbursts in this sample. Among these 111 [OIII] starbursts, only eight of them are in our sample of 22 [OIII] emitters. Nine of them don't satisfy Equation (\ref{eq1}) or (\ref{eq2}).
The other 94 are not selected into our sample because of 
large photometric uncertainties, either \texttt{err\_u} $>$ 0.25 mag, or/and \texttt{err\_z} $>$ 0.15 mag. We find that large photometry uncertainty indicates relatively low L$_{\rm [OIII]\lambda5007}$ -- only 7 [OIII] starbursts (6\%) with \texttt{z\_err} $>$ 0.25 mag have L$_{\rm [OIII]\lambda5007}$ $>$ 5$\times$10$^{42}$ erg $\rm s^{-1}$.

Since the photometric uncertainty turns out to be so critical to our sample selection, we repeated the photometric search, keeping all other criteria unchanged, but loosening our limit on \texttt{err\_z} to be 0.25 mag. This search increases our number of candidates from the SDSS DR14 \texttt{PhotoPrimary} catalog dramatically, up to 5169. Of these 5169 candidates, 771 have observed SDSS spectra. Our inspection shows that 66 of them are genuine [OIII] starbursts at z $\sim$ 0.5. 
The mean and median L$_{\rm [OIII]\lambda5007}$ of 49 newly found [OIII] starbursts are 4.6$\times$10$^{42}$ erg $\rm s^{-1}$ and 4.2$\times$10$^{42}$ erg $\rm s^{-1}$, respectively. Twenty of them have L$_{\rm [OIII]\lambda5007}$ $>$ 5$\times$10$^{42}$ erg $\rm s^{-1}$, and 33 of them have L$_{\rm [OIII]\lambda5007}$ $>$ 3$\times$10$^{42}$ erg $\rm s^{-1}$. The number of [OIII] starbursts could still increase if we continue to loosen the constraints in the photometry uncertainty, but doing so will not be likely to: (1) increase the `success rate' (defined as the number of [OIII] starbursts/the number of SDSS spectra available) of our photometric selection; (2) find more [OIII] starbursts with L$_{\rm [OIII]\lambda5007}$ $>$ 5$\times$10$^{42}$ erg $\rm s^{-1}$. Thus, we deduce from our searches that the space density of [OIII] emitters is $\sim$ 2$\times$10$^{-8}$ Mpc$^{-3}$, with L$_{\rm [OIII]\lambda5007}$ down to $\sim$ 3$\times$10$^{42}$\ erg $\rm s^{-1}$.


These are smaller volume number densities than with previous studies of extreme emission line galaxies at lower redshifts: 

(1) $\sim$ 3.0$\times$10$^{-6}$ Mpc$^{-3}$ for EELGs at z $\lesssim$ 0.05 (`Blueberries', \citealt{2017arXiv170602819Y});

(2) $\sim$ 7.0$\times$10$^{-6}$ Mpc$^{-3}$ for EELGs at 0.112 $\lesssim$ z $\lesssim$ 0.360 (`Green Peas', \citealt{2009MNRAS.399.1191C}).
\begin{figure}
\includegraphics[width=\linewidth, clip]{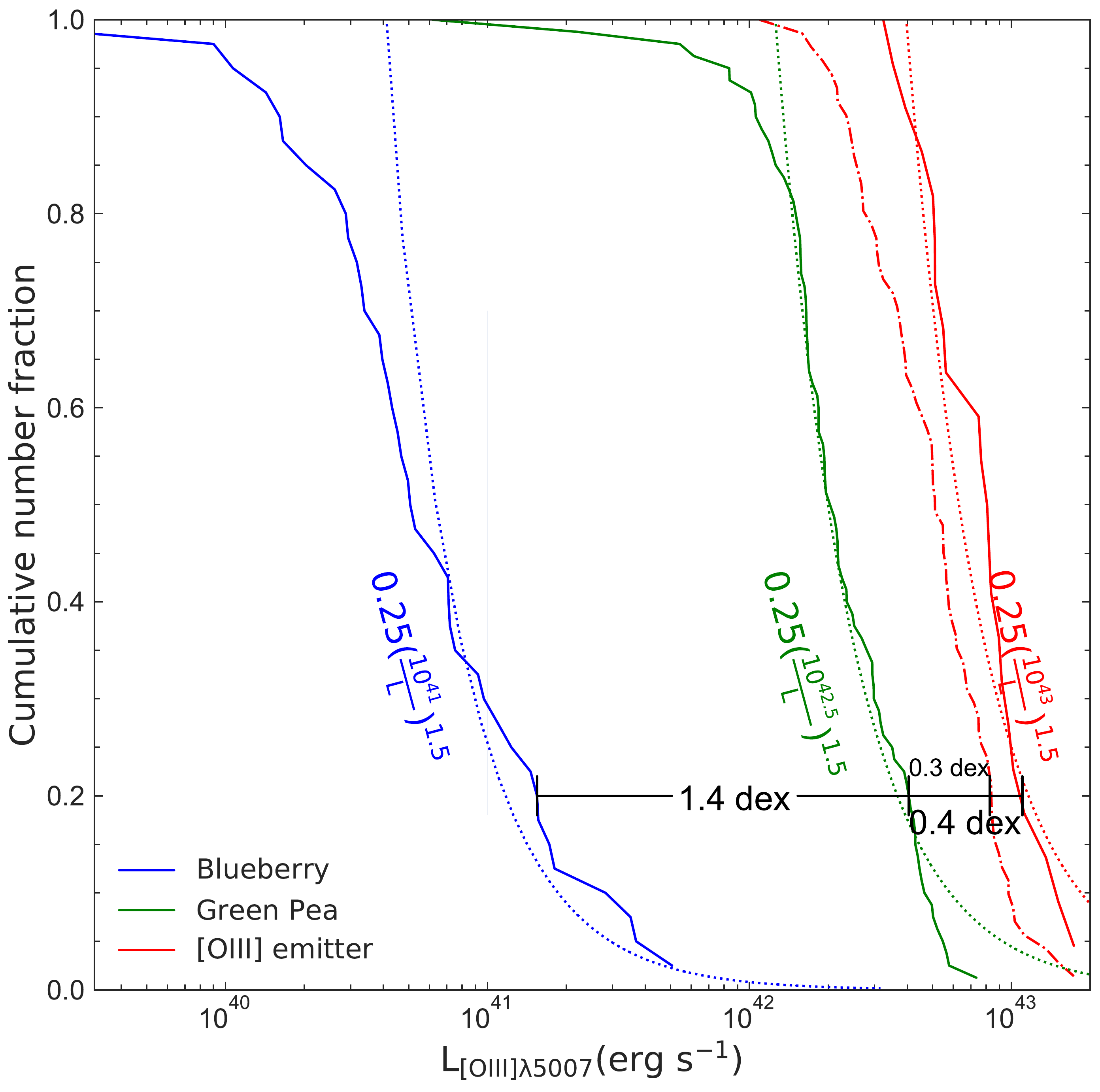}
\caption{The cumulative number fractions with L$_{\rm [OIII]\lambda5007}$ $>$ L$_0$ (L$_0$ is a certain luminosity) of Blueberries, Green Peas and [OIII] emitters. At a cumulative number fraction of 0.2, the difference in L$_{\rm [OIII]\lambda5007}$ is 1.4 dex between Blueberries and Green Peas, and 0.4 dex between Green Peas and 22 [OIII] emitters. Even when taking 49 [OIII] emitters with 0.15 mag $<$ \texttt{err\_z} $<$ 0.25 mag into account (the red dash-dotted line), the difference in the L$_{\rm [OIII]\lambda5007}$ cumulative distribution is still 0.3 dex, indicative of strong redshift evolution at the bright end of the [OIII] luminosity function. Note that the dashed lines are not fits to the data but L$^{-1.5}$ power laws, with coefficients given in the plot.}
\label{fig:LF-evolution}
\end{figure}

However, it is not very fair to directly compare these different number densities. As Figure \ref{fig:distribution} shows, our [OIII] emitters are substantially more luminous and would therefore be expected to be less common.  Although the distribution of the rest-frame EW$_{\rm [OIII]\lambda5007}$, which mainly range from 200 \AA\ to 600 \AA, and the brightness of continuum at corresponding wavelengths (\emph{g}/\emph{r}/\emph{i}-band, 18-20.5 mag) of our [OIII] emitters are similar to Blueberries and Green Peas, which indicates they have similar line flux F$_{\rm [OIII]\lambda5007}$, their line luminosities L$_{\rm [OIII]\lambda5007}$ differ a lot because of their different luminosity distances. In fact, the luminosities of our [OIII] emitters have no overlap at all with the Blueberries, which are typically $\sim$ 30 times fainter. Although there is a modest overlap between the luminosities of our [OIII] emitters and those of Green Peas, it is apparent in Figure \ref{fig:distribution} that our sample extends to [OIII] luminosities several times higher than {\it any} Green Peas (above $10^{43}$\ erg $\rm s^{-1}$). \textbf{We interpret this as redshift evolution of the bright end of the [OIII] luminosity function.}


The lack of overlap between our sample and the Blueberries means that no meaningful comparison at a given L$_{\rm [OIII]\lambda5007}$ can be made. In spite of our incompleteness at L$_{\rm [OIII]\lambda5007}$ below $3 \times 10^{42}$\ erg $\rm s^{-1}$, we are able to compare our [OIII] emitters with the Green Peas in the overlapping region. We compare the luminous ends of the [OIII] LFs in the cumulative histograms (Figure \ref{fig:LF-evolution}), showing all three samples of EELGs.

To help interpret these histograms, we heuristically assume the cumulative number fraction of EELGs with L$_{\rm [OIII]\lambda5007}$ $>$ L$_0$ (L$_0$ is a certain luminosity) $\propto$ L$^{-\alpha}$. We find that $\alpha$ = 1.5 power laws can fairly well match these three cumulative distributions simultaneously (shown by the dotted blue, green and red lines).  At a cumulative number fraction of 0.2, the difference in L$_{\rm [OIII]\lambda5007}$ is 1.4 dex between Blueberries (solid blue line) and Green Peas (solid green line), and 0.4 dex between Green Peas and our [OIII] emitters (solid red line). Even when taking 49 [OIII] emitters with 0.15 mag $<$ \texttt{err\_z} $<$ 0.25 mag into account (shown by the dash-dotted red line), the difference in the L$_{\rm [OIII]\lambda5007}$ cumulative distribution is still 0.3 dex. 

Since there is no reason that the most luminous Green Peas are missing at z $\sim$ 0.2, the fact that our $z \sim 0.5$ [OIII] emitter sample extends to significantly higher luminosities indicates that the \textbf{bright end of the [OIII] luminosity function evolves with redshift.}  At L$_{\rm [OIII]\lambda5007}$ above $ 5 \times 10^{42}$\ erg $\rm s^{-1}$, this evolution appears to be a factor of 2 to 2.5. One (non-unique) interpretation, of pure luminosity evolution, could be that the L$_{\rm [OIII]\lambda5007}$ increase as $\propto (1+z)^{3\sim4}$.  Our high-z search here, however, is too incomplete at lower luminosities to draw more definitive conclusions.


\begin{deluxetable*}{cccccc}
\tabletypesize{\footnotesize}
\tablewidth{\textwidth}
\tablecaption{{\centering}Balmer decrements and derived extinction\label{table:e_bv}}

\tablehead{\colhead{ID}              &
           \colhead{H$\alpha$/H$\beta$}                          &        
           \colhead{H$\gamma$/H$\beta$}              &
           \colhead{\emph{E(B--V)}$_{\rm H\alpha, H\beta}$}       &
           \colhead{\emph{E(B--V)}$_{\rm H\gamma, H\beta}$}              &\\
           \colhead{(1)}                          &
           \colhead{(2)}              &
           \colhead{(3)}              &  
           \colhead{(4)}              &
           \colhead{(5)}              &   
           }
\startdata 
  OIII-1& ... & 0.43$^{+0.01}_{-0.00}$ & ... & 0.15$^{+0.03}_{-0.02}$\\
  OIII-2  & ... & 0.45$^{+0.02}_{-0.02}$ & ... & 0.09$^{+0.08}_{-0.09}$\\
  OIII-3& ... & 0.44$^{+0.02}_{-0.01}$ & ... & 0.13$^{+0.06}_{-0.07}$\\
  OIII-4& ... & 0.45$^{+0.02}_{-0.02}$ & ... & 0.09$^{+0.08}_{-0.09}$\\
  OIII-5&3.61$^{+0.10}_{-0.11}$ & 0.39$^{+0.03}_{-0.01}$ & 0.22$^{+0.04}_{-0.01}$ & 0.29$^{+0.07}_{-0.11}$\\
  OIII-6& ... & 0.37$^{+0.03}_{-0.03}$ & ... & 0.44$^{+0.15}_{-0.17}$\\
  OIII-7&3.68$^{+0.12}_{-0.06}$ & 0.35$^{+0.02}_{-0.00}$ & 0.28$^{+0.01}_{-0.04}$ & 0.51$^{+0.07}_{-0.06}$\\
  OIII-8&3.66$^{+0.07}_{-0.06}$ & 0.39$^{+0.01}_{-0.01}$ & 0.24$^{+0.03}_{-0.01}$ & 0.36$^{+0.06}_{-0.05}$\\
  OIII-9&3.46$^{+0.12}_{-0.01}$ & 0.40$^{+0.01}_{-0.02}$ & 0.21$^{+0.02}_{-0.01}$ & 0.30$^{+0.08}_{-0.03}$\\
  OIII-10&3.34$^{+0.09}_{-0.04}$ & 0.44$^{+0.01}_{-0.01}$ & 0.16$^{+0.02}_{-0.02}$ & 0.12$^{+0.03}_{-0.05}$\\
  OIII-11&3.18$^{+0.02}_{-0.12}$ & 0.43$^{+0.01}_{-0.01}$ & 0.10$^{+0.01}_{-0.03}$ & 0.15$^{+0.09}_{-0.01}$\\
  OIII-12& ... & 0.41$^{+0.02}_{-0.01}$ & ... & 0.26$^{+0.07}_{-0.09}$\\
  OIII-13& ... & 0.38$^{+0.02}_{-0.02}$ & ... & 0.35$^{+0.12}_{-0.05}$\\
  OIII-14& ... & 0.40$^{+0.01}_{-0.01}$ & ... & 0.30$^{+0.05}_{-0.08}$\\
  OIII-15&3.05$^{+0.05}_{-0.08}$ & 0.42$^{+0.01}_{-0.01}$ & 0.06$^{+0.03}_{-0.02}$ & 0.19$^{+0.06}_{-0.03}$\\
  OIII-16& ... & 0.41$^{+0.01}_{-0.01}$ & ... & 0.27$^{+0.07}_{-0.06}$\\
  OIII-17& ... & 0.42$^{+0.02}_{-0.00}$ & ... & 0.18$^{+0.06}_{-0.04}$\\
  OIII-18& ... & 0.45$^{+0.02}_{-0.01}$ & ... & 0.09$^{+0.06}_{-0.10}$\\
  OIII-19& ... & 0.34$^{+0.02}_{-0.00}$ & ... & 0.55$^{+0.11}_{-0.05}$\\
  OIII-20 & ... & 0.40$^{+0.03}_{-0.02}$ & ... & 0.20$^{+0.19}_{-0.07}$\\
  OIII-21& ... & 0.33$^{+0.02}_{-0.02}$ & ... & 0.61$^{+0.15}_{-0.06}$\\
  OIII-22& ... & 0.43$^{+0.01}_{-0.02}$ & ... & 0.13$^{+0.12}_{-0.01}$\\
\enddata


\tablenotetext{}{\textbf{Notes.} (1) Object ID. (2) Flux ratio of H$\alpha$ and H$\beta$. (3) Flux ratio of H$\gamma$ and H$\beta$. (4) Color excess derived from H$\alpha$/H$\beta$. (5) Color excess derived from H$\gamma$/H$\beta$. The uncertainties correspond to 68\% confidence levels. }
\end{deluxetable*}

\begin{deluxetable*}{ccccccccccc}
\tabletypesize{\footnotesize}
\tablewidth{\textwidth}
\tablecaption{{\centering}Emission line properties of 22 [OIII] emitters\label{table:OIII}}

\tablehead{\colhead{ID}              &
           \colhead{RA}                          &        
           \colhead{Dec}              &
           \colhead{z}       &
           \colhead{$\rm log\left(\frac{[NII]}{H\alpha}\right)$}  &
           \colhead{$\rm log\left(\frac{[SII]}{H\alpha}\right)$}  &
           \colhead{$\rm log\left(\frac{[OI]}{H\alpha}\right)$}  &
           \colhead{EW$_{\rm [OIII]\lambda5007,\ obs}$}              &
           \colhead{F$_{\rm [OIII]\lambda5007}$}       &
           \colhead{L$_{\rm [OIII]\lambda5007}$}       &
           \\
           \colhead{}                              &
           \colhead{(J2000 deg)}                              &
           \colhead{(J2000 deg)}              &
           \colhead{}                              &
           \colhead{}                              &
           \colhead{}                              &
           \colhead{}                              &
           \colhead{(\ \AA\ )}                      &
           \colhead{($\rm 10^{-17}$erg $\rm s^{-1}$ $\rm cm^{-2}$)} &
           \colhead{(erg $\rm s^{-1}$)}\\
           \colhead{(1)}                          &
           \colhead{(2)}              &
           \colhead{(3)}              &  
           \colhead{(4)}              &
           \colhead{(5)}              & 
           \colhead{(6)}              &
           \colhead{(7)}              &
           \colhead{(8)}              & 
           \colhead{(9)}              &
           \colhead{(10)}              &          
           }
\startdata 
OIII-1&124.27618 & 31.98355 & 0.408 &...&...&...& 691&626&3.9$\times 10^{42}$ \\
OIII-2&28.87035   & -1.12973 & 0.415 &...&...&...&1252 &  781&5.1$\times 10^{42}$\\
OIII-3&237.06111  & 7.45281 & 0.417 &...&...&...& 492&1221&8.1$\times 10^{42}$\\
OIII-4&1.92919     & 0.04127 & 0.426 &...&...&...& 339 &507&3.5$\times 10^{42}$ \\
OIII-5&144.25380 & 36.43289 & 0.436 &-1.59$\pm$0.07&-1.00$\pm$0.03&-1.72$\pm$0.08& 672&619&4.5$\times 10^{42}$ \\
OIII-6&238.81838 & 35.72351 & 0.452 &...&...&...& 384&1695&1.4$\times 10^{43}$ \\
OIII-7&182.09547 & 15.40173 & 0.454 &-1.05$\pm$0.02&-0.83$\pm$0.03&-1.48$\pm$0.03& 597&694&5.6$\times 10^{42}$\\
OIII-8&153.74986 & 3.16105 & 0.468 &-1.05$\pm$0.02&-0.97$\pm$0.03&-1.68$\pm$0.04& 941&860&7.5$\times 10^{42}$ \\
OIII-9&142.94537 & 60.62296 & 0.474 &-1.00$\pm$0.03&-0.87$\pm$0.03&-1.57$\pm$0.04& 606&857&7.7$\times 10^{42}$ \\
OIII-10&135.89416 & 17.78826 & 0.478 &-1.10$\pm$0.02&-0.86$\pm$0.02&-2.1$\pm$0.9& 849&1224&1.1$\times 10^{43}$ \\
OIII-11&244.89253 & 22.32658 & 0.480 &-1.11$\pm$0.03&-0.95$\pm$0.08&$<$-2.7& 743&906&8.4$\times 10^{42}$ \\
OIII-12&48.98032   & -1.07757 & 0.485 &...&...&...&450 & 531&5.0$\times 10^{42}$ \\
OIII-13&16.18940   & 0.22369 & 0.491 &...&...&...& 269&564&5.5$\times 10^{42}$ \\
OIII-14&211.53847   & 49.04090 & 0.501 &...&...&...&383& 500&5.1$\times 10^{42}$ \\
OIII-15&350.74701   &23.02511 & 0.507 &-1.16$\pm$0.05&-0.86$\pm$0.10&-1.66$\pm$0.07&888& 927&9.8$\times 10^{42}$ \\
OIII-16&25.73972   & 1.14013 & 0.524 &...&...&...& 333&890&1.0$\times 10^{43}$ \\
OIII-17&31.94996   & 0.79318 & 0.540 &...&...&...& 700& 1413&1.7$\times 10^{43}$\\
OIII-18&27.44835   & 1.00270 & 0.566 &...&...&...& 488& 670&9.2$\times 10^{42}$ \\
OIII-19&132.92699 & 50.61637 & 0.591 &...&...&...& 269&540&8.2$\times 10^{42}$\\
OIII-20&3.57051     & 0.23331& 0.594 &...&...&...&132&210&3.2$\times 10^{42}$ \\
OIII-21&244.17766 & 20.95559 & 0.599 &...&...&...& 320&569&9.0$\times 10^{42}$ \\
OIII-22&49.99925   & -1.08055 & 0.630&...&...&... & 522&850&1.5$\times 10^{43}$ \\
\enddata


\tablenotetext{}{\textbf{Notes.} (2)-(3) Right ascension and declination. (4) Redshift. (5)-(7) Emission line ratios of $\rm {[NII]}/{H\alpha}$, $\rm {[SII]}/{H\alpha}$ and $\rm {[OI]}/{H\alpha}$. (8) Observed-frame equivalent width of [OIII]$\lambda$5007. (9) Emission line flux of [OIII]$\lambda$5007. (10) Line luminosity of [OIII]$\lambda$5007. Column (8) and (9) are taken from the default emission line fits in SDSS DR14. The uncertainties correspond to 68\% confidence levels. The median relative uncertainties of EW$_{\rm [OIII]\lambda5007,\ obs}$ and F$_{\rm [OIII]\lambda5007}$ are 9$\permil$ and 4$\permil$, with a maximum of 2\% and 1\% respectively, according to the SDSS DR14 catalog.}
\end{deluxetable*}

\begin{deluxetable*}{cccccccccccccccc}
\centering
\tabletypesize{\tiny}
\tablewidth{\textwidth}
\tablecaption{{\centering}Emission-line ratios, electron temperatures and metallicities\label{table:metal}}

\tablehead{\colhead{ID}              &
            \multicolumn{2}{c}{$\rm log\left(\frac{[OIII]\lambda\lambda4959, 5007}{[OIII]\lambda4363}\right)$}                &
           \multicolumn{2}{c}{$\rm log(T_{e}/K)$}                &
           \multicolumn{2}{c}{$\rm log\left(\frac{[OII]}{H\beta}\right)$}                &
            \colhead{$\rm log\left(\frac{[OIII]}{H\beta}\right)$}                          &
           \multicolumn{2}{c}{$\rm log\left(\frac{O^{+}}{H^{+}}\right)$}                &
          \multicolumn{2}{c}{$\rm log\left(\frac{O^{++}}{H^{+}}\right)$}                &
           \multicolumn{2}{c}{$\rm 12+log\left(\frac{O}{H}\right)$}               &\\
           \colhead{(1)}                          &
           \colhead{(2)}              &
           \colhead{(3)}              &  
           \colhead{(4)}              &
           \colhead{(5)}              &
           \colhead{(6)}              &
           \colhead{(7)}              &
           \colhead{(8)}              &
           \colhead{(9)}              &
           \colhead{(10)}              &
           \colhead{(11)}              &
           \colhead{(12)}              &
           \colhead{(13)}              &
          \colhead{(14)}              &
           }
\startdata 
  OIII-1 & 2.28$^{+0.07}_{-0.03}$ & 2.25$^{+0.06}_{-0.03}$ & 3.99$^{+0.01}_{-0.02}$ & 4.00$^{+0.01}_{-0.02}$ & -0.17$^{+0.01}_{-0.00}$ & -0.10$^{+0.01}_{-0.00}$ & 0.87$^{+0.00}_{-0.00}$
& -4.17$^{+0.01}_{-0.11}$ & -4.22$^{+0.08}_{-0.04}$ & -3.68$^{+0.10}_{-0.00}$ & -3.71$^{+0.08}_{-0.02}$
& 8.41$^{+0.11}_{-0.00}$ & 8.44$^{+0.05}_{-0.05}$\\
  OIII-2 & 1.99$^{+0.06}_{-0.05}$ & 1.96$^{+0.06}_{-0.05}$ & 4.08$^{+0.03}_{-0.01}$ & 4.10$^{+0.02}_{-0.02}$ & -0.26$^{+0.01}_{-0.03}$ &-0.23$^{+0.03}_{-0.02}$ & 0.90$^{+0.02}_{-0.01}$ 
& -4.78$^{+0.13}_{-0.03}$ & -4.74$^{+0.10}_{-0.06}$ & -3.95$^{+0.05}_{-0.08}$ & -4.01$^{+0.09}_{-0.04}$
& 8.06$^{+0.11}_{-0.02}$ & 8.06$^{+0.09}_{-0.04}$\\
  OIII-3 & 2.12$^{+0.15}_{-0.03}$ & 2.07$^{+0.16}_{-0.02}$ & 4.04$^{+0.02}_{-0.04}$ & 4.04$^{+0.03}_{-0.03}$ & -0.18$^{+0.03}_{-0.01}$ &-0.11$^{+0.01}_{-0.02}$ & 0.81$^{+0.00}_{-0.01}$ 
& -4.48$^{+0.23}_{-0.02}$ & -4.31$^{+0.07}_{-0.17}$ & -3.82$^{+0.09}_{-0.12}$ &-3.88$^{+0.09}_{-0.11}$
& 8.22$^{+0.16}_{-0.06}$ & 8.21$^{+0.14}_{-0.07}$\\
  OIII-4 & >2.65 & >2.63 & $<$3.89 & $<$3.89 &-0.13$^{+0.02}_{-0.02}$ &-0.08$^{+0.02}_{-0.02}$ &0.71$^{+0.01}_{-0.00}$ 
& >-3.76 & >-3.75 & >-3.42& >-3.44
& >8.74 & >8.74\\
  OIII-5 & 1.95$^{+0.05}_{-0.06}$ & 1.84$^{+0.08}_{-0.03}$ & 4.11$^{+0.02}_{-0.02}$ & 4.16$^{+0.01}_{-0.04}$ & -0.08$^{+0.01}_{-0.01}$ &+0.06$^{+0.01}_{-0.02}$ & 0.91$^{+0.01}_{-0.01}$
& -4.64$^{+0.12}_{-0.04}$ & -4.60$^{+0.10}_{-0.07}$ & -3.99$^{+0.05}_{-0.08}$&-4.14$^{+0.12}_{-0.02}$ 
& 8.03$^{+0.13}_{-0.01}$ & 7.98$^{+0.12}_{-0.02}$\\
  OIII-6 & 2.32$^{+0.07}_{-0.07}$ & 2.20$^{+0.06}_{-0.06}$ & 3.98$^{+0.02}_{-0.02}$ & 4.02$^{+0.02}_{-0.02}$ &-0.12$^{+0.01}_{-0.00}$ &+0.09$^{+0.01}_{-0.01}$ & 0.77$^{+0.00}_{-0.01}$
& -4.16$^{+0.10}_{-0.07}$ & -4.12$^{+0.11}_{-0.05}$ & -3.72$^{+0.06}_{-0.08}$&-3.90$^{+0.11}_{-0.03}$
& 8.41$^{+0.08}_{-0.07}$ & 8.30$^{+0.12}_{-0.03}$\\
  OIII-7 & 2.22$^{+0.10}_{-0.06}$ & 2.10$^{+0.07}_{-0.08}$ & 4.01$^{+0.02}_{-0.03}$ & 4.05$^{+0.03}_{-0.03}$ &-0.07$^{+0.01}_{-0.01}$ &+0.17$^{+0.01}_{-0.02}$ & 0.80$^{+0.01}_{-0.01}$
& -4.25$^{+0.15}_{-0.06}$ & -4.07$^{+0.03}_{-0.18}$ & -3.82$^{+0.13}_{-0.05}$& -3.92$^{+0.07}_{-0.10}$ 
& 8.32$^{+0.13}_{-0.06}$ & 8.26$^{+0.11}_{-0.08}$\\
  OIII-8 & 2.22$^{+0.10}_{-0.04}$ & 2.11$^{+0.11}_{-0.03}$ & 4.00$^{+0.02}_{-0.02}$ &4.05$^{+0.01}_{-0.04}$ &-0.25$^{+0.03}_{-0.00}$ &-0.08$^{+0.03}_{-0.00}$ & 0.82$^{+0.01}_{-0.01}$
& -4.37$^{+0.10}_{-0.07}$ & -4.36$^{+0.15}_{-0.04}$ & -3.78$^{+0.10}_{-0.05}$& -3.88$^{+0.10}_{-0.06}$
 & 8.34$^{+0.08}_{-0.07}$ & 8.24$^{+0.11}_{-0.05}$\\
  OIII-9 & 2.31$^{+0.08}_{-0.10}$ & 2.15$^{+0.16}_{-0.01}$ & 3.98$^{+0.03}_{-0.03}$ & 4.00$^{+0.03}_{-0.02}$ &-0.13$^{+0.01}_{-0.01}$ &+0.00$^{+0.01}_{-0.01}$ & 0.79$^{+0.01}_{-0.00}$ 
& -4.11$^{+0.04}_{-0.19}$ & -4.22$^{+0.18}_{-0.03}$ & -3.78$^{+0.18}_{-0.03}$& -3.89$^{+0.18}_{-0.00}$
& 8.38$^{+0.15}_{-0.06}$ & 8.29$^{+0.17}_{-0.03}$\\
  OIII-10 & 2.12$^{+0.03}_{-0.03}$ & 2.06$^{+0.06}_{-0.01}$ & 4.04$^{+0.01}_{-0.01}$ & 4.06$^{+0.01}_{-0.01}$ &-0.02$^{+0.01}_{-0.01}$ &+0.04$^{+0.01}_{-0.01}$ & 0.85$^{+0.01}_{-0.00}$
& -4.31$^{+0.05}_{-0.04}$ & -4.30$^{+0.05}_{-0.04}$ & -3.87$^{+0.04}_{-0.03}$& -3.90$^{+0.04}_{-0.04}$
& 8.25$^{+0.06}_{-0.02}$ & 8.24$^{+0.05}_{-0.04}$\\
  OIII-11 & 2.43$^{+0.11}_{-0.10}$ & 2.33$^{+0.14}_{-0.06}$ & 3.95$^{+0.03}_{-0.03}$ & 3.97$^{+0.02}_{-0.04}$ &-0.07$^{+0.02}_{-0.01}$ &+0.04$^{+0.01}_{-0.01}$ & 0.79$^{+0.00}_{-0.01}$ 
& -3.97$^{+0.14}_{-0.11}$ & -4.01$^{+0.20}_{-0.05}$ & -3.65$^{+0.19}_{-0.03}$& -3.72$^{+0.19}_{-0.04}$
& 8.53$^{+0.16}_{-0.07}$ & 8.54$^{+0.12}_{-0.12}$\\
  OIII-12 & >2.45 & >2.38 & $<$3.94 & $<$3.96 &-0.07$^{+0.01}_{-0.02}$ &+0.05$^{+0.01}_{-0.02}$ &0.72$^{+0.01}_{-0.01}$
& >-3.94 & >-3.90 & >-3.63& >-3.70
& >8.54 & >8.51\\
  OIII-13 & >2.46 & >2.36 & $<$3.94 & $<$3.97 &-0.05$^{+0.03}_{-0.00}$ &+0.12$^{+0.01}_{-0.02}$ &0.78$^{+0.01}_{-0.01}$
& >-3.91 & >-3.85 & >-3.56& >-3.67
& >8.60 & >8.55\\
  OIII-14 & >2.34 & >2.26 & $<$3.97 & $<$4.00 &-0.00$^{+0.02}_{-0.01}$ &+0.15$^{+0.01}_{-0.02}$ &0.71$^{+0.02}_{-0.00}$
& >-4.01 & >-3.96 & >-3.76&>-3.83
& >8.44 & >8.41\\
  OIII-15 & 2.25$^{+0.05}_{-0.07}$ & 2.14$^{+0.12}_{-0.01}$ & 4.00$^{+0.02}_{-0.02}$ & 4.02$^{+0.02}_{-0.02}$ &-0.08$^{+0.01}_{-0.01}$ &+0.02$^{+0.01}_{-0.01}$ & 0.84$^{+0.01}_{-0.00}$
& -4.24$^{+0.12}_{-0.04}$ & -4.19$^{+0.09}_{-0.07}$ & -3.71$^{+0.05}_{-0.09}$& -3.80$^{+0.08}_{-0.06}$
& 8.36$^{+0.11}_{-0.04}$ & 8.35$^{+0.08}_{-0.07}$\\
  OIII-16 & >2.44 & >2.36 & $<$3.94 & $<$3.97 &-0.15$^{+0.02}_{-0.02}$ &-0.03$^{+0.03}_{-0.01}$ & 0.71$^{+0.01}_{-0.00}$
& >-4.03 & >-4.00 & >-3.65 & >-3.73
& >8.50 & >8.46\\
  OIII-17 & 2.16$^{+0.09}_{-0.02}$ & 2.12$^{+0.07}_{-0.03}$ & 4.02$^{+0.02}_{-0.02}$ & 4.06$^{+0.00}_{-0.04}$ &-0.09$^{+0.01}_{-0.02}$ &-0.01$^{+0.02}_{-0.01}$ & 0.86$^{+0.01}_{-0.01}$
& -4.32$^{+0.11}_{-0.04}$ & -4.33$^{+0.14}_{-0.00}$ & -3.78$^{+0.07}_{-0.05}$& -3.85$^{+0.09}_{-0.03}$
& 8.32$^{+0.09}_{-0.03}$ & 8.33$^{+0.03}_{-0.08}$\\
  OIII-18 & 2.00$^{+0.24}_{-0.03}$ & 2.04$^{+0.17}_{-0.09}$ & 4.09$^{+0.01}_{-0.08}$ & 4.07$^{+0.04}_{-0.06}$ &-0.11$^{+0.03}_{-0.02}$ &-0.06$^{+0.02}_{-0.02}$ & 0.83$^{+0.01}_{-0.02}$
& -4.56$^{+0.32}_{-0.04}$ & -4.46$^{+0.23}_{-0.13}$ & -3.97$^{+0.21}_{-0.10}$& -4.00$^{+0.21}_{-0.09}$
& 8.11$^{+0.25}_{-0.07}$ & 8.09$^{+0.26}_{-0.06}$\\
  OIII-19 & >2.41 & >2.27 & $<$3.95 & $<$3.99 &-0.03$^{+0.01}_{-0.02}$ &+0.23$^{+0.01}_{-0.01}$ & 0.77$^{+0.01}_{-0.01}$
& >-3.94 & >-3.86 & >-3.62& >-3.78
& >8.55& >8.48\\
  OIII-20 & >2.01 & >1.96 & $<$4.09 & $<$4.11 &-0.06$^{+0.05}_{-0.00}$ &+0.06$^{+0.03}_{-0.02}$ & 0.47$^{+0.03}_{-0.01}$ 
& >-4.50 & >-4.46 & >-4.37& >-4.42
 & >7.87 & >7.86\\
  OIII-21 & 1.87$^{+0.20}_{-0.03}$ & 2.01$^{+0.28}_{-0.12}$ & 4.14$^{+0.01}_{-0.08}$ & 4.02$^{+0.11}_{-0.04}$ &-0.12$^{+0.03}_{-0.02}$ &+0.18$^{+0.01}_{-0.04}$ & 0.78$^{+0.01}_{-0.01}$
& -4.53$^{+0.29}_{-0.03}$ & -4.38$^{+0.20}_{-0.14}$ &-3.96$^{+0.16}_{-0.11}$&-4.21$^{+0.22}_{-0.06}$
& 8.09$^{+0.24}_{-0.04}$ & 8.04$^{+0.19}_{-0.11}$\\
  OIII-22 & 2.14$^{+0.10}_{-0.05}$ & 2.07$^{+0.12}_{-0.02}$ & 4.03$^{+0.02}_{-0.03}$ & 4.04$^{+0.03}_{-0.03}$ &-0.09$^{+0.02}_{-0.01}$ &-0.04$^{+0.03}_{-0.00}$ & 0.82$^{+0.01}_{-0.01}$
& -4.34$^{+0.12}_{-0.07}$ &-4.29$^{+0.08}_{-0.11}$ & -3.87$^{+0.11}_{-0.06}$ &-3.90$^{+0.09}_{-0.07}$& 8.26$^{+0.11}_{-0.06}$
& 8.19$^{+0.15}_{-0.02}$\\
\enddata

\tablenotetext{}{\textbf{Notes.} (2)-(3) Flux ratio of [OIII]$\lambda\lambda$4959, 5007 to [OIII]$\lambda$4363. (4)-(5) Electron temperature. (6)-(7) Flux ratio of [OII]$\lambda$3727 to H$\beta$. (8) Flux ratio of [OIII]$\lambda\lambda$4959, 5007 to H$\beta$. (9)-(10) Gas-phase abundance of ${\rm O^{+}/H^{+}}$. (11) Gas-phase abundance of ${\rm O^{+}/H^{++}}$. (12)-(13) Gas-phase abundance of oxygen. Columns (3), (5), (7), (10) and (13) have been corrected for internal dust extinction using \citet{1989ApJ...345..245C} law.}
\end{deluxetable*}

\section{Summary}\label{sec:sum}
We have searched for [OIII] emitters at z $\sim$ 0.5 from SDSS broadband photometry, and confirmed a subset of them with SDSS spectroscopy. Our main results are as follows:

(1) We have established a simple way to select EELGs with \emph{i}-band excess based on broadband photometry. Our selection criteria not only provide us with 22 spectroscopically confirmed [OIII] emitters at z $\sim$ 0.5, but also produces a list of 2658 [OIII] candidates;

(2) Our [OIII] emitters have fairly blue \emph{r}-W2 and red W1-W4 colors, compared to most other objects that also satisfy our selection criteria. All of our [OIII] emitters have W1-W2 $\geqslant$1.6 mag, strongly indicative of warm dust emission of 400-600 K. Their strong H$\beta$ emission implies that they have very young star-forming regions, which could serve as the heating source of dust;

(3) The rest-frame [OIII]$\lambda$5007 equivalent widths of our [OIII] emitters mainly range from 200 \AA\ to 600 \AA \ and their high [OIII]$\lambda$5007/H$\beta$ ratios put them at the boundary of star-forming galaxies and AGNs on BPT diagrams;

(4) The typical \emph{E(B--V)} and electron temperature of [OIII] emitters is $\sim$0.1-0.3 mag and $\sim$10$^4$ K, respectively;

(5) For our [OIII]$\lambda$4363-detected sources, the lowest and median metallicities are 12 + log(O/H) = 7.98$^{+0.12}_{-0.02}$ and 8.24$^{+0.05}_{-0.04}$. For our [OIII]$\lambda$4363 non-detected sources, the lowest and median metallicities are 12 + log(O/H) = 7.86 and 8.48. Our results are comparable to the values in \citet{2016ApJ...828...67L};

(6) By performing SED fitting using CIGALE, we derived stellar masses of our [OIII] emitters, which range from 10$^{9.2}$ \emph{M$_{\odot}$} to 10$^{10.15}$ \emph{M$_{\odot}$}, with an average value of 10$^{9.71}$ \emph{M$_{\odot}$} and a median value of 10$^{9.78}$ \emph{M$_{\odot}$}. By converting H$\alpha$ luminosity to SFR, we derived SFR of our [OIII] emitters, which range from 9 \emph{M$_{\odot}$} yr$^{-1}$ to 129 \emph{M$_{\odot}$} yr$^{-1}$, with an average value of 32.4 \emph{M$_{\odot}$} yr$^{-1}$ and a median value of 30.2 \emph{M$_{\odot}$} yr$^{-1}$. Our [OIII] emitters lie above the normal SFR-\emph M$_{\star}$ relation and below the \emph M$_{\star}$-metallicity relation;

(7) The \emph{F}$_{\rm NUV}$/\emph{F}$_{\rm FUV}$ ratios of our [OIII] emitters are positively correlated with redshift, as a result of Lyman break absorption;

(8) Our [OIII] emitters exhibit remarkably high line luminosity -- 18/22 have L$_{\rm [OIII]\lambda5007}$ $>$ 5$\times$10$^{42}$ erg $\rm s^{-1}$ and 5/22 have L$_{\rm [OIII]\lambda5007}$ $>$ 10$^{43}$ erg $\rm s^{-1}$;

(9) The estimated volume number density of [OIII] emitters at z $\sim$ 0.5 is $\sim$ 2$\times$10$^{-8}$ Mpc$^{-3}$, with L$_{\rm [OIII]\lambda5007}$ down to $\sim$ 3$\times$10$^{42}$\ erg $\rm s^{-1}$. The cumulative number fraction distribution of EELGs across different redshifts could be indicative of a strong redshift evolution at the bright end of the [OIII] luminosity function.

\acknowledgments

We thank Chun Ly for constructive suggestions and helpful discussions. We are also grateful to the anonymous referee for carefully reading our manuscript and constructive feedback, which substantially helped improving the quality of this paper. Z. L is grateful for support from UCLA-CSST Program for undergraduate research.

This work has made use of the data obtained by SDSS and WISE. Funding for the Sloan Digital Sky Survey IV has been provided by the Alfred P. Sloan Foundation, the U.S. Department of Energy Office of Science, and the Participating Institutions. SDSS-IV acknowledges
support and resources from the Center for High-Performance Computing at
the University of Utah. The SDSS web site is http://www.sdss.org/. SDSS-IV is managed by the Astrophysical Research Consortium for the 
Participating Institutions of the SDSS Collaboration including the 
Brazilian Participation Group, the Carnegie Institution for Science, 
Carnegie Mellon University, the Chilean Participation Group, the French Participation Group, Harvard-Smithsonian Center for Astrophysics, 
Instituto de Astrof\'isica de Canarias, The Johns Hopkins University, 
Kavli Institute for the Physics and Mathematics of the Universe (IPMU) / 
University of Tokyo, Lawrence Berkeley National Laboratory, 
Leibniz Institut f\"ur Astrophysik Potsdam (AIP),  
Max-Planck-Institut f\"ur Astronomie (MPIA Heidelberg), 
Max-Planck-Institut f\"ur Astrophysik (MPA Garching), 
Max-Planck-Institut f\"ur Extraterrestrische Physik (MPE), 
National Astronomical Observatories of China, New Mexico State University, 
New York University, University of Notre Dame, 
Observat\'ario Nacional / MCTI, The Ohio State University, 
Pennsylvania State University, Shanghai Astronomical Observatory, 
United Kingdom Participation Group,
Universidad Nacional Aut\'onoma de M\'exico, University of Arizona, 
University of Colorado Boulder, University of Oxford, University of Portsmouth, 
University of Utah, University of Virginia, University of Washington, University of Wisconsin, 
Vanderbilt University, and Yale University.  WISE is a joint project of the University
of California, Los Angeles, and the Jet Propulsion Laboratory
of California Institute of Technology, funded by the National
Aeronautics and Space Administration. The WISE web site
is http://wise.astro.ucla.edu/.


\bibliography{extreme-oiii-emitters}
\appendix

\section{Broadband Photometric Selection Criteria}\label{sec:criterion}
We present our broadband photometric selection criteria in SQL query form here. To make our search as exhaustive as possible, we only use model magnitudes instead of other types of magnitudes. We also did not require clean photometry in the initial search. Our search derived a list of 2658 unique objects in total, which is available in a machine-readable format online.

\texttt{\scriptsize
SELECT ra, dec, field, mode, type, clean, probPSF, modelmag\_u, modelmag\_g, modelMmag\_r, modelmag\_i, modelmag\_z into mydb.OIII FROM PhotoPrimary}

\texttt{\scriptsize
WHERE modelmag\_r$>$=18.5 and modelmag\_z$>$=18.5}

\texttt{\scriptsize
and modelmag\_i-modelmag\_z$<$=modelmag\_r-modelmag\_i-0.7}

\texttt{\scriptsize
and modelmag\_u-modelmag\_g$<$=0.3 and modelmag\_g-modelmag\_r$<$=0.45 and modelmag\_r-modelmag\_z$<$=0.8}

\texttt{\scriptsize
and modelmagerr\_u$<$0.25 and modelmagerr\_g$<$0.15 modelmagerr\_r$<$0.15 and modelmagerr\_i$<$0.15 and modelmagerr\_z$<$0.15}

\texttt{\scriptsize
and modelmagerr\_u$>$0 and modelmagerr\_g$>$0 and modelmagerr\_r$>$0 and modelmagerr\_i$>$0 and modelmagerr\_z$>$0}

\section{SEDS and SDSS Spectra of 22 [OIII] emitters}\label{sec:SED}

\begin{figure}[htbp]
\centering
\figurenum{14a}
\includegraphics[width=\linewidth, clip]{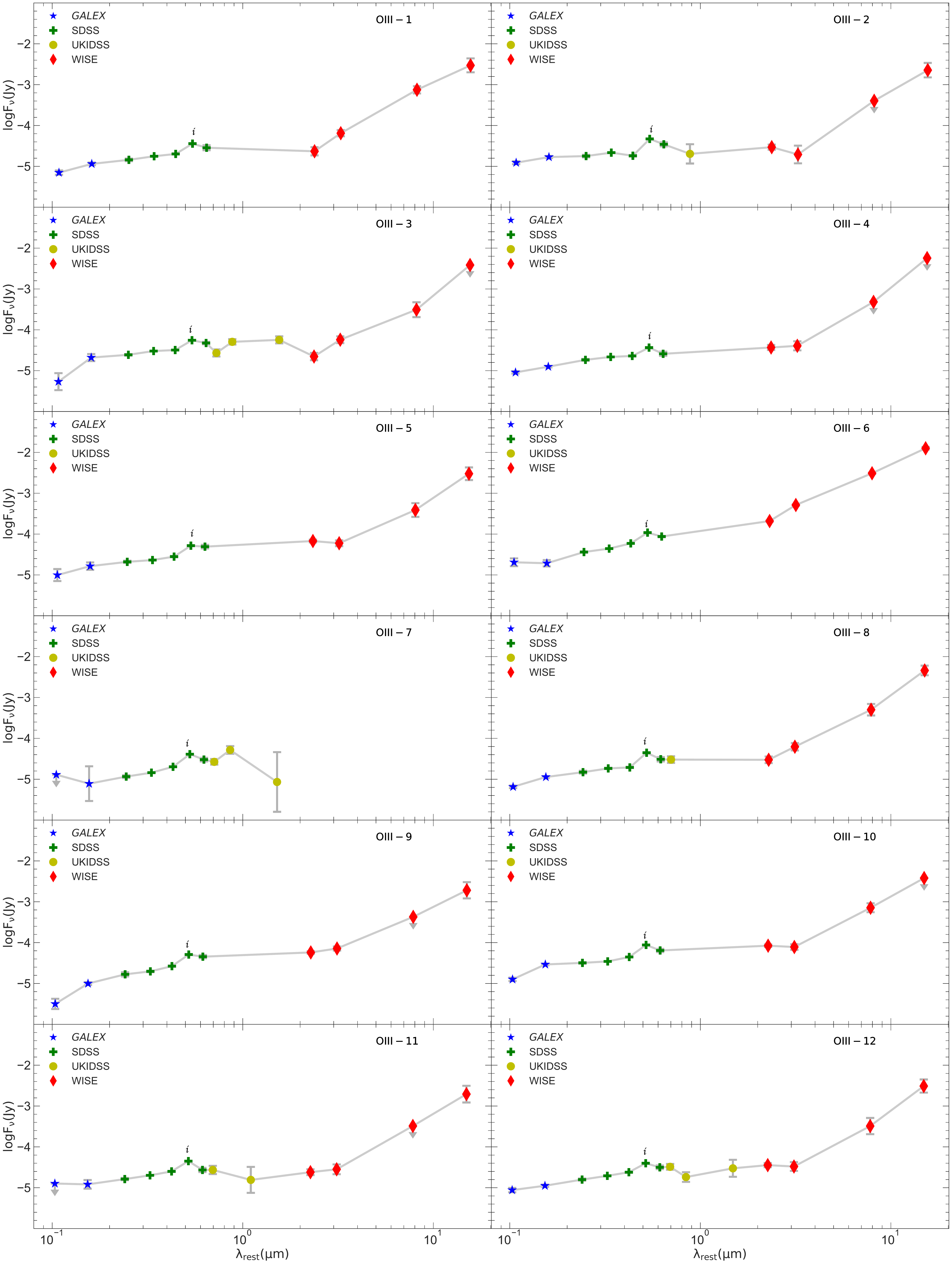}

\caption{SEDs of 22 [OIII] emitters.}
\label{fig:SED-1}
\end{figure}

\begin{figure}[htbp]
\centering
\figurenum{14b}
\includegraphics[width=\linewidth, clip]{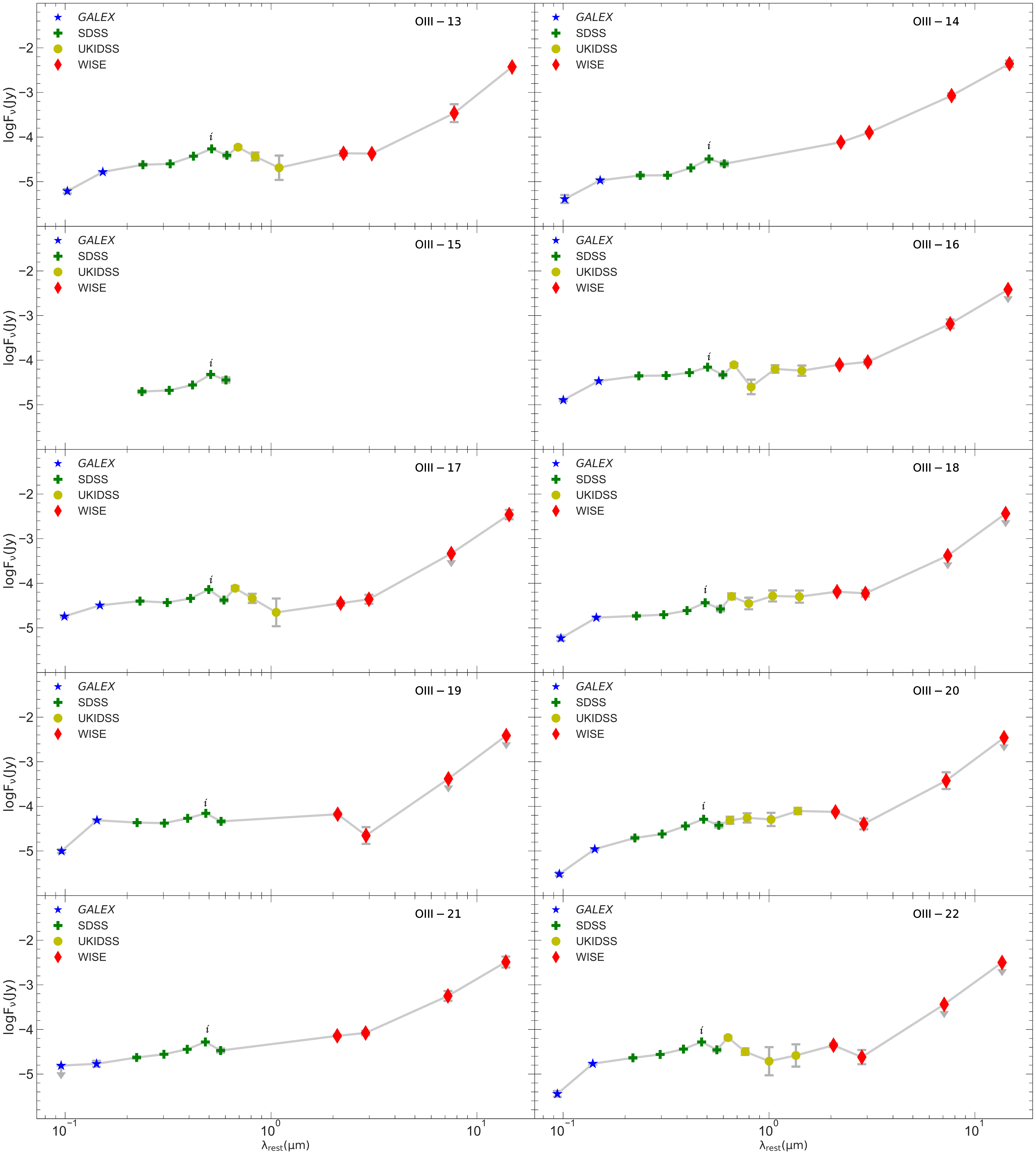}

\caption{SEDs of 22 [OIII] emitters.}
\label{fig:SED-2}
\end{figure}

\begin{figure}
\centering
\figurenum{15a}
\includegraphics[width=0.49\linewidth, clip]{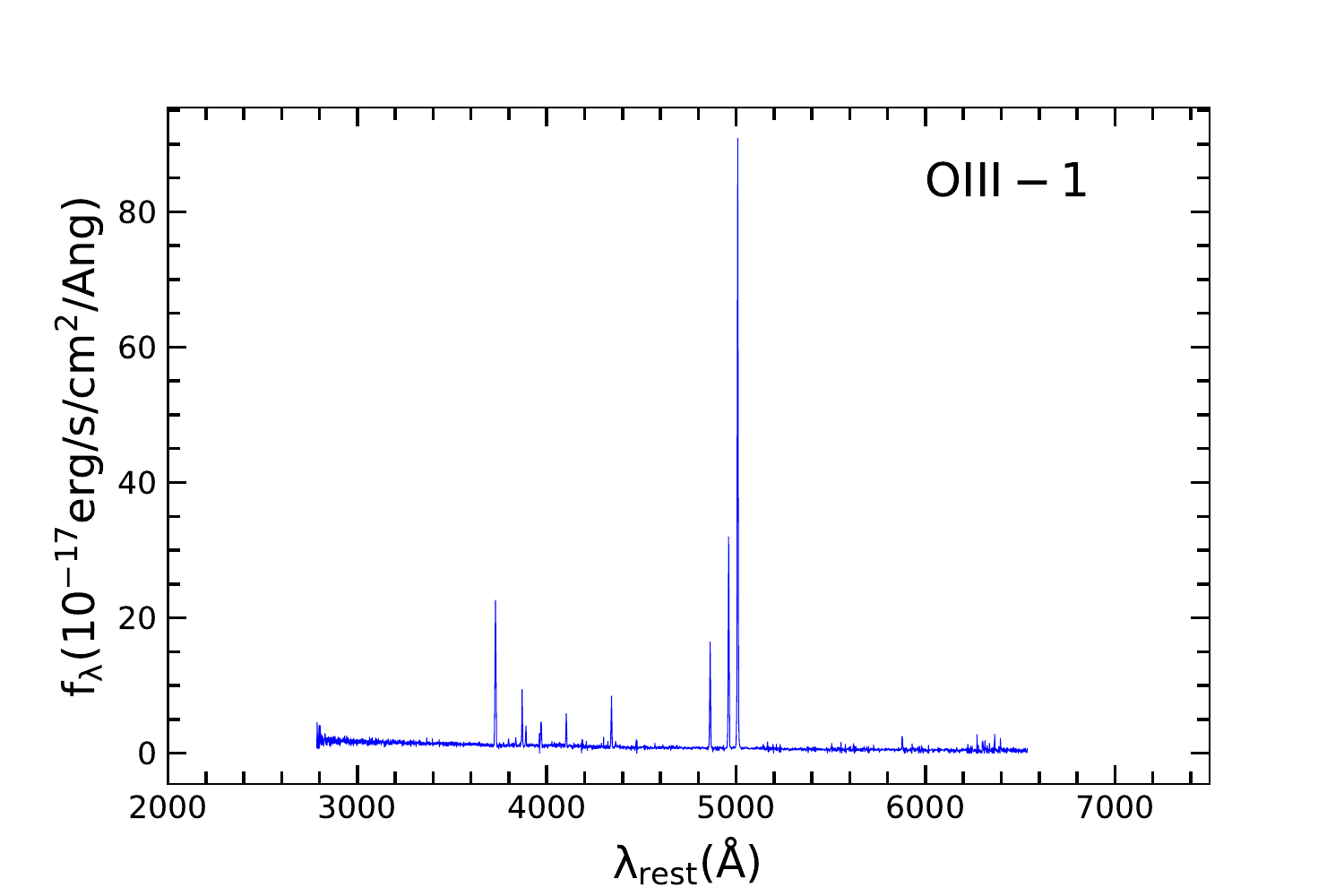}
\includegraphics[width=0.49\linewidth, clip]{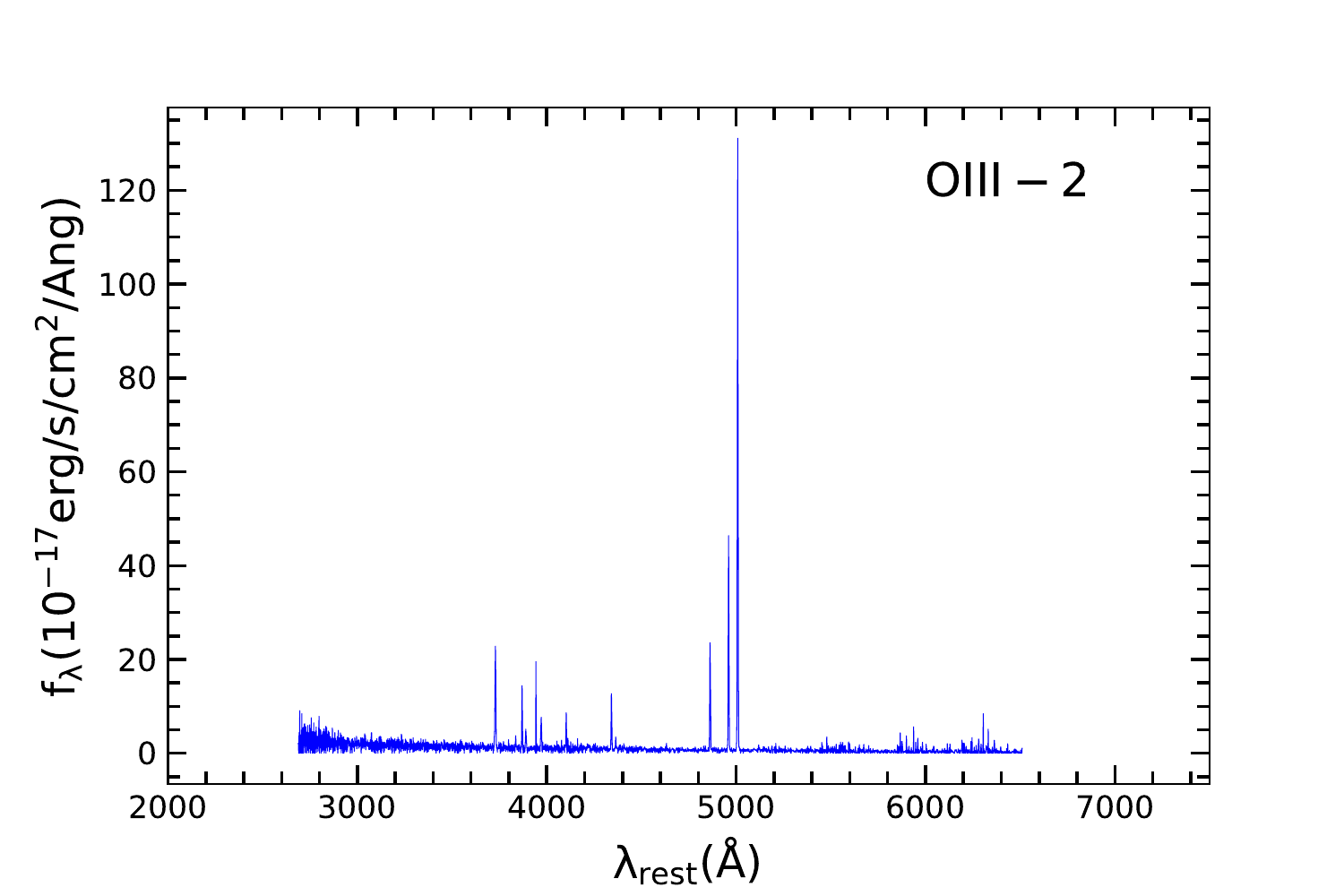}
\includegraphics[width=0.49\linewidth, clip]{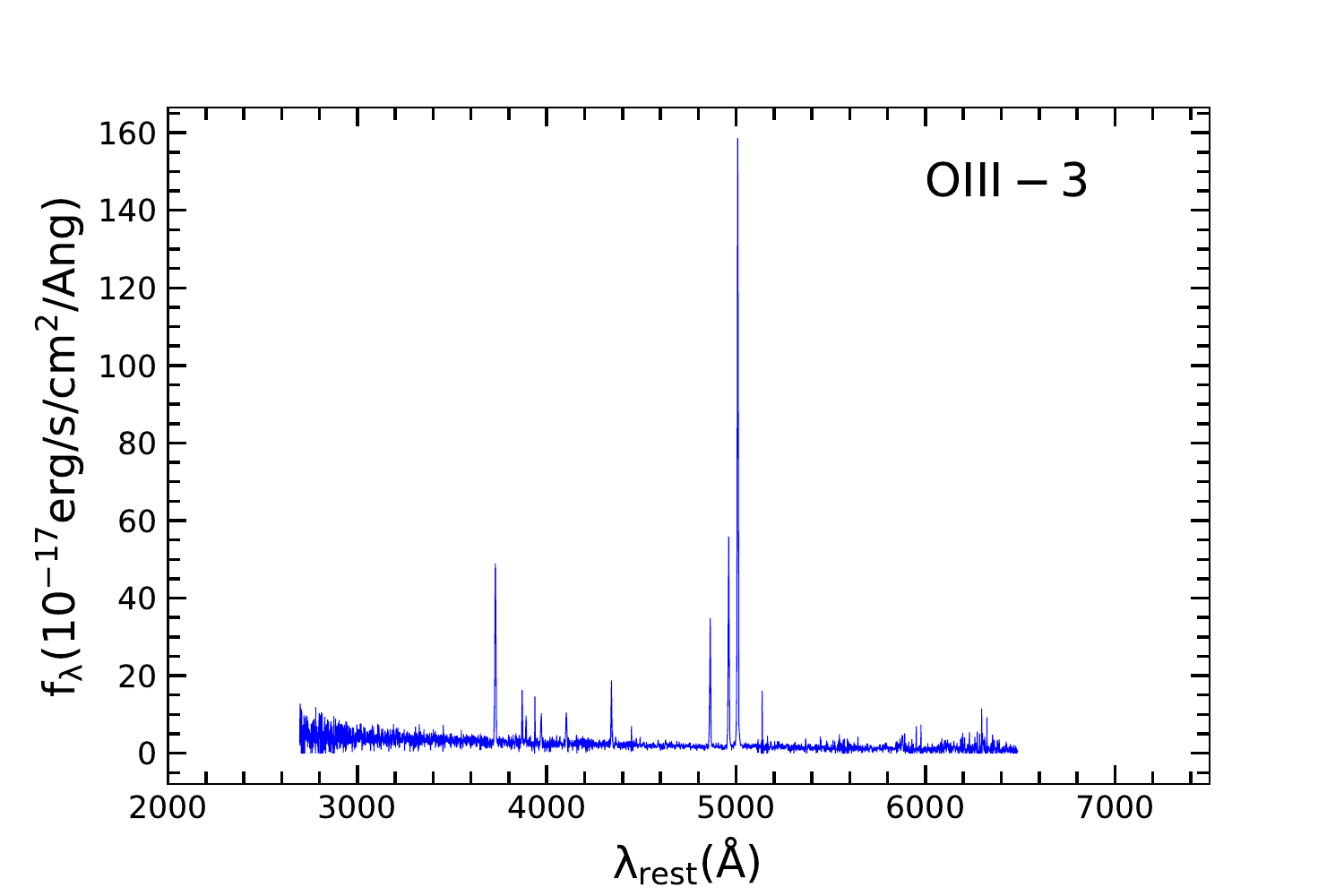}
\includegraphics[width=0.49\linewidth, clip]{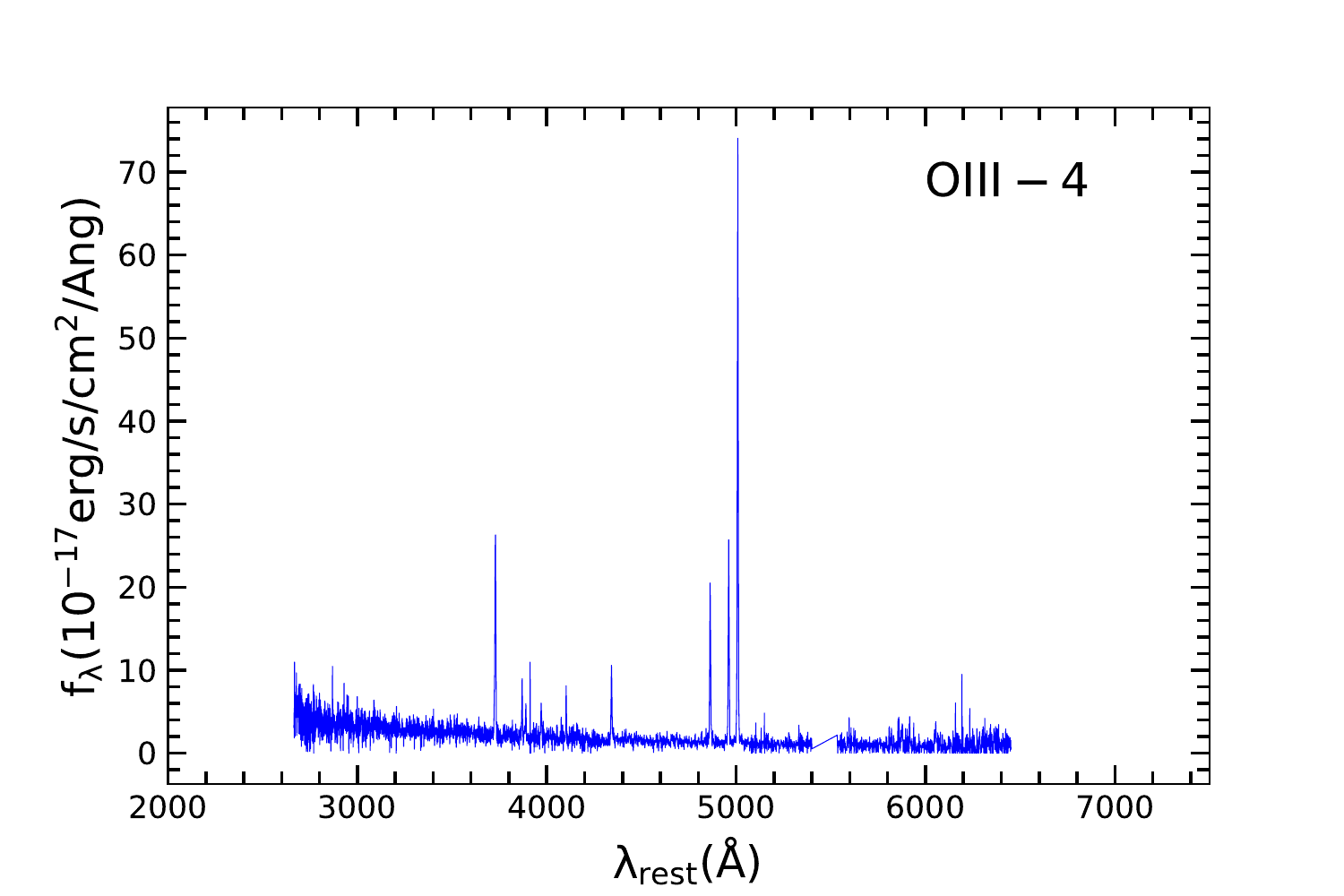}
\includegraphics[width=0.49\linewidth, clip]{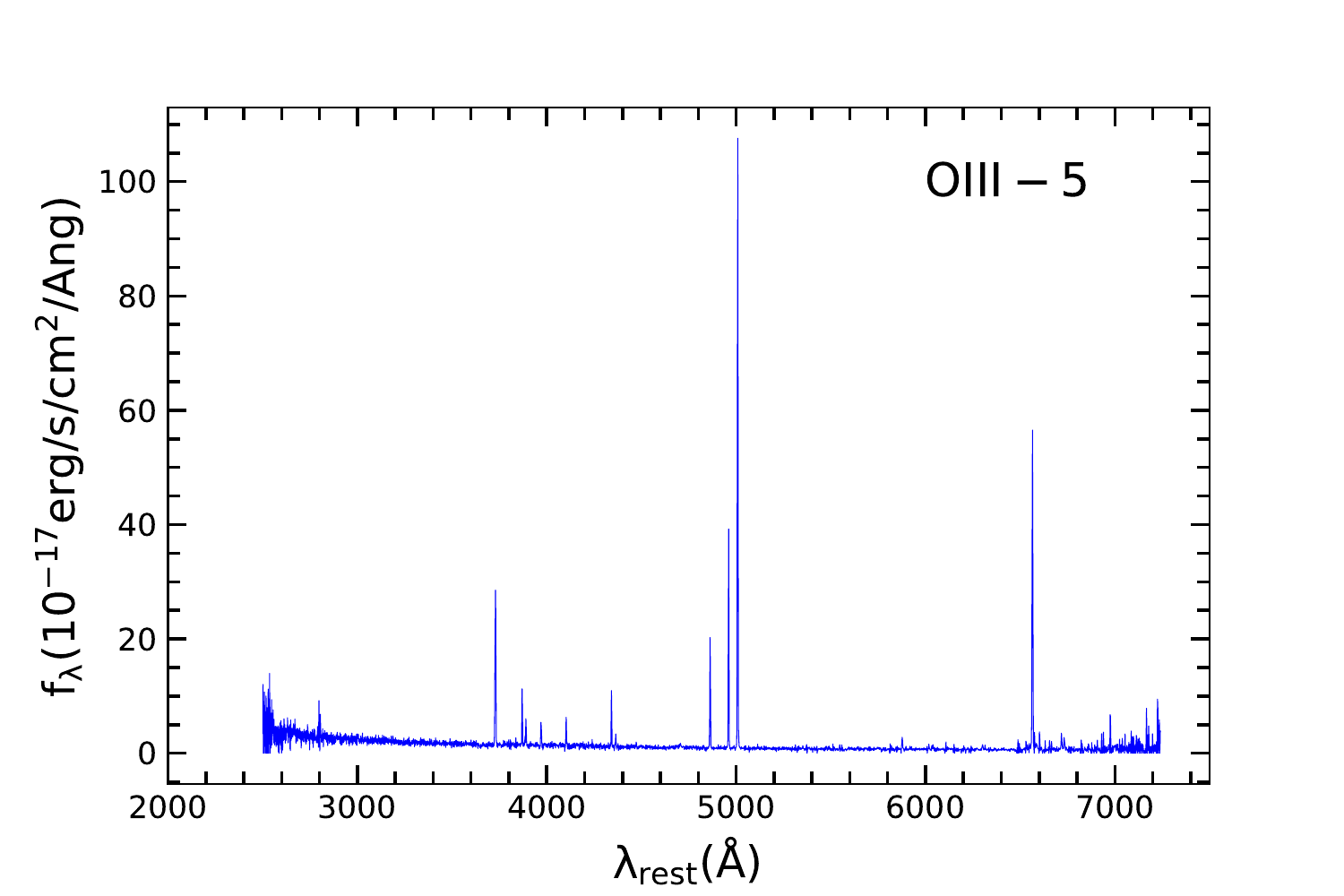}
\includegraphics[width=0.49\linewidth, clip]{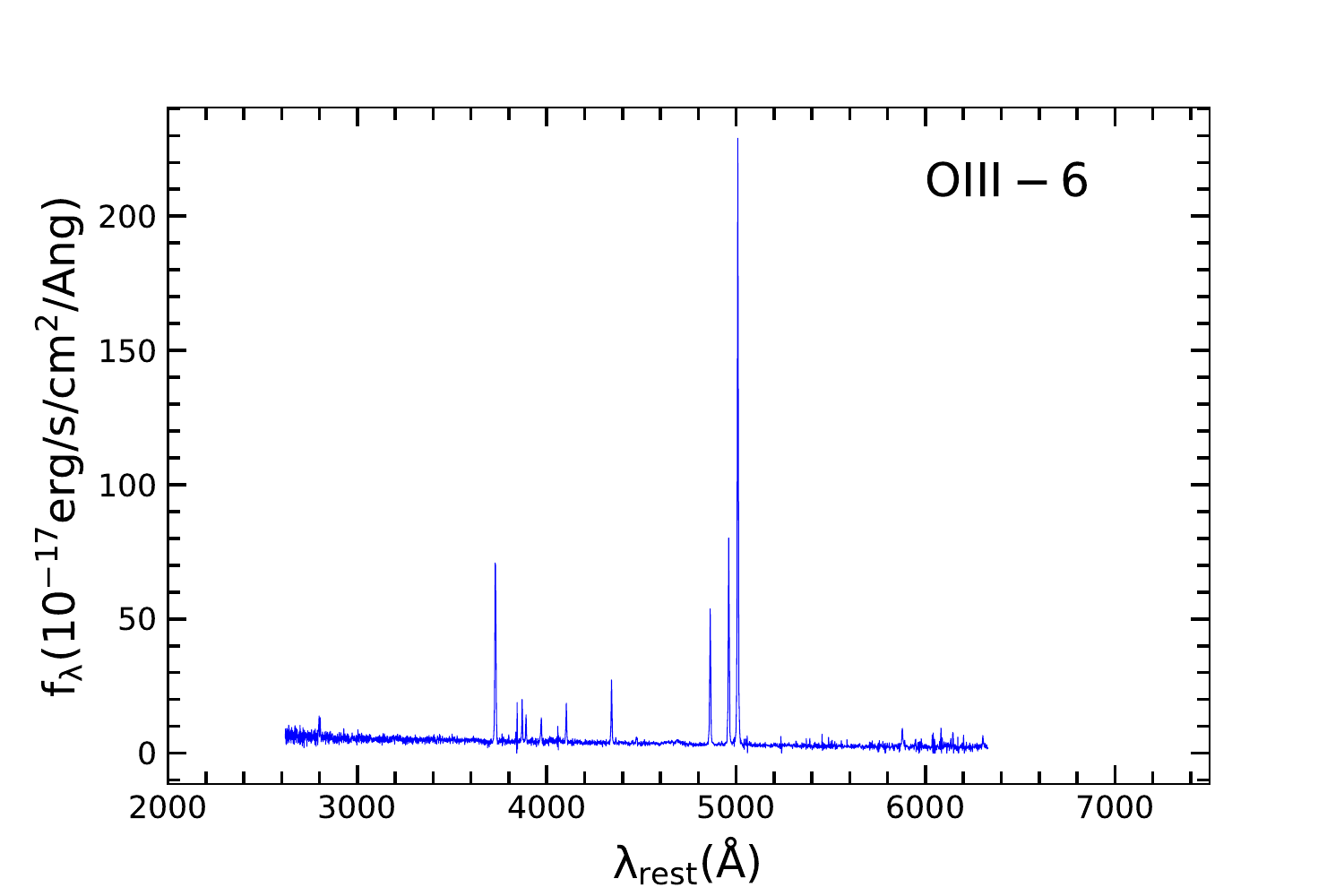}
\includegraphics[width=0.49\linewidth, clip]{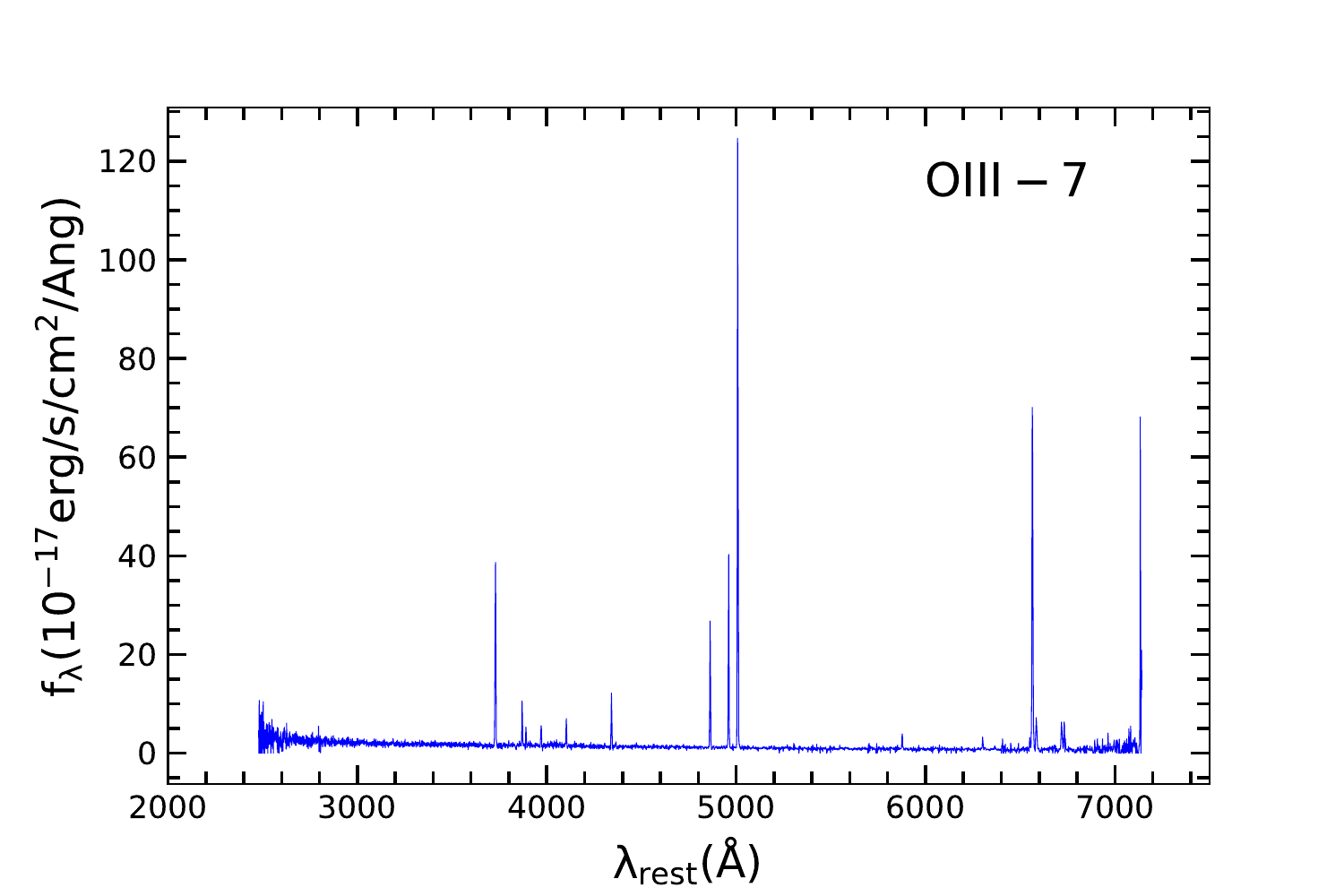}
\includegraphics[width=0.49\linewidth, clip]{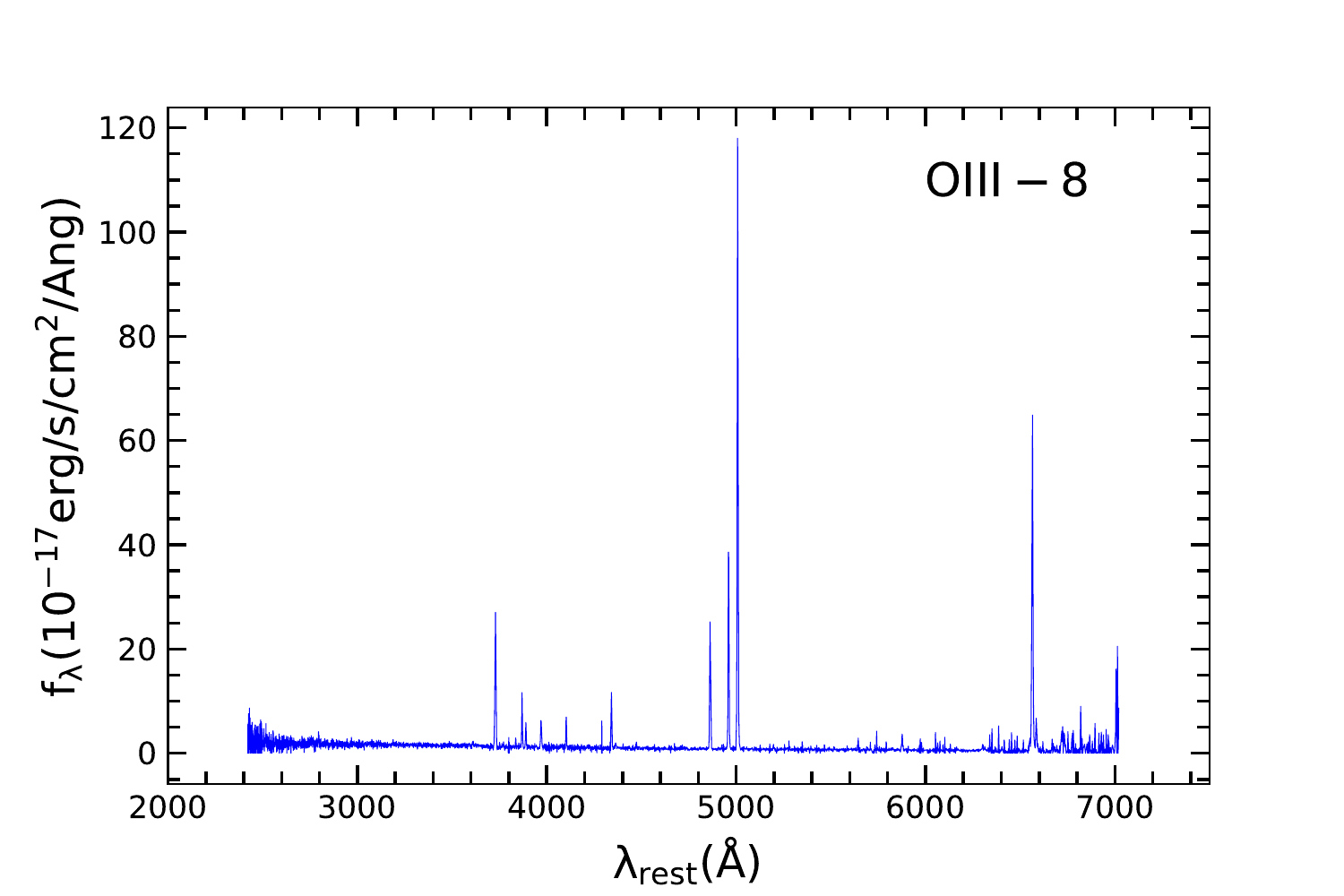}

\caption{Rest-frame spectra of 22 [OIII] emitters. The spectra are taken from SDSS DR14. }
\label{fig:spec1}
\end{figure}

\begin{figure}
\begin{center}
\figurenum{15b}
\includegraphics[width=0.49\linewidth, clip]{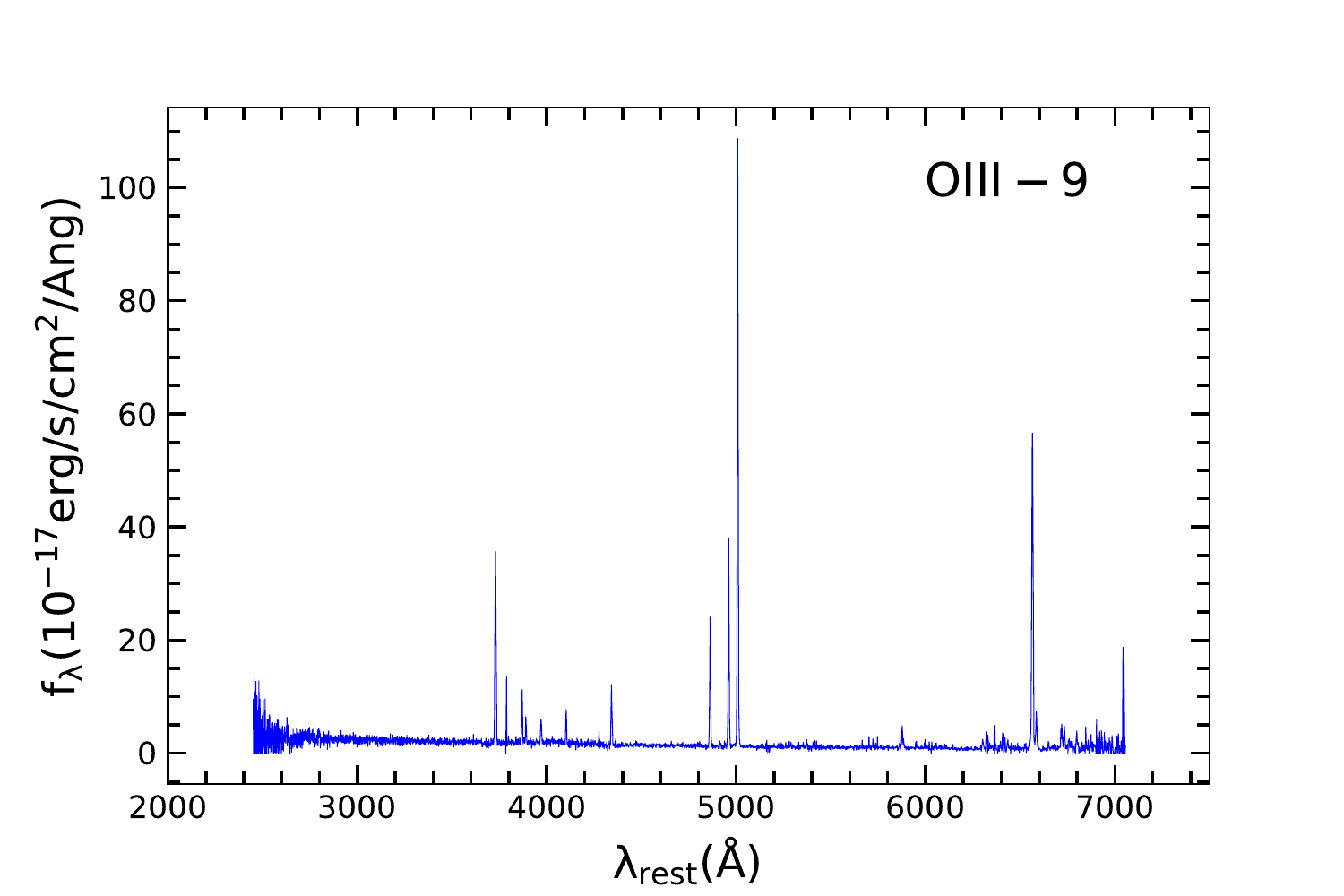}
\includegraphics[width=0.49\linewidth, clip]{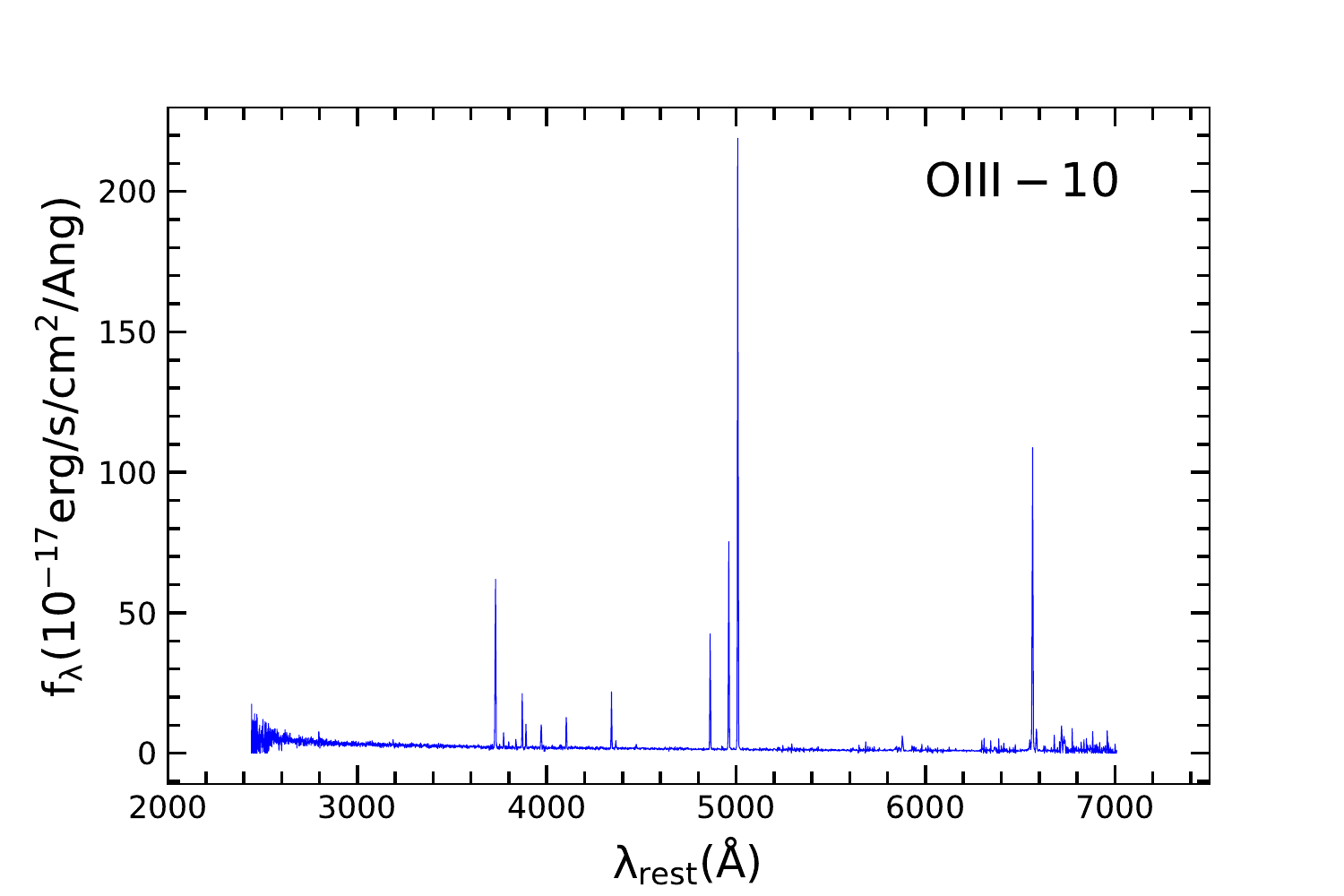}
\includegraphics[width=0.49\linewidth, clip]{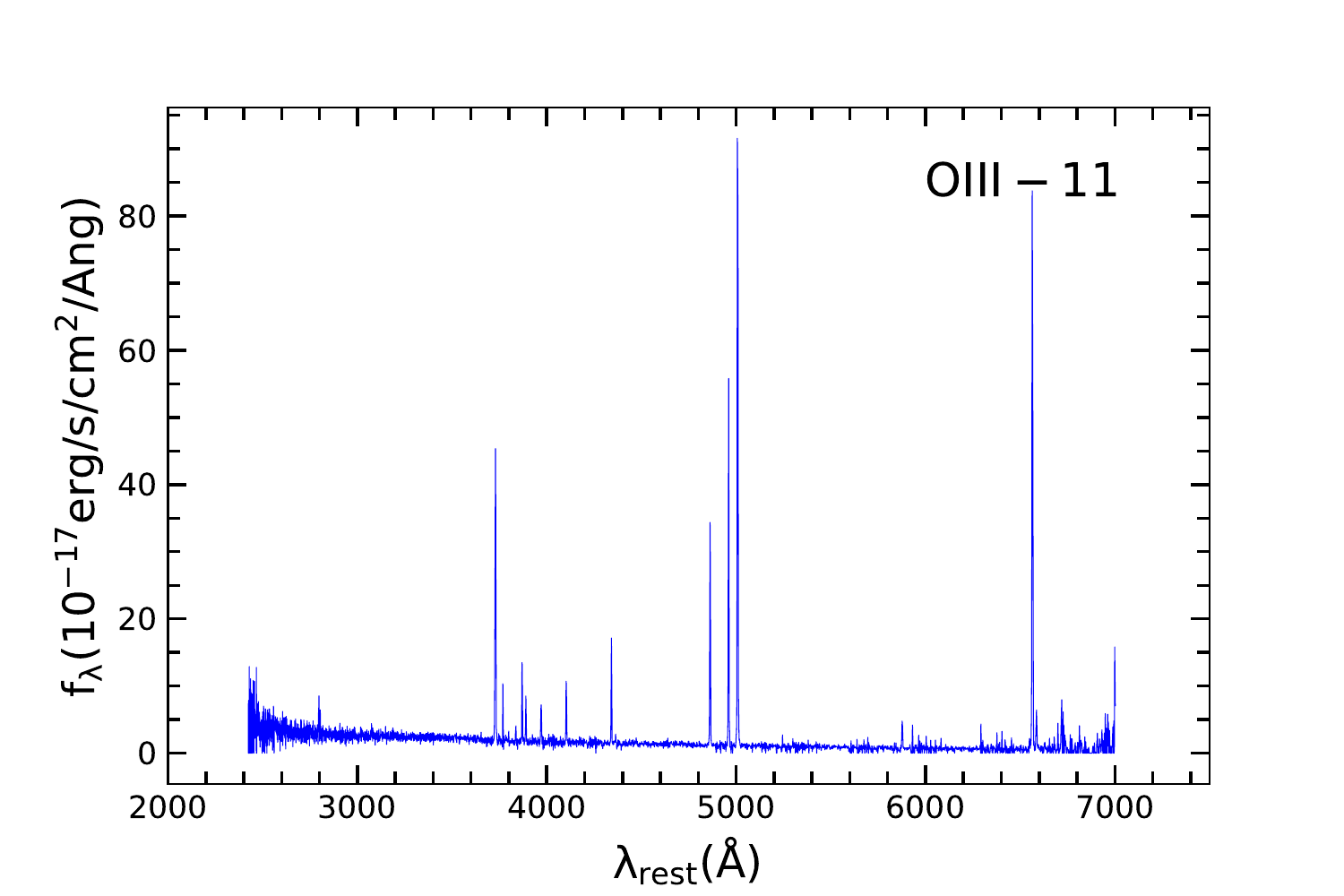}
\includegraphics[width=0.49\linewidth, clip]{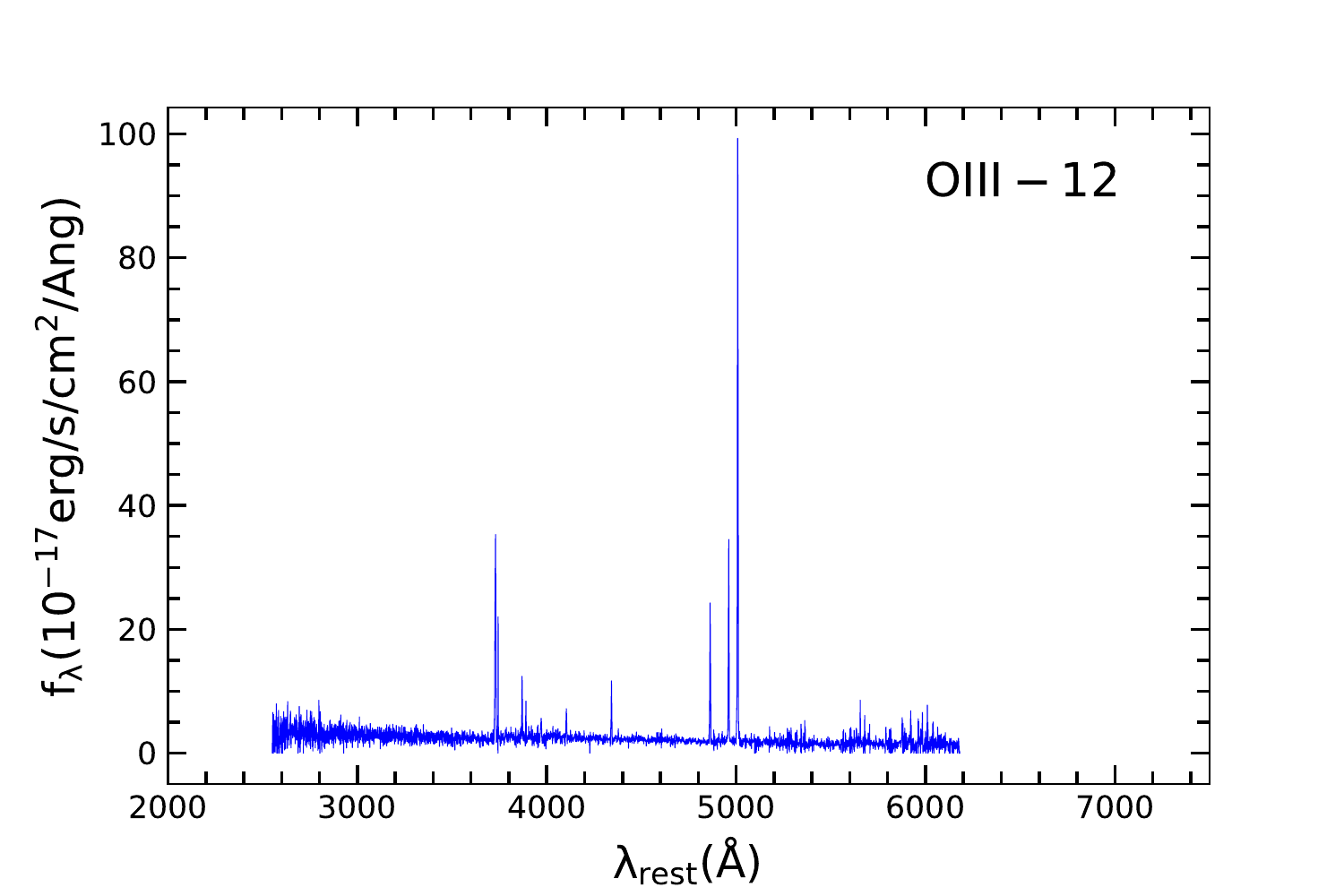}
\includegraphics[width=0.49\linewidth, clip]{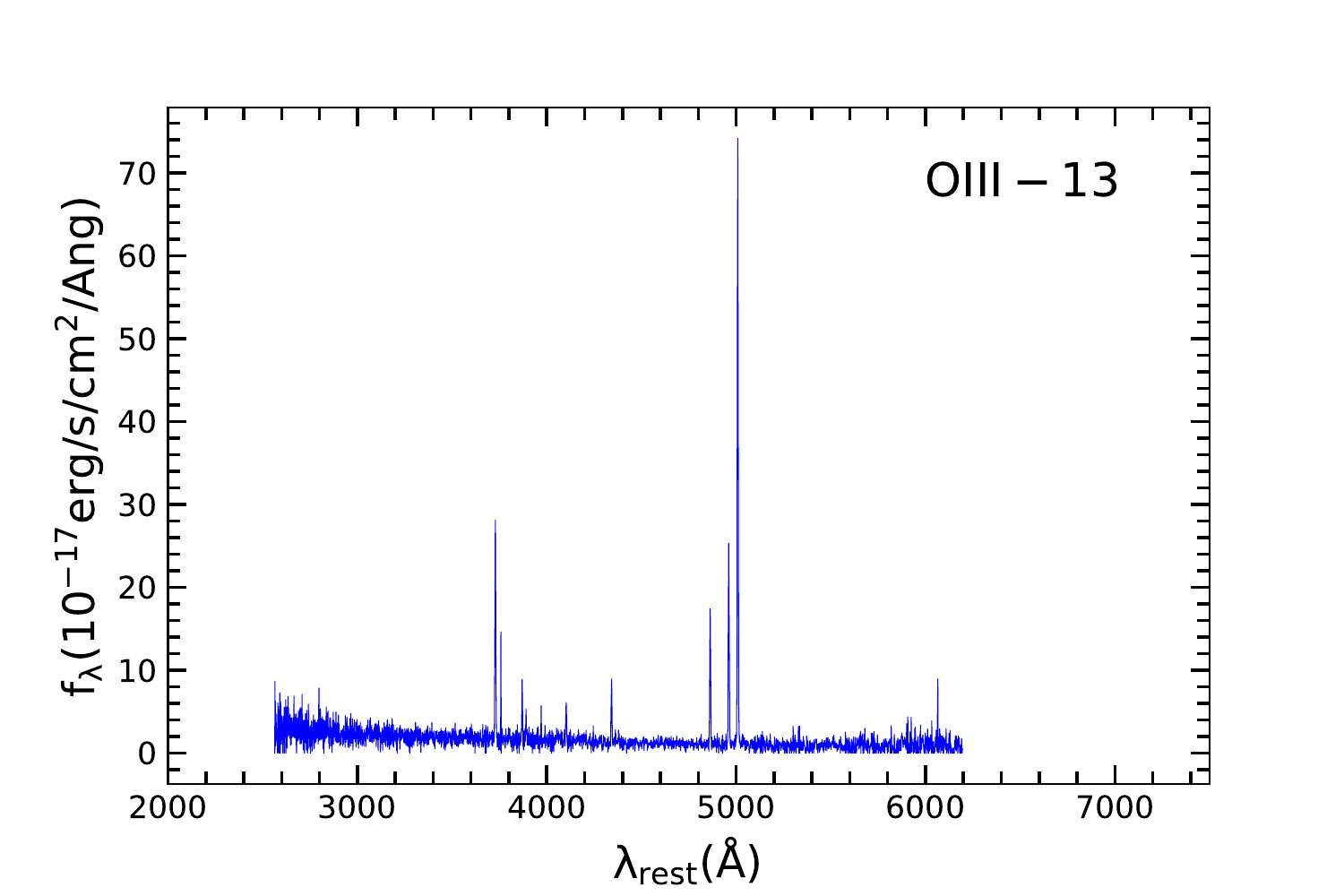}
\includegraphics[width=0.49\linewidth, clip]{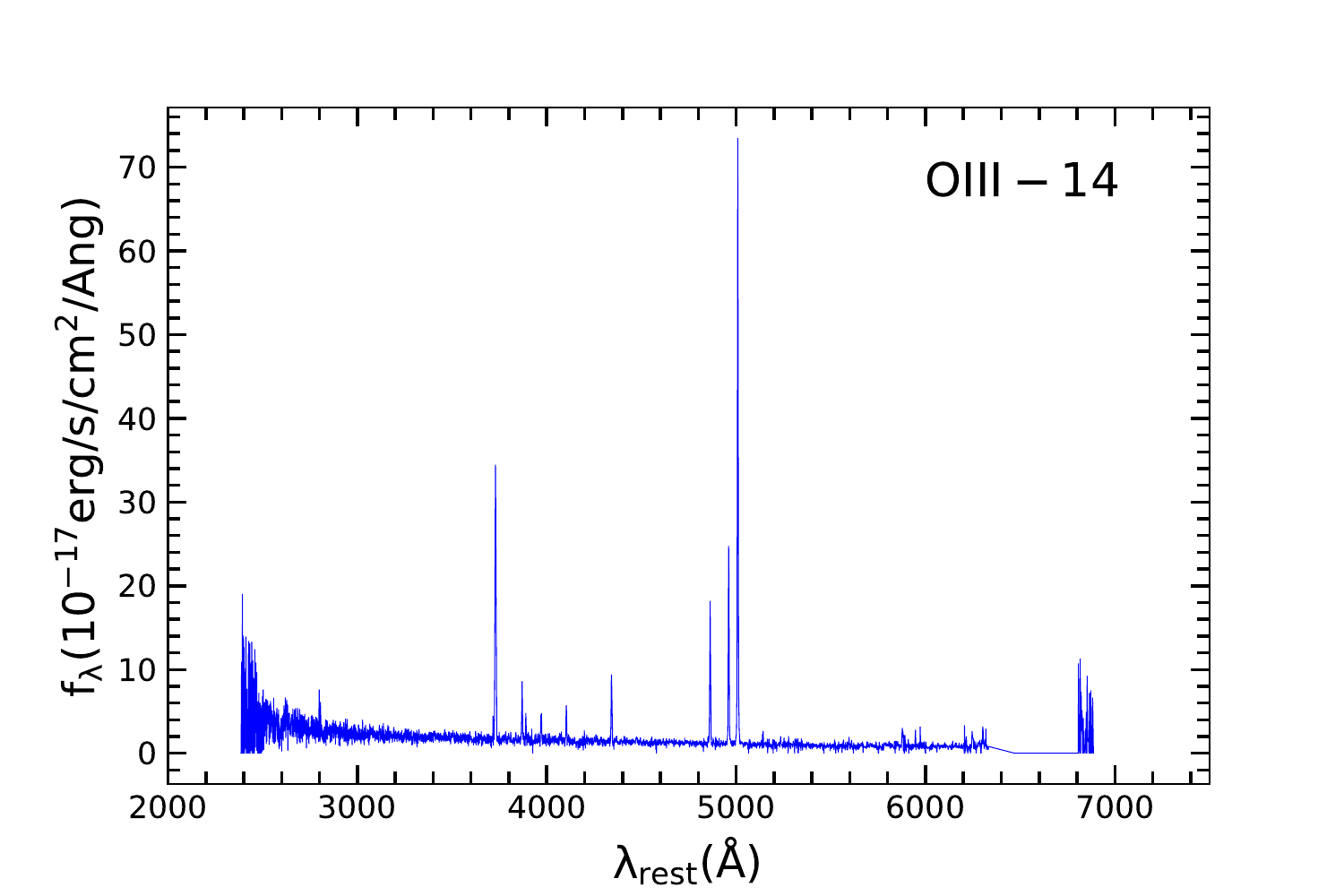}
\includegraphics[width=0.49\linewidth, clip]{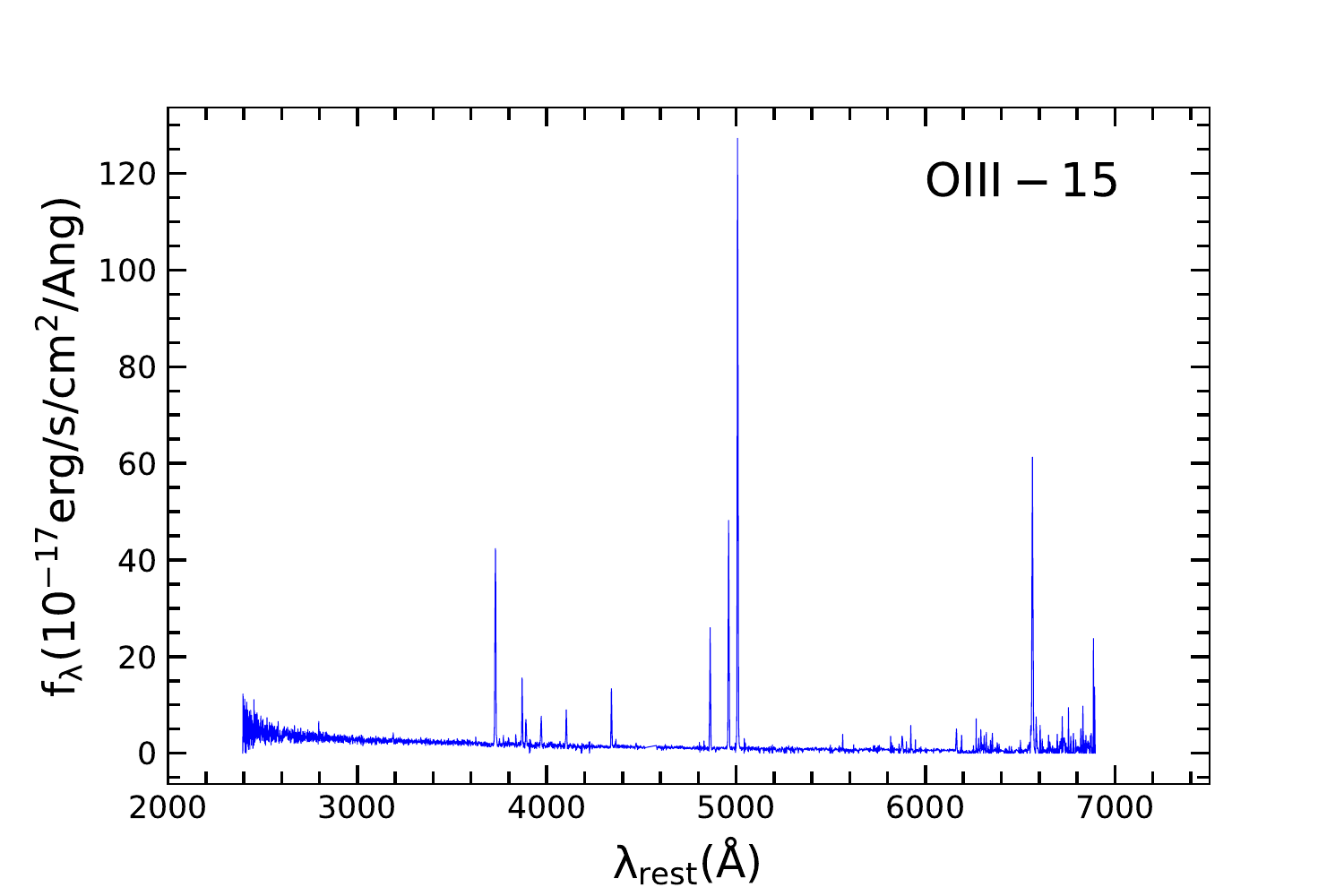}
\includegraphics[width=0.49\linewidth, clip]{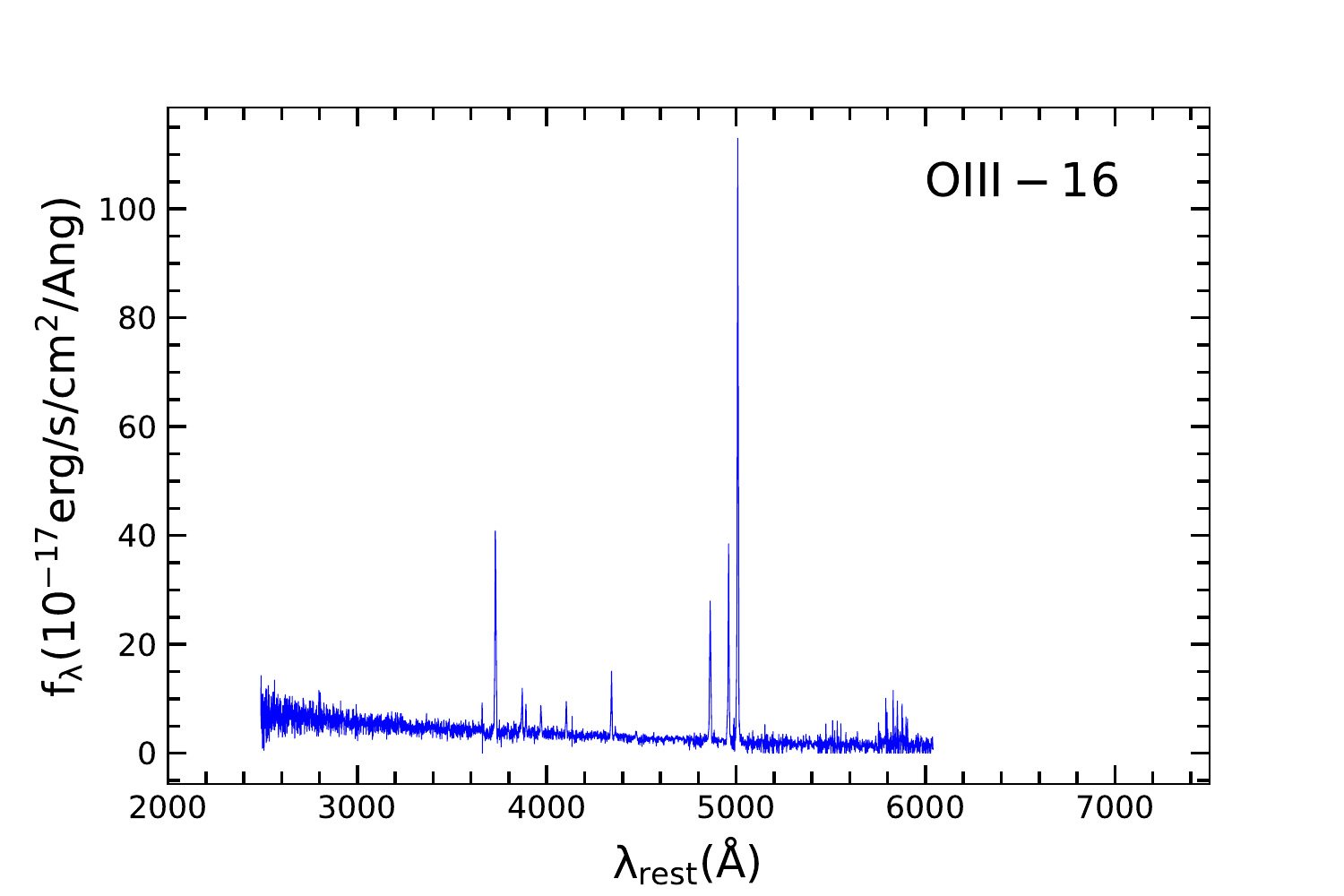}
\caption{Rest-frame spectra of 22 [OIII] emitters. The spectra are taken from SDSS DR14. }
\end{center}
\label{fig:spec2}
\end{figure}

\begin{figure}
\centering
\figurenum{15c}
\includegraphics[width=0.49\linewidth, clip]{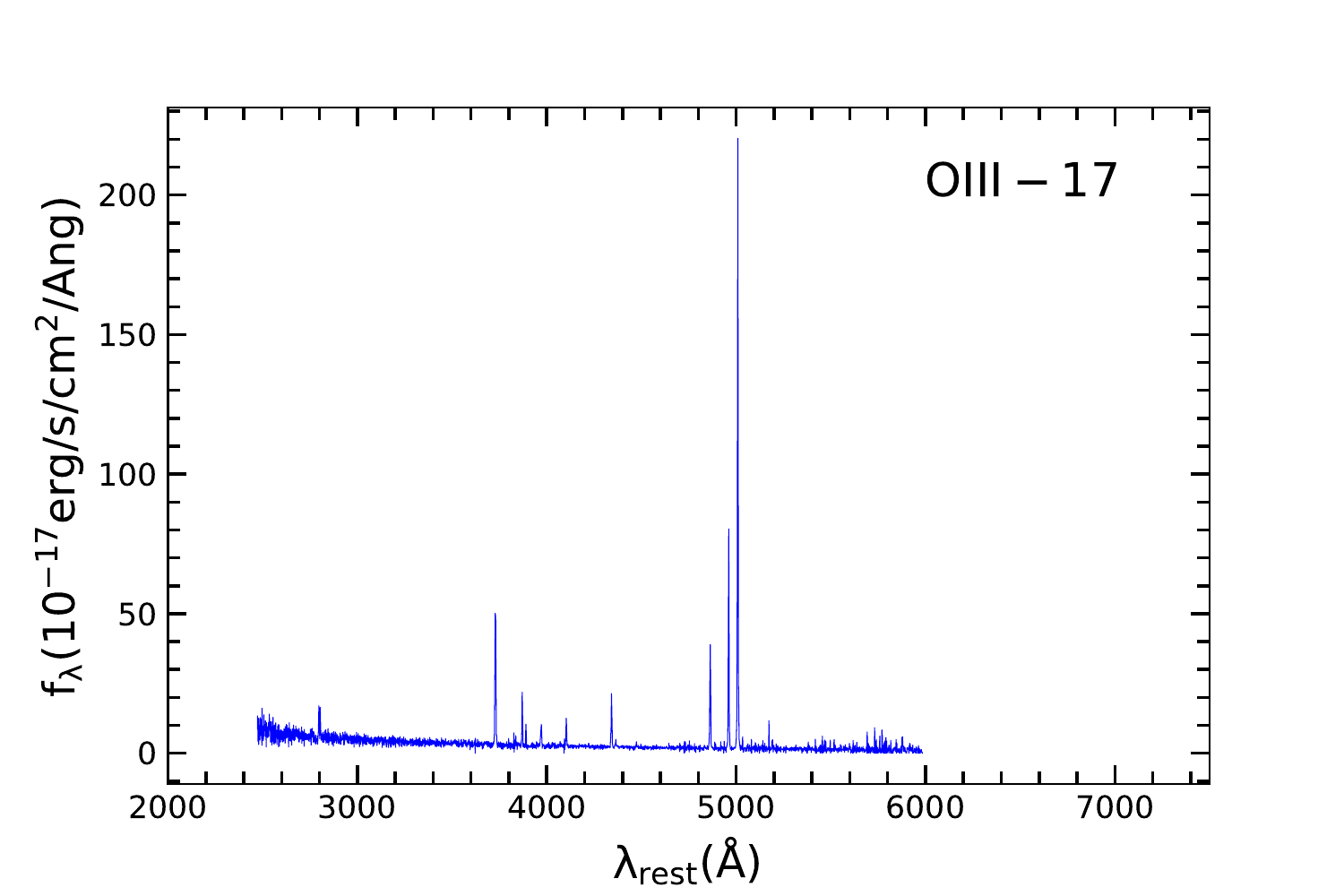}
\includegraphics[width=0.49\linewidth, clip]{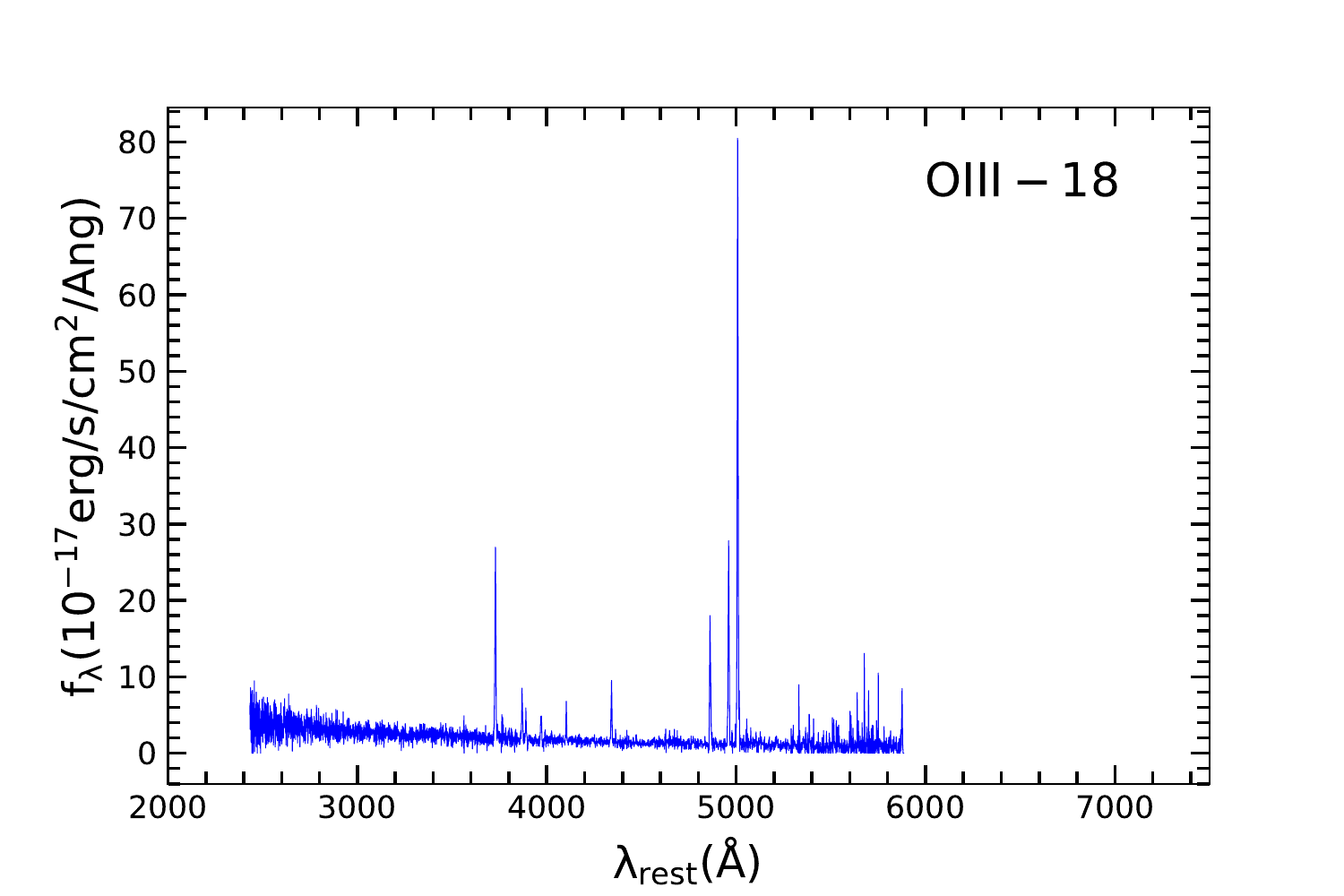}
\includegraphics[width=0.49\linewidth, clip]{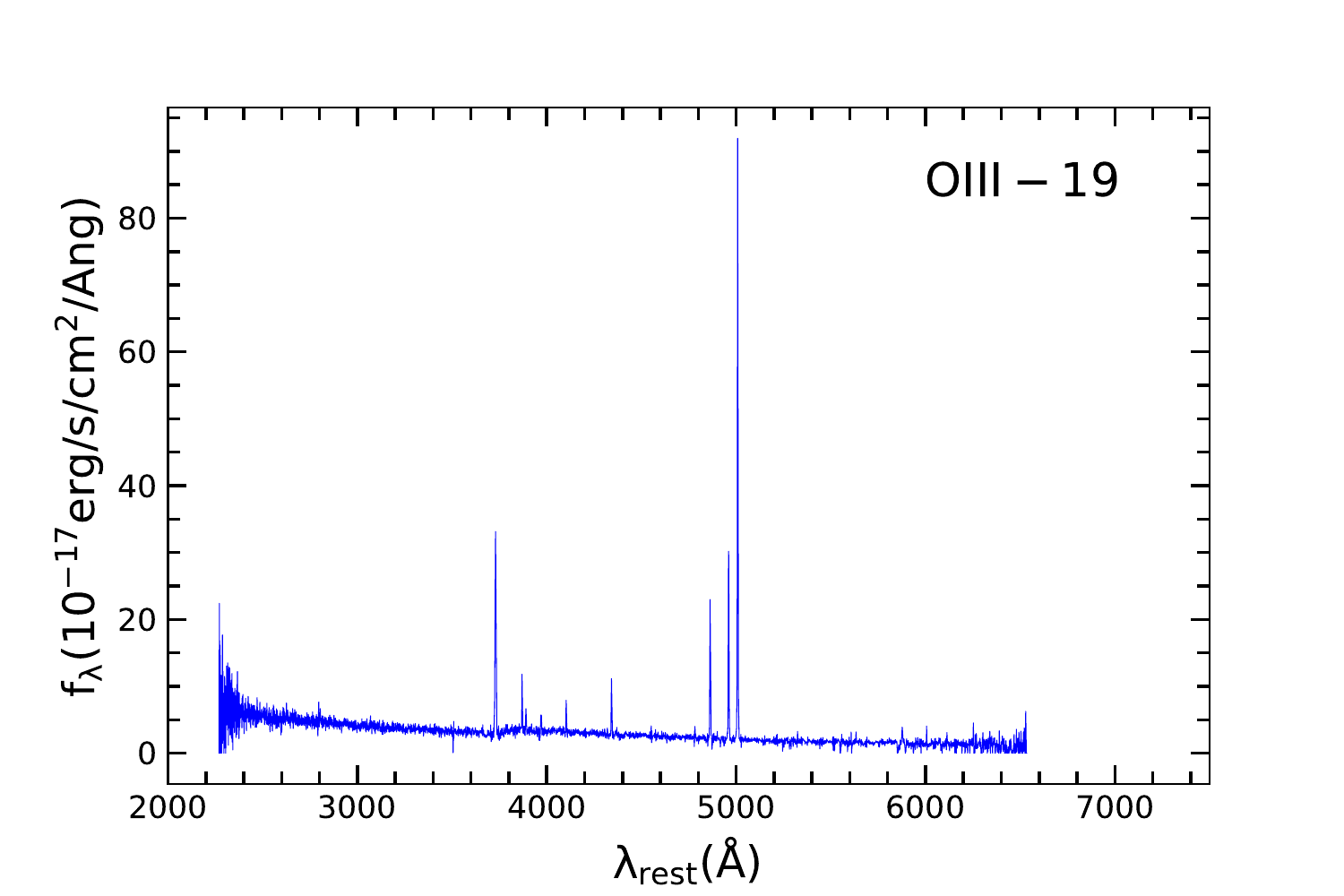}
\includegraphics[width=0.49\linewidth, clip]{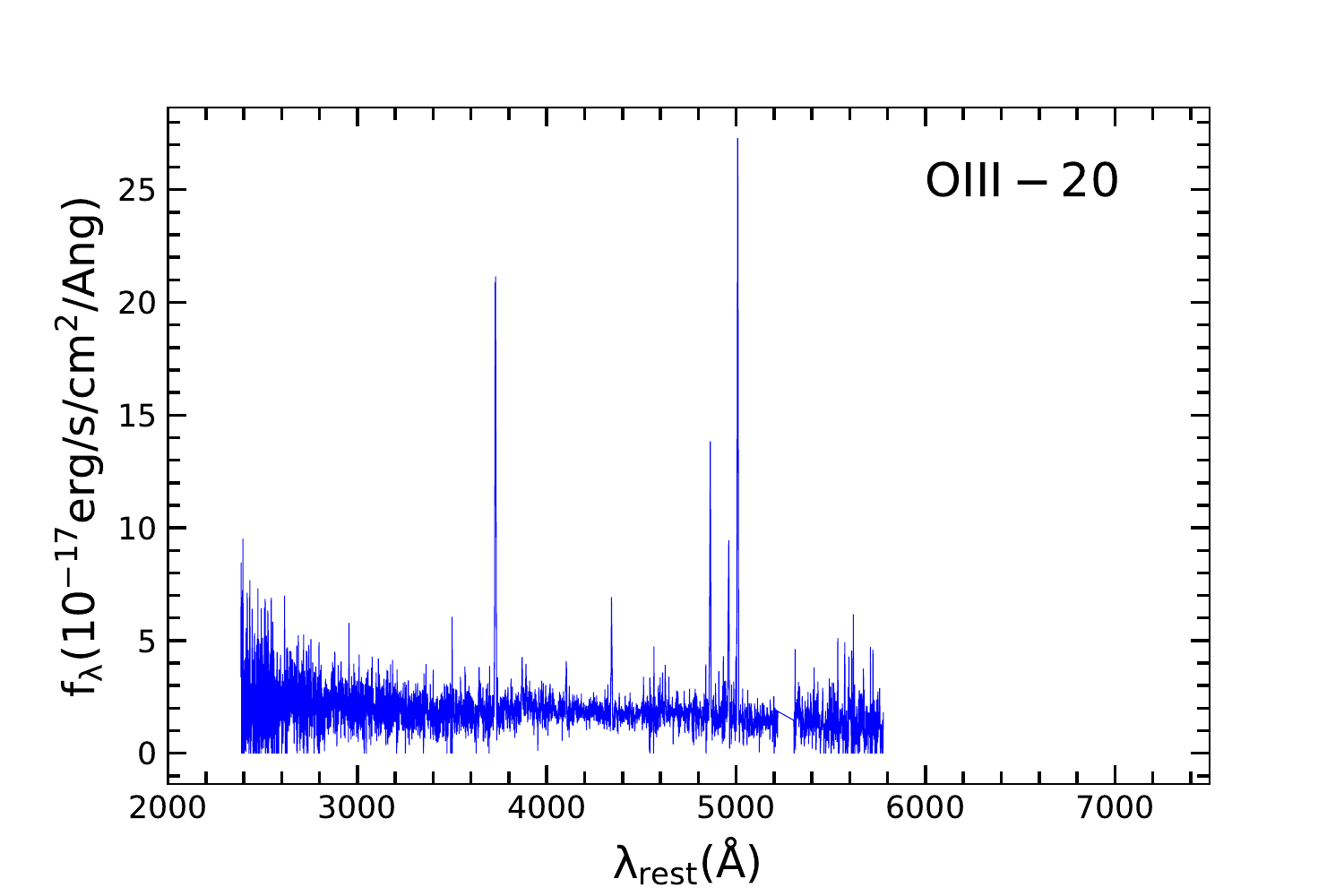}
\includegraphics[width=0.49\linewidth, clip]{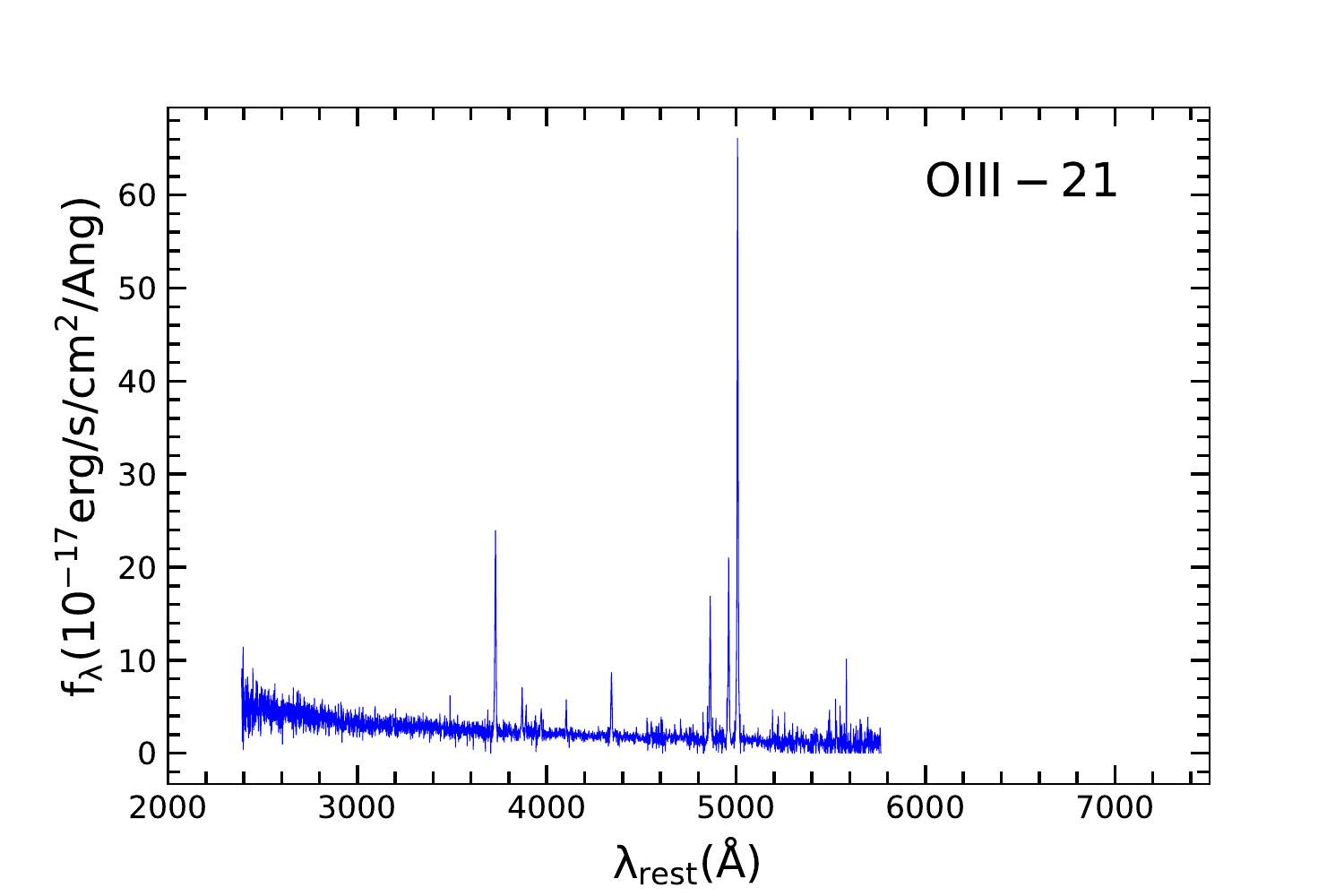}
\includegraphics[width=0.49\linewidth, clip]{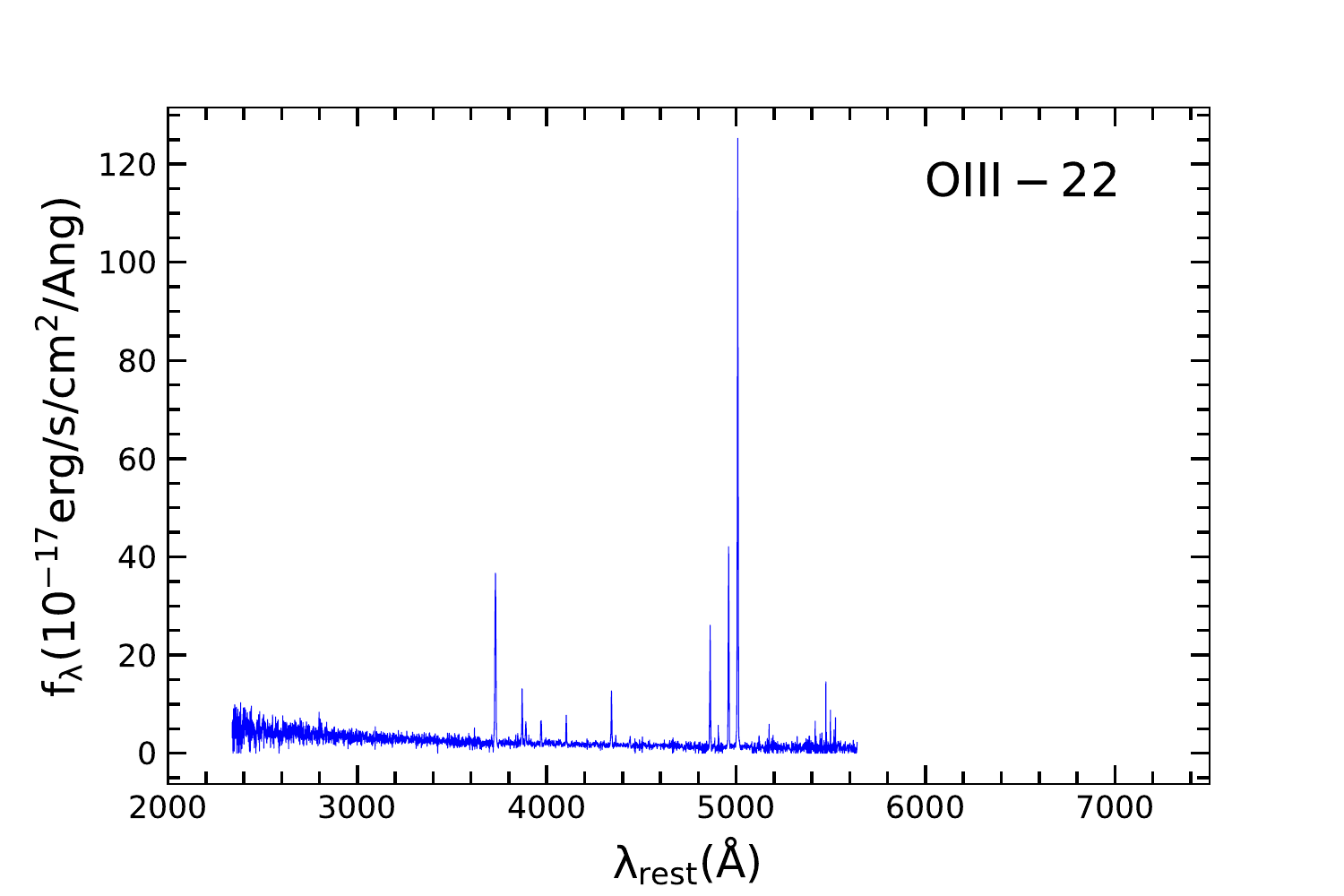}

\caption{Rest-frame spectra of 22 [OIII] emitters. The spectra are taken from SDSS DR14. }
\label{fig:spec3}
\end{figure}

\end{document}